\documentclass[12pt,preprint]{aastex}

\usepackage{amsmath}
\usepackage{apjfonts}
\usepackage{graphicx}
\usepackage[countmax]{subfloat}

\newcommand{\acet}{C$_{\rm 2}$H$_{\rm 2}$}
\newcommand{\hmol}{H$_{2}$}

\newcommand{\cop}{HCO$^{+}$}

\slugcomment{}

\shorttitle{Molecules in an AGN disk}
\shortauthors{Harada et al.}

\begin{document}
\bibliographystyle{apj}

\title{Modeling The Molecular Composition in an AGN Disk}


\author{Nanase Harada}
\affil{Department of Physics, The Ohio State University, Columbus, Ohio, 43210 U. S. A.}
\affil{Max Planck Institute for Radio Astronomy, Bonn 53121, Germany}
\author{Todd A.~Thompson}
\affil{Department of Astronomy, The Ohio State University, Columbus, Ohio, 43210 U. S. A.}
\affil{Center for Cosmology and Astro-Particle Physics, The Ohio State University, Columbus, Ohio, 43210 U. S. A.}

\and
\author{Eric Herbst}
\affil{Departments of Chemistry, Astronomy, and Physics, University of Virginia, Charlottesville, Virginia 22904 U. S. A.}


\begin{abstract}
We use a high-temperature chemical 
network to derive the molecular abundances in axisymmetric accretion
disk models around active galactic nuclei (AGNs) within 100 pc using
simple radial and vertical density and temperature distributions 
motivated by more detailed physical models. 
We explore the effects of X-ray irradiation and cosmic ray ionization
on the spatial distribution of the molecular abundances of CO, CN, CS, HCN, HCO$^{+}$, 
HC$_{3}$N, C$_{2}$H, and c-C$_{3}$H$_{2}$ using a variety of plausible disk structures.
These simple models have molecular regions with a layer of X-ray dominated regions, a midplane without the strong influence of X-rays, 
and a high-temperature region in the inner portion with moderate X-ray flux where families of polyynes (C$_{\rm n}$H$_{2}$) and cyanopolyynes can be enhanced.
For the high midplane density disks we explore, we find that cosmic rays produced by supernovae do not significantly
affect the regions unless the star formation efficiency significantly exceeds that of the Milky Way. 
We highlight molecular abundance observations and ratios that may distinguish among theoretical models of the density distribution in  AGN disks.  
Finally, we assess the importance of the shock crossing time and the accretion time  relative
to the formation time for various chemical species.  
 Vertical column densities are tabulated for a number of molecular species at both the characteristic
 shock crossing time and steady state.  
Although we do not attempt to fit any particular system or set of observations, 
we discuss our models and results in the context of 
the nearby AGN NGC 1068.
\end{abstract}

\section{Introduction}
The growth of active galactic nuclei (AGNs) is an important key to the evolution of galaxies, and its connection with star formation inside  galaxies remains of great interest in astrophysics. 
\begin{sloppypar}
There are many unknowns in our knowledge of the accretion process and the structures of disks, such as mass accretion rates and disk thicknesses on scales ranging from fractions of a parsec to tens of parsecs. Other unknowns include the conditions for  star formation in these environments, feedback from  star formation, and the rate of star formation on parsec scales.  Several models have been proposed to explain these quantities \citep[e.g.,][]{1990Natur.345..679S,2003MNRAS.339..937G,2003MNRAS.341..501S,2005ApJ...630..167T,2007ApJ...662...94C,2007ApJ...661...52K,2003astro.ph..7084L,2007ApJ...660..276W,2008A&A...491..441V,2008ApJ...681...73K,2010MNRAS.403.1801S,2010MNRAS.407.1529H}.  
\end{sloppypar}
 
The detection of molecules in external galaxies can help to characterize their star formation regions and the physical conditions of the gas.  Star formation takes place in a molecular region as opposed to one that is mainly atomic or ionized.  Carbon monoxide has been a main tool to obtain the molecular mass, whether in galactic clouds or extragalactic sources.  In addition, many other molecular species have been detected in extragalactic sources, and show different features in different types of galaxies such as AGN-dominated galaxies, (ultra-) luminous infrared galaxies ((U)LIRGs), and starburst galaxies.  For example, \citet{2005AIPC..783..203K}  introduced the HCN/\cop($1-0$)~line ratio as a diagnostic between AGN- and starburst-dominated galaxies. Based on the observation by \citet{1994ApJ...426L..77T}, \citet{2004A&A...419..897U} suggested a high HCN/CO intensity ratio in the AGN-containing galaxy NGC 1068.
Unlike AGNs and starburst galaxies, (U)LIRGs such as Arp 220 and NGC 4418  seem to have more abundant complex organic species such as C$_{2}$H$_{2}$ and HC$_{3}$N \citep{2007ApJ...659..296L,2007A&A...475..479A,2010A&A...515A..71C,2011A&A...528A..30C}.

The observed intensities are affected both by the abundances of molecular species and by the physical conditions, which in turn affect radiative transfer. The 3D-radiative transfer simulation of NGC 1068 by \citet{2007ApJ...671...73Y} suggested that the higher line ratio of HCN/\cop~in AGN-dominant galaxies is caused by the difference in chemical abundances rather than by excitation conditions. Chemical abundances are themselves affected by different physical parameters such as the ionization rate, temperature, and density. \citet{1996A&A...306L..21L} examined the effect of X-rays to explain the enhanced HCN/CO abundance ratio in NGC 1068 and the Galactic Center. \citet{2005A&A...436..397M} developed  steady-state gas-phase chemical models of  photon-dominated regions (PDRs) and X-ray dominated regions (XDRs), and worked out an extensive grid of such models to characterize extragalactic and galactic sources  \citep{2007A&A...461..793M}. \citet{2008A&A...488L...5L} examined the effect of mechanical heating to enhance the HCN abundance in (U)LIRGs. \citet{2010ApJ...720..226P}  proposed an  elevated cosmic-ray ionization rate as a possible dominant mechanism in ULIRGs to create cosmic-ray dominated regions (CRDRs),  and its effect is discussed in \citet{2011A&A...525A.119M} and \citet{2011MNRAS.tmp..539P}.  The spatial dependence of the effect of X-rays in NGC 1068 was studied by  \citet{2003A&A...412..615G}, who modeled the  H$_{2}$ ro-vibrational emission and the CO rotational emission using their X-ray irradiated model, which they find can explain the overall shape of the observed double-peaked emission.

 In this paper, we present detailed model calculations of the molecular abundances of a large number of species in the accretion disk of an AGN, using a black hole mass of 
$M_{\rm BH}=10^7 M_{\odot}$, and assuming the gas density estimated from a Toomre-stable disk (see Section \ref{section:density}~for details)  out to $\sim$ 100 pc from the AGN core. This type of analysis should help us to use molecules as better probes of the physical conditions in AGN disks.  In principle, when combined with chemical modeling and radiative transfer, molecular line observations can powerfully constrain the disk structure, density profile, and temperature profile on parsec to tens of parsecs scales in AGN disks, particularly when very high resolution images from ALMA become available.
 Unlike earlier authors,  we focus our analysis on the radial and height dependence of molecular abundances,  which to the best of our knowledge has never been done for AGN disks except for CO \citep{2003A&A...412..615G} and \hmol~\citep{2003A&A...412..615G,2009ApJ...702...63W,2011ApJ...730...48P}.  The radial dependences of molecular abundances in the disk midplane were examined in our previous paper \citep{2010ApJ...721.1570H}. Adopting parameterized vertical density structures, we add here the height dependence, an explicit calculation of X-ray  and cosmic-ray ionization rates, and a self-consistent determination of the gas temperature including molecular line cooling.  As a function of density and temperature throughout the disk we provide chemical abundances and average column densities for input to future radiative transfer calculations, and as a guide to interpreting current and future molecular line observations of AGN disks.

The remainder of the paper is organized as follows.  In Section \ref{model}, we discuss the range of physical conditions and parameters such as the temperature, density, and ionization rate used in our models, while in Section \ref{results}, we present our results. Section \ref{discussion} contains a comparison between our results and observations, especially for molecules such as HCN, CN, and HCO$^{+}$, which have been observed at higher spatial resolution than other molecular species in NGC 1068.  In addition, we include a discussion on how our results change as a function of different AGN disk models, and in different types of galaxies.    We summarize our results in Section \ref{conclusion}.

\section{Model}
\label{model}

The chemical network used here is the OSU high-temperature network, which can be employed up to a temperature of 800 K.  With certain exceptions, such as the formation of H$_{2}$ on granular surfaces, this is a gas-phase network.  This network contains X-ray and cosmic-ray ionization and photodissociation from Lyman-Werner photons and Lyman-$\alpha$ photons. Doubly ionized species are not included since we are interested in regions that are mostly molecular, where these species are assumed to have very low abundance. The details of this network are described in \citet{2010ApJ...721.1570H} and \citet[][; for erratum]{2012ApJ...756..104H}, and the network is publicly available at  http://www.physics.ohio-state.edu/$\sim$eric, and it will be updated on KIDA (KInetics Database for Astrochemistry)\footnote{http://kida.obs.u-bordeaux1.fr/models}, kida.uva.2012. 
As initial conditions, we used  the so-called ``low metal'' elemental abundances in which some depletion on grains is assumed for metals.  There is also the possibility of higher abundances  of heavier elements, approaching those of solar abundances. The initial abundances both in the low-metal and solar abundance cases are shown in Table \ref{init}.  Physical conditions are estimated via methods described in the following sections. Important symbols are summarized in Table \ref{symbols}.

\subsection{Density Distributions}
\label{section:density}

 As discussed in our earlier paper \citep{2010ApJ...721.1570H}, the gas density, $\rho$, at the midplane is assumed to be  \citep{2005ApJ...630..167T} 
\begin{equation}
\label{density}
\rho = \frac{\Omega^{2}}{\sqrt{2} \pi G Q},
\end{equation} 
where the Keplerian rotation frequency $\Omega$ is given by
\begin{equation}
\Omega = \sqrt{\frac{GM_{\rm BH}}{r^{3}}+\frac{2\sigma^{2}}{r^{2}}}.
\end{equation}
Here $M_{\rm BH}$ is the mass of the central black hole, $\sigma$ is the velocity dispersion of the stellar bulge, and  $r$ is the distance from the AGN core. We use $M_{\rm BH}$ = 1.5$\times$10$^{7}$M$_{\odot}$ and $\sigma$ = 150 km s$^{-1}$ \citep{1997Ap&SS.248..261G,2002ApJ...574..740T}. For a gaseous disk, the Toomre stability parameter $Q$ \citep{1964ApJ...139.1217T}  is expressed as
\begin{equation}
Q \equiv \frac{v_{s} \kappa}{\pi G \Sigma},
\end{equation}
where $v_{s}$ is the sound speed, $\kappa$ is the epicyclic frequency, and $\Sigma$ is the surface density. For $Q \lesssim$ 1 the disk is gravitationally unstable, while for $Q \gtrsim$ 1, the disk is stable. Many authors have proposed that the value of $Q$ should self-regulate to a value close to unity \citep[e.g.,][]{1978AcA....28...91P, 2001ApJ...553..174G}. 
Observationally,  $Q$ does not differ from unity by more than a factor of a few over the entire galactic disk according to 
\citet{1972ApJ...176L...9Q}, \citet{1989ApJ...344..685K}, and \cite{2001ApJ...555..301M}.
For {the accretion disk of} NGC 1068, the value of the Toomre parameter that \cite{2008A&A...491..441V} suggest is $\sim5$ based on the observations of \citet{2009ApJ...696..448H} and \citet{2007ApJ...671.1388D} on 30\,pc scales. There are also opposing views in the literature suggesting that $Q$ can be much higher than unity \citep[e.g., $Q\sim100$ in the Galactic Center;][]{2008A&A...491..441V}.  
 Additionally, \citet{2007ApJ...660..276W} questioned the use of the $Q$-parameter as a criterion at all.
Nevertheless, we use two values of $Q$ for our models, $Q$ = 1 and 5.  
The midplane densities for $Q$ = 1 are $n = 2 \times10^{8}$\,cm$^{-3}$, $1\times10^{6}$\,cm$^{-3}$, and $1\times10^{4}$\,cm$^{-3}$ at $r $= 1, 10, and 100 pc, respectively, while for $Q$ = 5, they are simply a factor of 5 lower (see equation \ref{density}). 

The density at height $z$ from the midplane for a thin disk in hydrostatic equilibrium is given by \citet{2002apa..book.....F}\footnote{
The vertical density
distribution adopted in equation (\ref{bar}) is only a rough
approximation, and self-consistent models may yield a different
functional form for $\rho(z)$ \citep[see e.g., ][]{2005ApJ...630..167T,2007ApJ...661...52K,2007ApJ...662...94C}.} to be
\begin{equation}
\label{bar}
 \rho(r,z) = \rho(r,0)\exp\left[-\frac{1}{2} \left (\frac{z}{h} \right)^{2} \right],
\end{equation}
where $h$ is the scale height, which can be defined by the relation $h \equiv \Sigma(r)/\rho(r,z=0),$\footnote{For Model 4 and the torus portion in Model 1, discussed later in this section, $h$ is defined in a different way, and the surface density is lower than in other models.} with $\Sigma$  the surface  density as a function of radius.    The ratio of scale height to radius can be obtained \citep{2005ApJ...630..167T} from the expression
\begin{equation}\label{hr_q}
\frac{h}{r} = \frac{f_{\rm g} Q}{2^{3/2}},
\end{equation}
where $f_{\rm g}$ is the fraction of the mass of gas over the total mass (gas + stellar).  Since the gas fraction is uncertain, so is the disk scale height, and we varied $h/r$ in our calculations, in developing four different disk models.   These models, shown in Table~\ref{diskst}, are defined in three radial regions:  $<1$\,pc, $1-3$\,pc, and $3-100$\,pc.

 Model 2 is a relatively thin disk as in \citet{2005ApJ...630..167T}, in which equation~(\ref{bar}) is used for the dependence of density on height.  In this model, the value of $h/r$ first increases quadratically with radius then increases linearly.   This model is motivated in part by the recent hydrodynamic simulations of star-forming disks with a central black hole by \citet{2010MNRAS.407.1529H}, who show $h/r$ in their models as a function of various parameters, including the gas fraction, surface density, and the central black hole mass. Figure 10 in their paper shows that $h/r\lesssim0.1$ on $1-100$\,pc scales in most cases, although there are fairly large fluctuations. 

\begin{figure*}
\centerline{\includegraphics[angle=90,width=8.5cm]{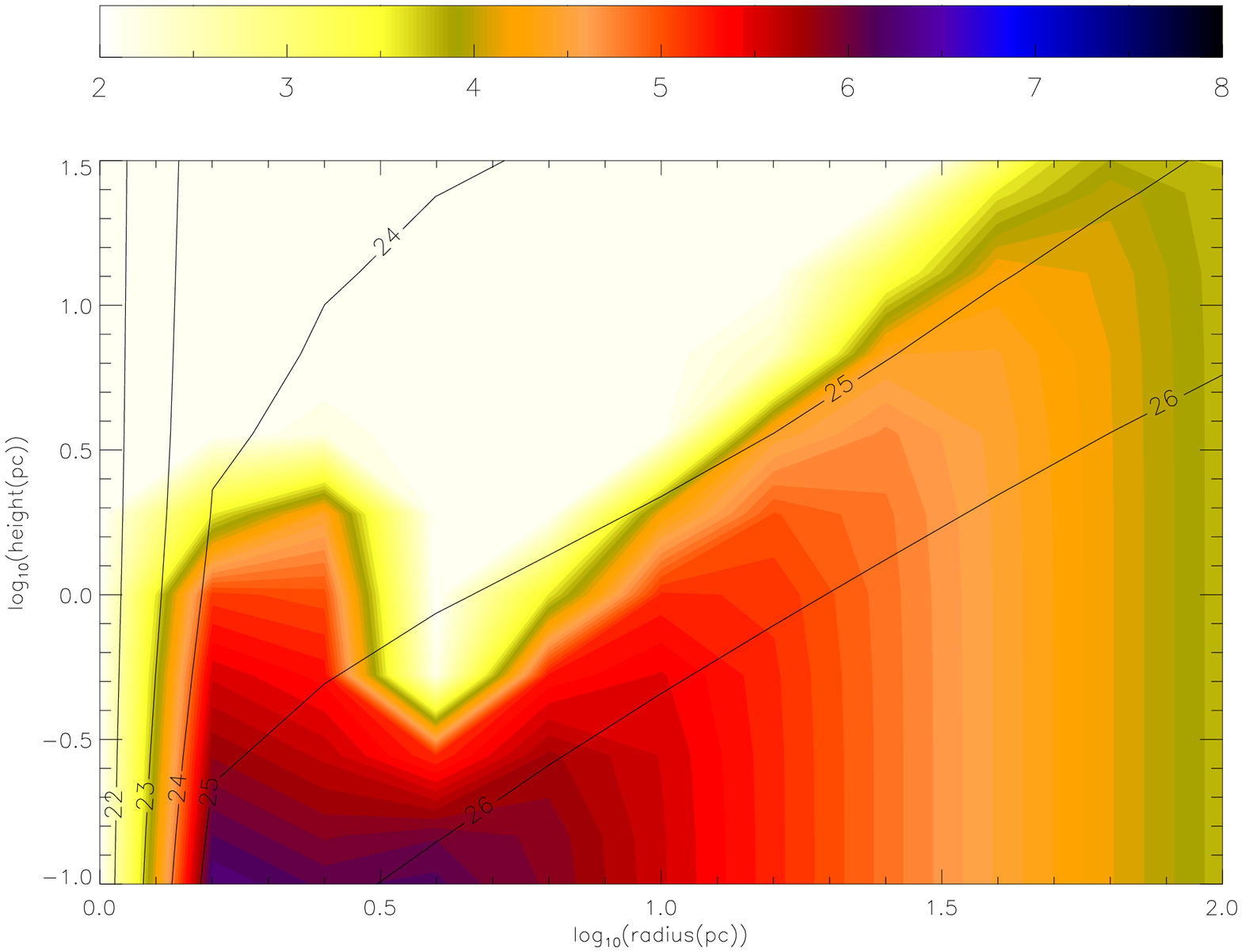}
\includegraphics[angle=90,width=8.5cm]{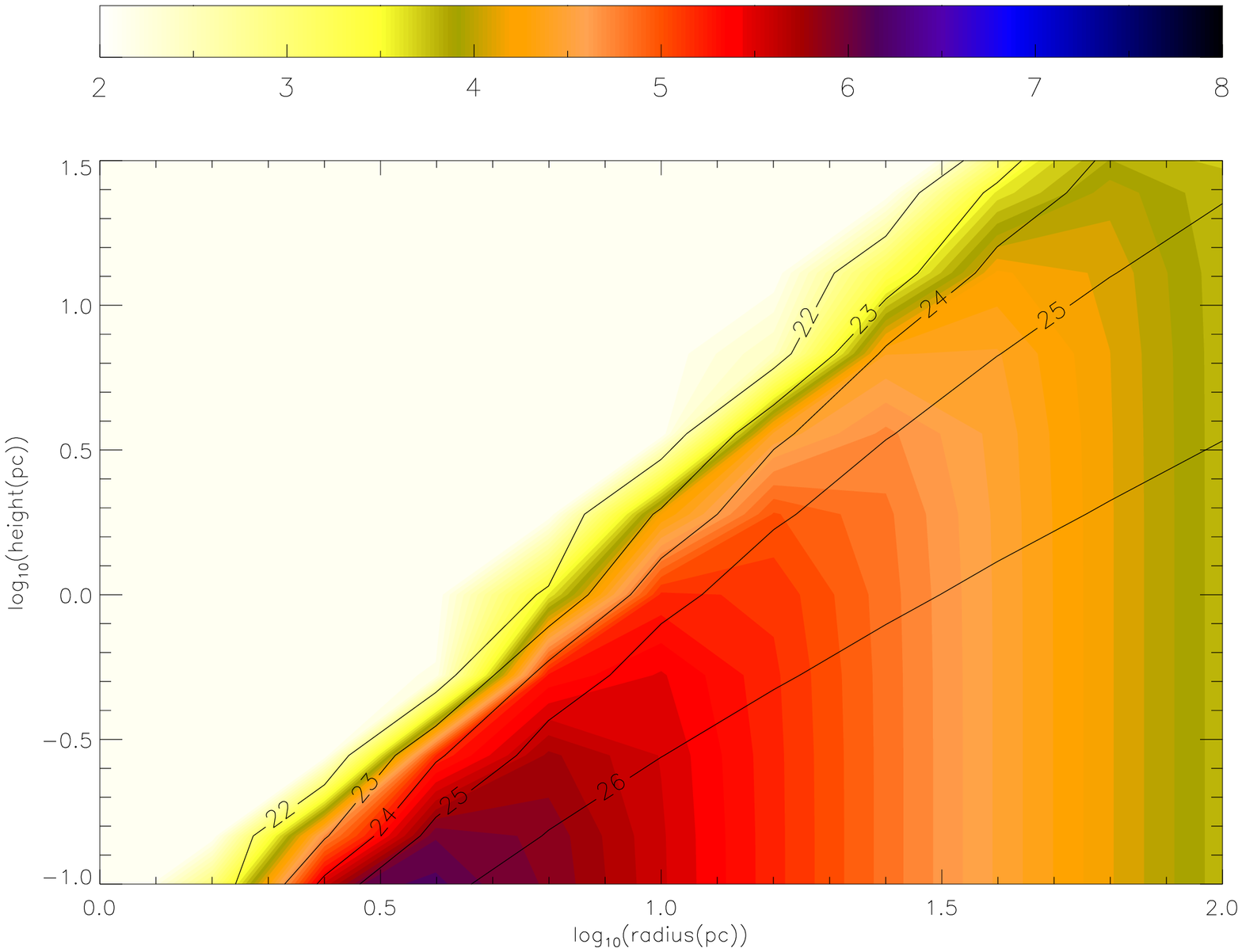}}
\centerline{\includegraphics[angle=90,width=8.5cm]{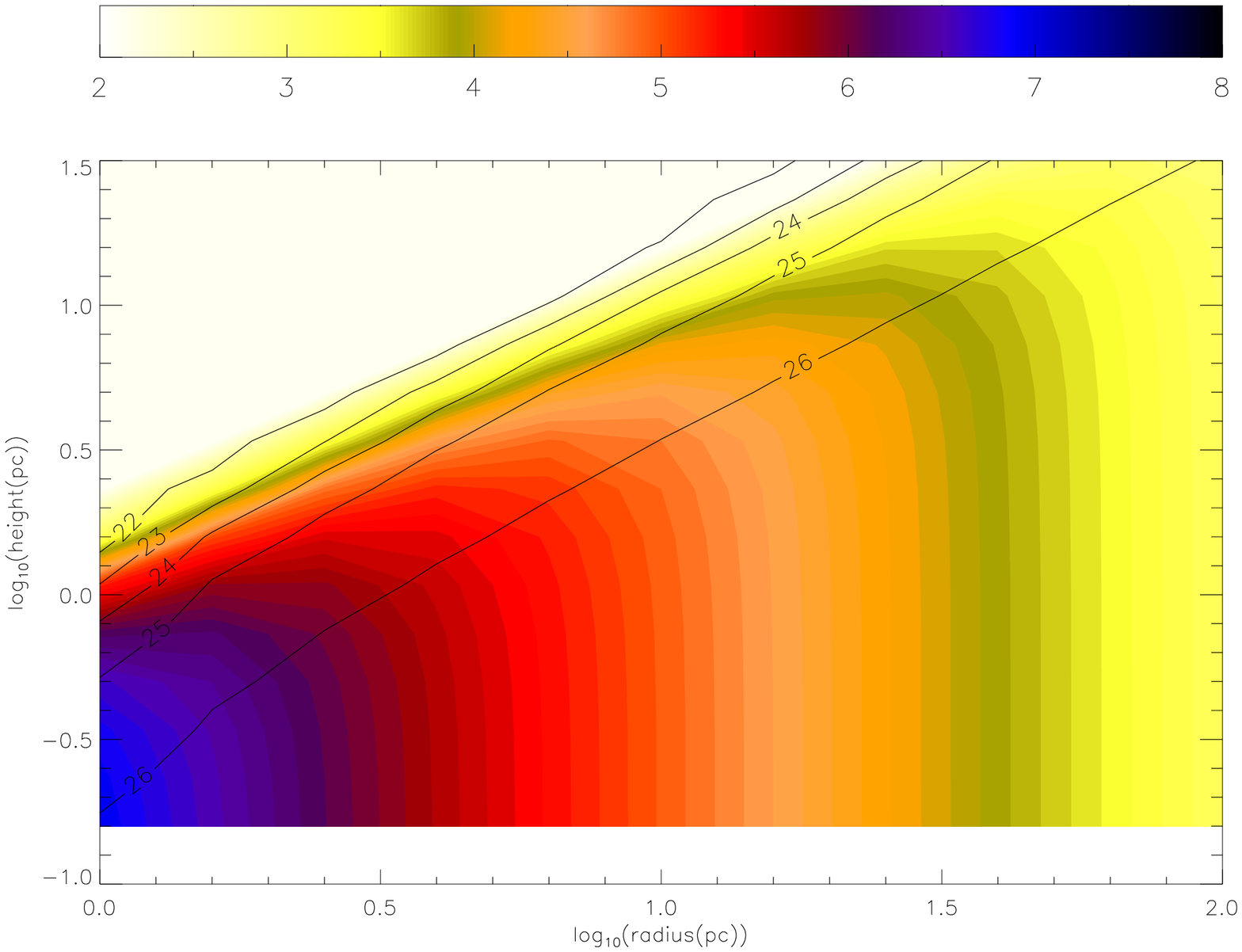}
\includegraphics[angle=90,width=8.5cm]{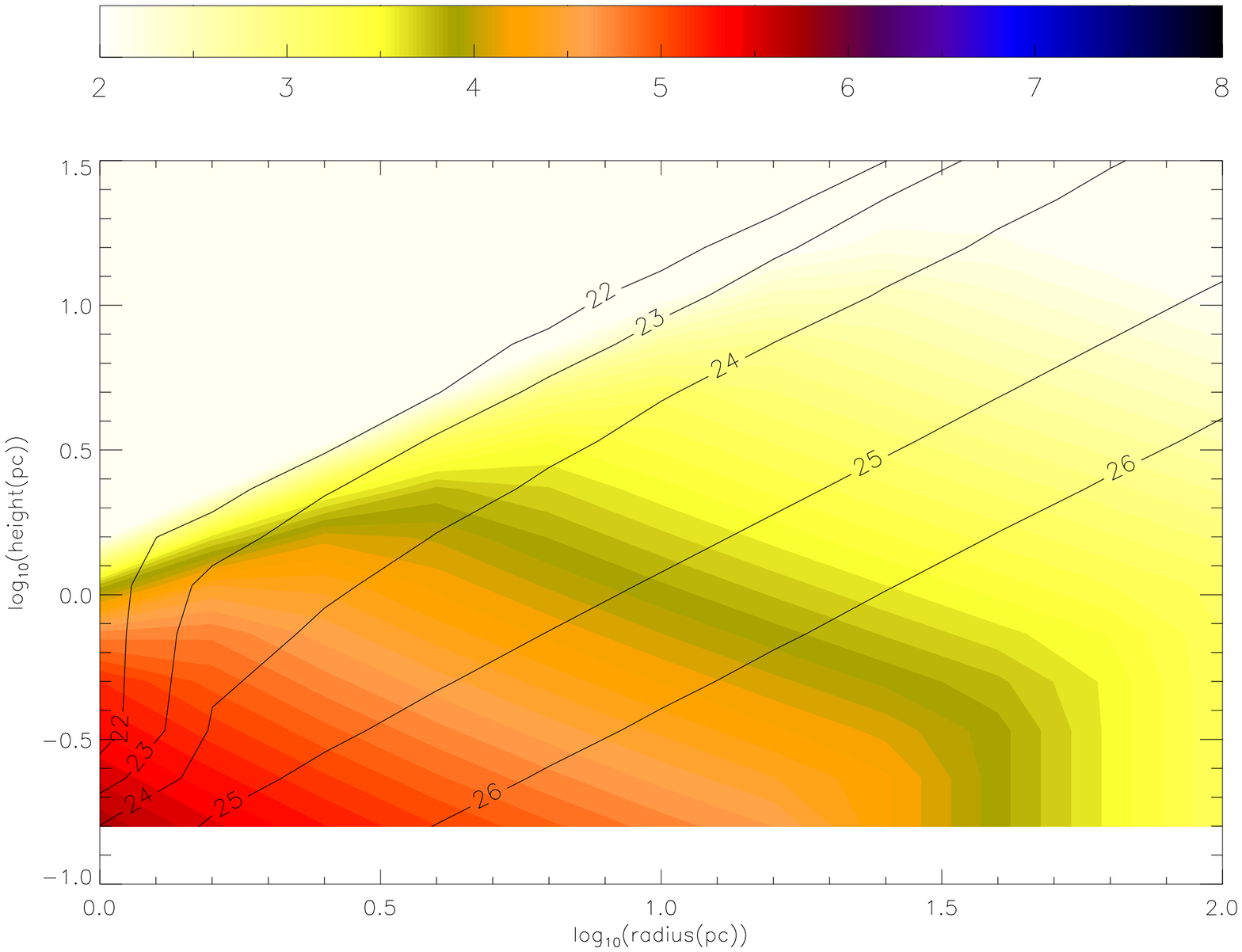}}
\caption{ The radial and height dependence of density (color scale) and logarithms of column densities from the X-ray source log$_{10}N(cm^{-2})$ (contours). The radius, height, and the densities are all shown in logarithmic scale. {\it Upper left panel:}
Model 1, $Q$ =1; {\it Upper right panel:} Model 2, $Q$ = 1;
{\it Lower left panel:} Model 3, $Q$ = 5; {\it Lower right panel:} Model 4, $Q$ = 5.}
\label{fig:n}
\end{figure*}

The other three models we consider contain tori of assorted radial widths.   \citet{2004Natur.429...47J} observed a torus of $ h/r \sim 0.6$ in NGC 1068 on a few pc scales,  and our Model 1 is a thin disk model analogous to Model 2  but with a torus of $h/r =0.5$ located at $1-3$\,pc.  A relatively thick torus has also been suggested by \citet{2009ApJ...696..448H}, with $h/r>1$ on 30 pc scales based on their studies of  ro-vibrational H$_{2}$ emission. This warm H$_{2}$ traces molecular gas at a few thousand Kelvin, and it is not certain if the warm gas is tightly coupled to the colder midplane of the disk.  Nonetheless, we also consider geometrically thick torus cases in Models 3 and 4, with $h/r = 0.5$ for radii of $1-100$ pc. We use the height dependence expressed in eq.~(\ref{bar})~for Model 3. However, this is a thin disk approximation, and may not express the correct height dependence of density in a thick torus, which could also be affected by the irradiation from the AGN core.  We therefore employed the height dependence of density from \citet{2007ApJ...662...94C} by fitting their numerical expression to $\rho(z)$ = $\rho(0)$ for $z <0.02h$,  $\rho(z) \propto $ $r^{-1.23}$ for $0.02h < z < h$, and $\rho(z) \propto \exp(-\frac{z^{2}}{2 h^{2}})$ for $z > h$ (see also footnote 2).   Fig \ref{fig:n} shows the densities of all the models plotted with contours representing column densities from the X-ray source. The total molecular mass is $\sim 7\times 10^{7}M_{\odot}$ for Model 1 and 2, $\sim 3\times 10^{6}M_{\odot}$ for Model 4 within 30 pc. The total masses of Model 1 and 2 are larger than those in the hydrodynamic models of \citet{2009ApJ...702...63W}, where the total molecular mass is $\sim 1.5 \times 10^6 M_{\odot}$when the black hole mass is 1.3$\times 10^7 M_{\odot}$.

\subsection{Temperature}
\label{section:temp}
We calculate the gas temperature from the heating and cooling rate; the coupling with the dust temperature is also included. For the range of  dust temperatures in the AGN disk, we used the blackbody approximation, determined by the luminosity of the central AGN:
\begin{equation}\label{agn_T}
T_{\rm dust}(r)\sim 750 \left(\frac{r}{\rm pc} \right)^{-1/2}\left (\frac{L_{\rm AGN}}{2\times 10^{45}~{\rm erg~s^{-1}}}\right)^{1/4} {\rm K},
\end{equation}
where the normalizing luminosity of 2$\times$10$^{45}$ erg s$^{-1}$ derives from the observation by \citet{2006ApJ...640..612M} of NGC 1068.   Using this formula, $T_{\rm dust}=750$, 240, and 75 K at $r = 1$, 10, and 100\,pc, respectively. 
 To calculate the gas temperature, we include heating from both X-rays and cosmic rays. Dust-gas collisions provide gas heating or cooling, depending on the dust-gas temperature ratio. The cooling rate is determined by molecular line cooling. Although photons produced by molecular lines can in principle escape in any direction, we make the simplifying assumption that the cooling line photons always escape from the AGN disk vertically, which reduces the problem to a plane-parallel geometry at every radius. Even so, the calculation of the gas temperature is complicated by the fact that the chemical abundances, which determine the cooling rate, and temperature are interdependent.   Our procedure is as follows.  We first calculate the chemical abundances assuming that the gas temperature is $T(r)=T_{dust}(r)$ everywhere.  Given these abundances, we then calculate the temperature in thermal equilibrium, with the heating and cooling rates described in Appendix \ref{ap_temp}.  Once the temperature is determined, from this initial step, the abundances and cooling rates are updated, and then the temperature is recomputed.  This procedure is iterated until convergence.

\begin{figure*}
\centerline{\includegraphics[angle=90,width=8.5cm]{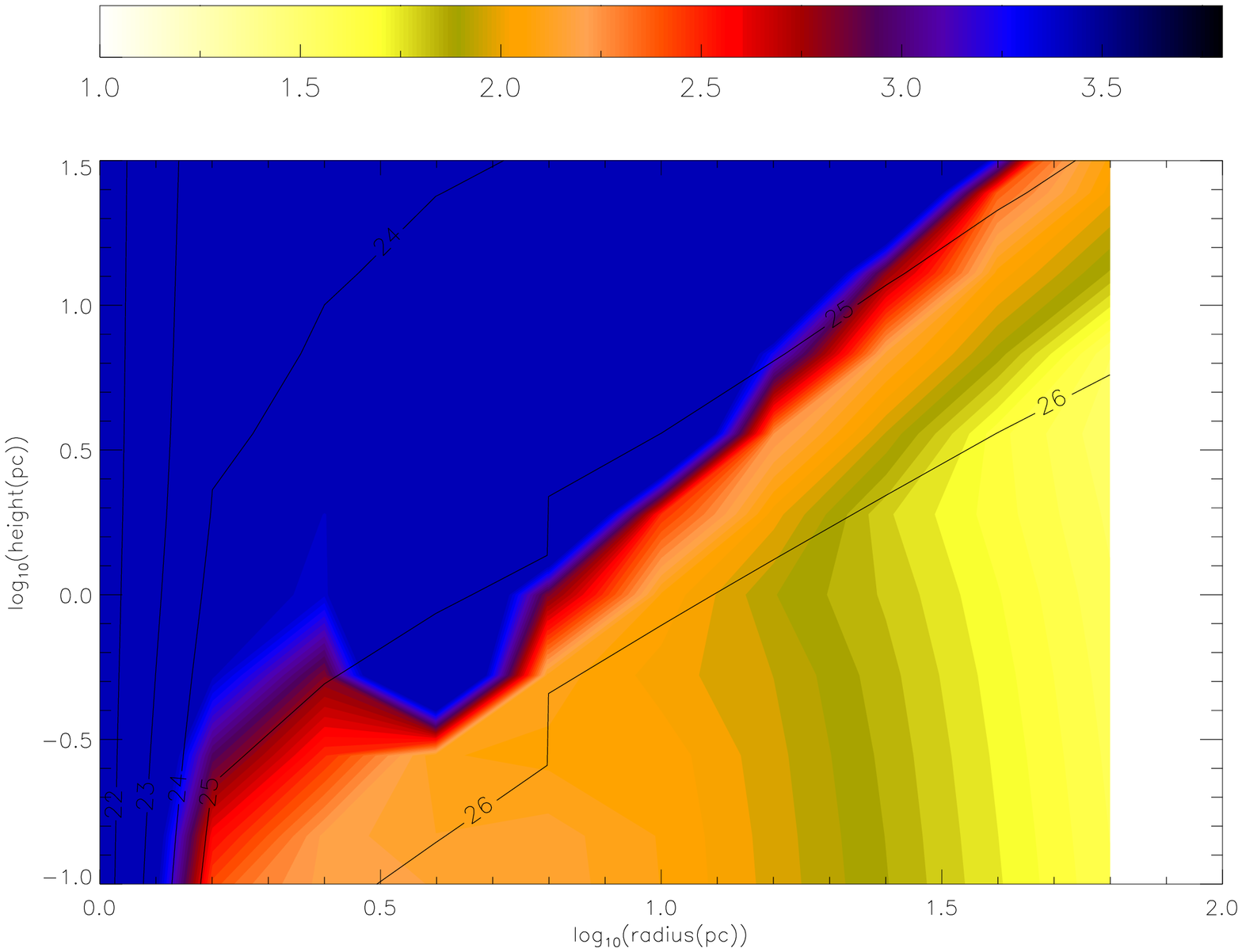}
\includegraphics[angle=90,width=8.5cm]{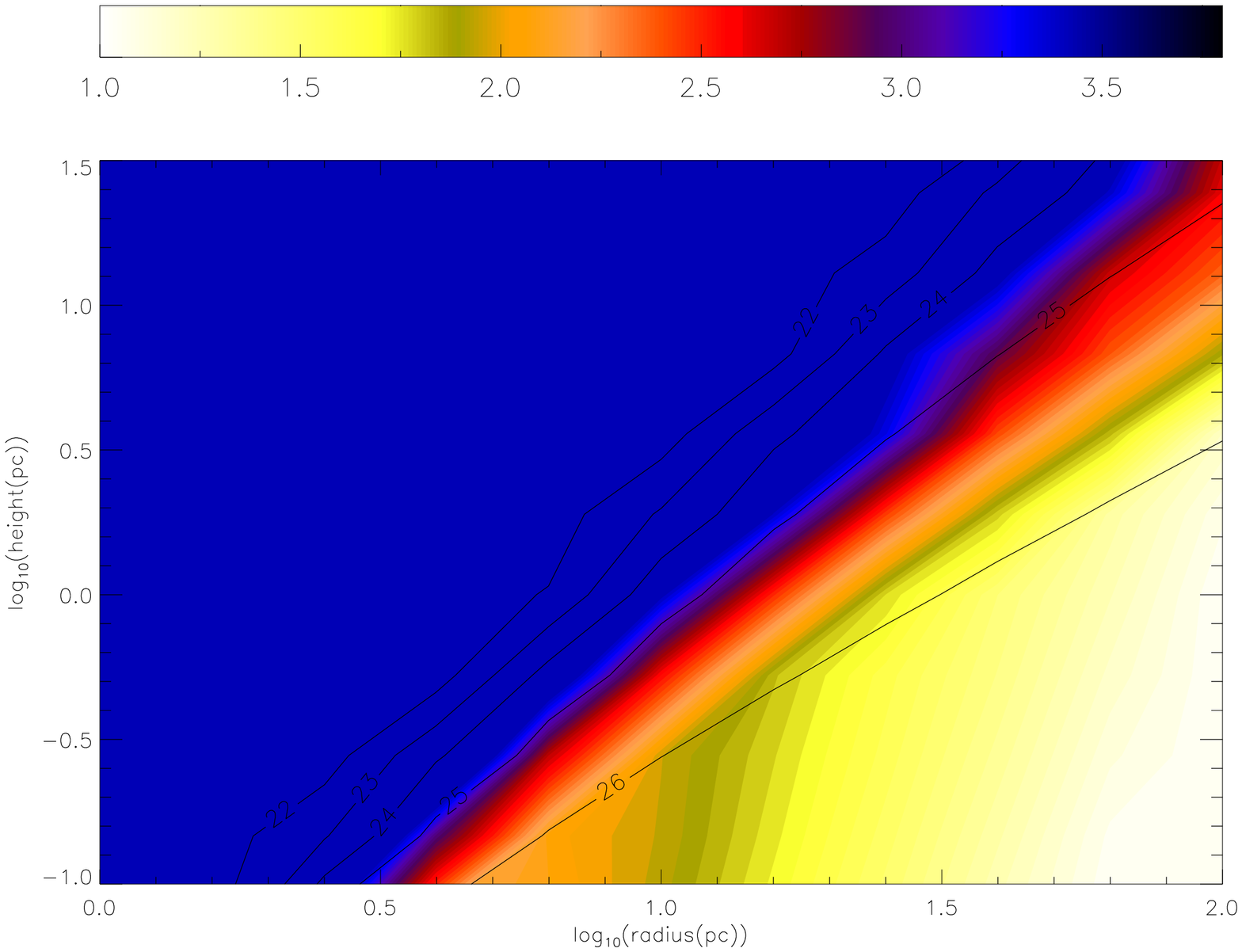}}
\centerline{\includegraphics[angle=90,width=8.5cm]{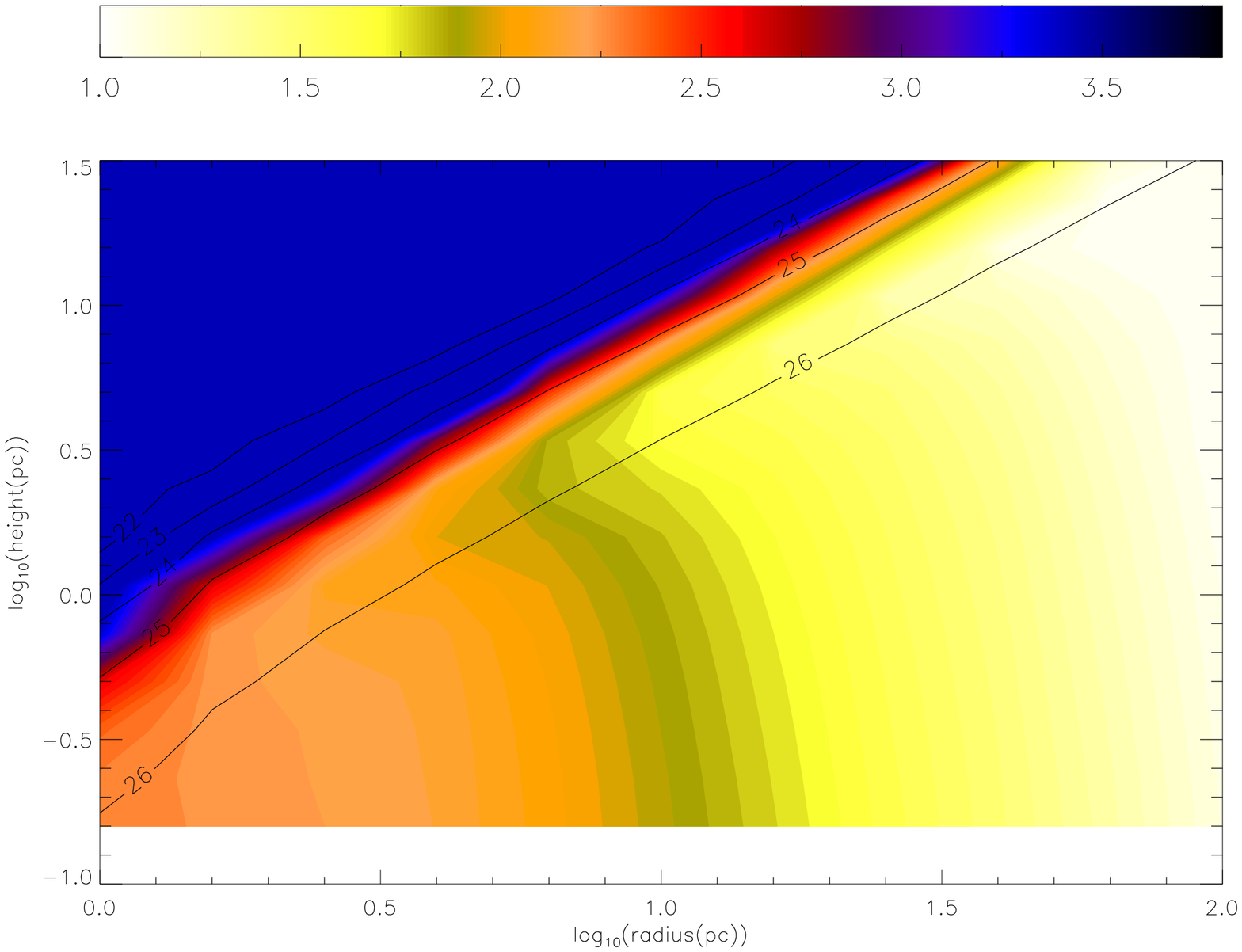}
\includegraphics[angle=90,width=8.5cm]{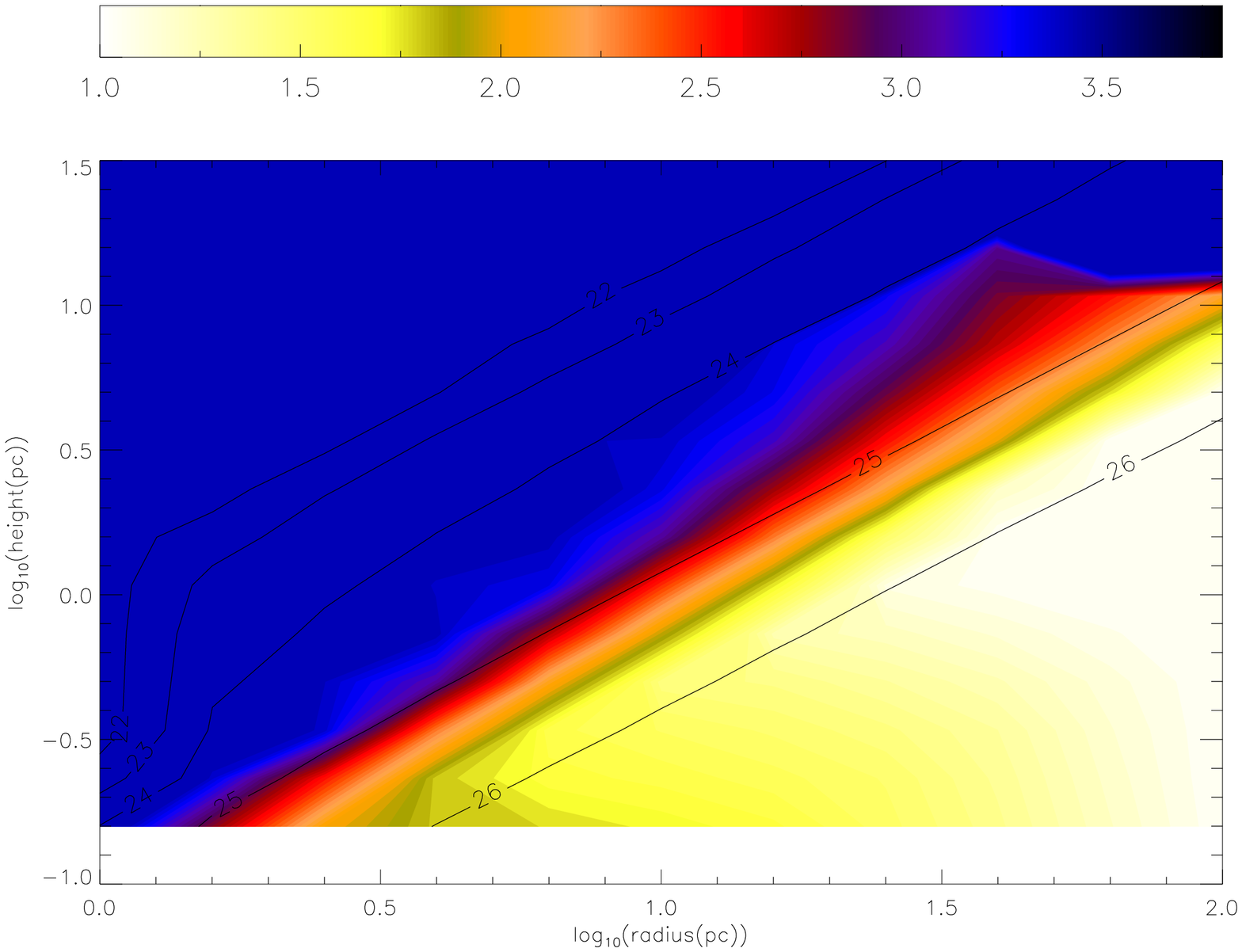}}
\caption{ The radial and height dependence of temperature $T$(K) (color scale) and logarithms of column densities from the X-ray source log$_{10}N(cm^{-2})$ (contours). The radius, height, and temperatures are all shown in logarithmic scale. {\it Upper left panel:}
Model 1, $Q$ =1; {\it Upper right panel:} Model 2, $Q$ = 1;
{\it Lower left panel:} Model 3, $Q$ = 5; {\it Lower right panel:} Model 4, $Q$ = 5.}
\label{fig:T}
\end{figure*}

 Figure~\ref{fig:T} shows the gas temperature at each part of the disk in Models 1-4. In the regions that are irradiated strongly by X-rays, the temperature can reach $T>10^4\,$K. For columns of $\sim10^{24}\,$cm$^{-2}$, X-rays are attenuated, and the gas is predominantly heated by other sources. When $r\lesssim5\,$pc, the gas temperature is coupled with the dust temperature because of the high density. Farther away from the AGN core, the temperature slowly decouples from the dust temperature and goes down to 10\,K at $r=100\,$pc. We calculated molecular abundances and followed the iterative procedure described above only when $T<5000\,$K since few molecules exist at higher temperatures. Observations of ammonia by \citet{2011A&A...529A.154A}  suggest warm $\sim 80\,$K and  $\sim140\,$K molecular gas components on a 1.2 kpc scale, which implies that other sources of heating such as turbulence/shock heating may be dominant on these scales, as has been suggested by many authors. 

 Our temperature calculation was checked against the results by \citet{2005A&A...436..397M}. We reproduce the general behavior of the run of temperature in XDRs as a function of column, although \citet{2005A&A...436..397M} use a more detailed calculation of the X-ray penetration cross section than employed here, which makes X-rays in our results attenuate somewhat faster.

\subsection{Timescales}\label{section:timescale}

Although the density and temperature of our disk models are independent of time, 
several timescales, and their ratios, determine in part the abundances that might be observed in AGN disks.
The relevant timescales are the accretion time $t_{\rm acc}$, the shock crossing time $t_{\rm cross}$, and the chemistry time $t_{\rm chem}$. 
The accretion time, $t_{\rm acc}$, is the time for the medium at radius $r$  to be accreted. 
In viscous disks, $t_{\rm acc}\sim(\alpha \Omega)^{-1}(r/h)^{2}$ where $\alpha$ is the viscosity parameter. 
With $\alpha \sim 0.01 - 0.1$, $t_{\rm acc} \gtrsim$1000 $\Omega^{-1}$ for $h/r$ $\sim 0.1$. However, \citet{2010MNRAS.407.1529H} show that
the gravitational torque from the stars on the gas is likely to be the dominant angular momentum transport mechanism in disks that feed AGN.   
In such disks $t_{\rm acc} \sim |a|^{-1}\Omega^{-1}$, where $|a|$ is the fractional magnitude of the asymmetry \citep{1972MNRAS.157....1L}. 
In the simulation by \citet{2010MNRAS.407.1529H}, $|a|\sim0.01-0.3$, implying that $t_{\rm acc} \sim 3-100 \Omega^{-1}$.  

The accretion time in any model should be compared with the shock crossing time, $t_{\rm cross}$.  The medium is turbulent, and  the simulation by \citet{1998ApJ...508L..99S} shows that the dissipation time of shock waves should be on the order of $\Omega^{-1}$. The  velocity of these shocks can be  tens of km s$^{-1}$, and they can dissociate the molecules, effectively resetting the molecular abundances  \citep{1980ApJ...241L..47H}.  For the disk models we consider, the shock crossing time $t_{\rm cross} = 5\times10^{4}$\,yr\,\, $\times Q^{-1/2}(\frac{n}{10^{6} {\rm cm}^{-3}})^{-1/2}$ and the values in the model are 5$\times$10$^{5}$, 4$\times$10$^{4}$, and 3$\times$10$^{3}$ yr at $r$ = 100, 10, and 1 pc.

 There are three cases possible for the relationship of these three time scales: (1) $t_{\rm chem} < t_{\rm cross} < t_{\rm acc}$, (2) $t_{\rm cross} < t_{\rm chem} < t_{\rm acc}$, and (3) $t_{\rm cross} < t_{\rm acc} < t_{\rm chem}$. If the shock waves are dissociative, then starting from our initial abundances at $t=0$, the chemical abundances calculated at $t = t_{\rm cross}$ will represent average time-independent abundances.   Conversely, if the shocks are not dissociative and molecules survive shock passage, then the chemistry calculated to $t = t_{\rm chem}$ (i.e., the true steady-state time) should be representative of the abundances as long as $t_{\rm acc}~>~t_{\rm chem}$. This assumption is valid for cases 1 and 2, but the change in physical conditions due to accretion should be considered for case 3.\footnote{We are aware that these are two extreme cases. In between there is a C-shock, which dissociates only some of the molecules.}

Starting from our initial abundances at $t=0$,  choosing the so-called low-metal values, (see Table \ref{init}), we calculate the time evolution of the abundances at each point in density and temperature in our disk models until both $t_{\rm cross}$ and $t_{\rm chem}$. Depending on the value of $t_{\rm acc}/t_{\rm chem}$ for the molecule in question, the effect of accretion is sometimes important for small enough $r$.  We discuss these cases explicitly.  The accretion in the midplane was considered in \citet{2010ApJ...721.1570H}, and its effect is discussed in later sections. As discussed in models of protoplanetary disks \citep[see e.g., ][]{1984ApJ...287..371M,2001A&A...378..192G,2004A&A...415..643I,2006ApJ...644.1202W}, the effects of vertical and radial mixing in addition to  accretion might be important, but consideration of these factors is left to future work.

\subsection{Ionization Rate from Cosmic Rays}
\label{section:cr}

Primary cosmic ray protons are an important 
source of ionization in star-forming disks, and their secondary particles such as photons, gamma rays, electrons and positrons can also cause ionizations \citep[primary electrons are less energetically 
important than primary protons and their secondaries, and thus these are neglected here, 
as in standard analogous Galactic calculations;][]{1981PASJ...33..617U}.  Cosmic rays are likely produced by 
the supernova explosions and stellar winds of massive OB stars, and thus their production
rate is directly connected with the star formation rate.  Cosmic rays can also come from the 
central AGN.

\subsubsection{Volumetric Star Formation Rate}
The star formation rate per unit volume, $\dot{\rho_{*}}$, may be written as \citep{2005ApJ...630..250K}
\begin{eqnarray}
\label{sfrvol}
\dot{\rho_{*}} & = & \nu \frac{\rho}{t_{\rm dyn}} \sim \nu \rho^{3/2} G^{1/2}\nonumber     \\
 &\simeq &2\times 10^{-3} \frac{M_{\odot}}{{\rm pc^{3}~yr}}\left(\frac{\nu}{0.01}\right)\left(\frac{n}{{\rm 10^{6}~cm^{-3}}}\right)^{3/2},
\end{eqnarray}
where $t_{\rm dyn} \sim 1/\sqrt{G \rho}$ and $\nu$ is the star formation efficiency, whose value of $\sim0.01$ in the above is suggested by the results of \citet{1998ApJ...498..541K}.  Since the star formation efficiency and cosmic ray production rate in AGN disks is uncertain, we vary $\nu$ from $10^{- 4}-10^{-2}$. The star formation rate per unit area inferred by \citet{2007ApJ...671.1388D} within 35 pc from the AGN core  in NGC 1068 is 100M$_{\odot}$ yr$^{-1}$ kpc$^{-2}$.  For the disk density models we use (see eq. \ref{density}), a star formation efficiency of 10$^{-4}$ - 10$^{-3}$ corresponds to this value of the star formation rate per unit area. Such low values for $\nu$ at small radii may be consistent with models of feedback \citep{2005ApJ...630..167T,2007ApJ...671.1388D}.

\subsubsection{Cosmic Rays from Supernovae \& Stellar Winds}\label{ion_sn}
In the Milky Way, the supernova rate is a few per century \citep{2006Natur.439...45D}, and the estimated star formation rate is $\sim1$\,M$_{\odot}$ yr$^{-1}$ \citep{2001ASPC..230....3G,1997ApJ...476..144M,2010ApJ...710L..11R}. Assuming that every 100 M$_{\odot}$ of star formation produces 1 supernova, and that a fraction $f_{\rm CR}$ of the supernova explosion energy ($E_{\rm SN}=10^{51}$\,erg) is injected into the ISM per explosion, the cosmic ray luminosity per volume ($L_{\rm CR}/V$) is directly proportional to $\dot{\rho_*}$ (eq.~\ref{sfrvol}).  The energy density of cosmic rays, which is directly connected with the cosmic ray ionization rate \citep{2010ApJ...720..226P}, can be written roughly as 
\begin{equation}
U_{\rm CR}\sim\frac{L_{\rm CR}}{V}\min[t_{\rm escape},t_{\rm pp}]
\end{equation}
where $t_{\rm escape}$ is the escape timescale due to either diffusive (as in the Galaxy)
or advective losses (as perhaps in starburst winds), and $t_{\rm pp}$ is the timescale
for inelastic proton-proton collisions (pion production).  
If escape is dominated by an outflow, then
\begin{equation}
t_{\rm escape}=t_{\rm wind}
\sim \frac{h}{v}\sim3\times10^3\,\,{\rm yr}\left(\frac{h}{\rm pc}\right)
\left(\frac{300\,{\rm km/s}}{v}\right),
\end{equation}
where $h$ is the gas scale height and $v$ is an assumed wind velocity.
This should be compared with \citep[e.g.,][]{2002cra..book.....S,2004ApJ...617..966T}
\begin{equation}
t_{\rm pp}\sim50\,\,{\rm yr}\left(\frac{10^6\,{\rm cm^{-3}}}{n}\right).
\end{equation}
Although only a rough estimate, the fact that 
$t_{\rm pp}\ll t_{\rm wind}$ suggests that in the disks we consider cosmic ray losses 
are strongly dominated by pion production losses. Hence,

\begin{eqnarray}
U_{CR}&=& f_{CR} \dot{\rho_{*}} \,f_{SN}\,E_{SN}\,t_{\rm pp}\\
&\sim&4\times10^{-9}\, {\rm erg~cm}^{-3} \left(\frac{n}{10^{6} \,{\rm cm}^{-3}}\right)^{1/2} 
\left(\frac{\nu}{0.01} \right)
\end{eqnarray}
where we have taken $f_{\rm SN}=1$\,SN/100\,M$_\odot$, $E_{\rm SN}=10^{51}$\,erg,
and $f_{\rm CR}=0.1$.
Taking the cosmic-ray energy density of our Galaxy to be $\sim2\times10^{-12}$\,erg~cm$^{-3}$
\citep{1990ApJ...365..544B},
\begin{equation}
\frac{U_{CR}}{U_{CR, Gal}}\sim2\times10^{3}\left(\frac{n}{10^{6} \,{\rm cm}^{-3}}\right)^{1/2} 
\left(\frac{\nu}{0.01} \right).
\label{ucr_ratio}
\end{equation}
Although simplified, this result for $U_{\rm CR}$ as a function of $n$ is in
accord with the more detailed modelling of \citet{2010ApJ...717....1L} 
in high-density starburst galaxies (their Fig.~15).  
See \citet{2010ApJ...720..226P} and \citet{2011MNRAS.414.1705P} for a more detailed discussion of the effects of high cosmic ray ionization rates on the interstellar chemistry and thermodynamics in dense starburst galaxies.

Note that with the average density of the Milky Way 
$\langle n_{\rm MW}\rangle \sim1$\,cm$^{-3}$, equation (\ref{ucr_ratio})
implies $\frac{U_{CR}}{U_{CR, Gal}} \sim 2$. The value from this simple 
approximation is expected to be higher than the Galactic value since the 
majority of the cosmic rays escape from the Galaxy via diffusion before 
energetic losses \citep[e.g.,][]{2000ApJ...537..763S}
. Taking the galactic value of the
cosmic-ray ionization rate to be $1\times10^{-17}$\,s$^{-1}$, we have that
\begin{equation}\label{eq:ion_sn}
\zeta_{\rm CR} = 2\times 10^{-14}\, {\rm s^{-1}}\left( \frac{n}{10^{6} \,{\rm cm}^{-3}}\right)^{1/2} 
\left(\frac{\nu}{0.01} \right),
\end{equation}
where we have again assumed $f_{\rm CR}=0.1$ and $E_{\rm SN}=10^{51}$\,erg. 
The value of the cosmic-ray ionization rate in the midplane with $\nu=0.01$ is $\zeta_{\rm CR}=10^{-13}$~s$^{-1}$ at $r=1$\,pc , 
and $10^{-15}$~s$^{-1}$ at $r=100\,$pc,   and it linearly scales as $\nu$.

Stellar winds from massive OB stars can also be a source of cosmic rays in the dense ISM of AGN disks. The power input in the Milky Way is thought to be about an order of magnitude lower than that supplied by supernovae, and so we neglect them here \citep{1982ApJ...258..860C,2002cra..book.....S}. However, even if they were to be a significant source of cosmic rays comparable to supernovae, this would only contribute at order unity to the ionization rate, and result in an overall rescaling of the constant in equation (\ref{eq:ion_sn}) since the supernova rate and massive OB star birth rate scale directly with the star formation rate.

\subsubsection{Cosmic Rays in the AGN core}\label{ion_cr_agn}
The AGN core is another possible source of cosmic rays. 
Assuming that the gamma-ray flux reported by \citet{2010A&A...524A..72L}
is hadronic (from $\pi^0$ decay), the ratio of the gamma-ray flux to the bolometric 
flux indicates that NGC 1068 possesses near the maximum attainable ratio from star formation alone
\citep{2010arXiv1003.3257L}.  Thus,  a dominant contribution to the total cosmic ray
budget from the AGN core of 1068 is  neither supported or excluded.

 Finally, even without the contribution to cosmic ray production from  supernovae or OB star-formation, there should be some minimum flux of cosmic-rays produced by radioactive nuclei.  For a minimum value, we use $\zeta_{\rm CR, min}=10^{-18}\,$s$^{-1}$ based on $^{26}$Al, as discussed in \citet{2009ApJ...690...69U}.

\subsection{Effects of stellar UV-photons}\label{ion_uv}
 When gas is irradiated by a nearby OB star, the UV-photons can dissociate molecules. If the star formation rate is high, the UV-photons can affect the chemistry significantly.  At the same time, the penetration depth of UV-photons is much less than that of X-rays or cosmic rays; thus, its effect can be minimal in a dense medium. We use a simple approximation to determine if UV photons are important or not. The rate coefficient of photodissociation by UV-photons is $k_{\rm pd}\propto G_{0}\exp(-\gamma A_{\rm V}$) where $\gamma\sim 2-3$, which means that a region with high visual extinction surrounding the star is only affected significantly quite near to the star, where we refer to the conditions as those of a photon-dominated region, or PDR. The ratio of the volume of a PDR where $A_{\rm V}< 10$  to the entire volume (the inverse of the OB stellar density)  can be estimated by 
\begin{eqnarray}
\frac{V_{PDR}}{V}&=&\dot{\rho_{*}} \frac{4\pi (N_{A_{V}10})^{3}}{3n^{3}}f_{OB} \tau_{OB} \nonumber
\\&=&2 \times 10^{-4} \left(\frac{n}{10^4{\rm cm}^{-3}}\right)^{-3/2}\left(\frac{\nu}{0.01}\right)\left(\frac{f_{OB}}{0.01/M_{\odot}}\right)\left(\frac{\tau_{OB}}{10^{7} {\rm yr}}\right).
\end{eqnarray} 
 where $N_{A_{V}10} $ is the PDR column density, for which we use a value of  $2 \times 10^{22}$~cm$^{-2}$.
When $n = 10^{6}$ cm$^{-3}$, the PDR volume only accounts for a fraction of 2$\times$10$^{-7}$ of the entire region.  If the star formation efficiency is higher, PDRs may affect the observable molecular abundances more significantly, but here we choose not to include UV photons in our model. We also ignore the UV photons from the AGN core because of the high obscuration.

\subsection{X-rays from the AGN Core}
The ionization rate caused by X-rays interacting with H$_{2}$ is given by \citet{1996ApJ...466..561M} via the equation
\begin{equation}
\zeta_{\rm X}=N_{\rm sec}\int^{E_{max}}_{E_{min}}\sigma(E)F(E)dE
\end{equation}
where the flux per unit energy, $F(E) = F_{0}(E)e^{-\tau(E)}$, $N_{\rm sec}$ is the number of secondary ionization events  caused by electrons produced by  the primary ionization, $\tau(E)$ is the optical depth, and $\sigma(E)$ (cm$^{2}$) is the primary ionization cross section at energy $E$, which is given by
\begin{equation}\label{cross_sec_pa}
\sigma(E)=2.6\times 10^{-22} E^{-8/3} {\rm cm}^{2}~~~ {\rm (1~keV\leq E \leq 7~keV} ),
\end{equation}
\begin{equation}\label{cross_sec_pa2}
\sigma(E)=4.4\times 10^{-22} E^{-8/3} {\rm cm}^{2}~~~{\rm (7~keV < E)}.
\end{equation}
We assume that $F_{0}(E)\propto E^{-0.7}$
\citep{2006ApJ...638..642B}, and we normalize the flux at each energy
from the total X-ray flux.  The penetration of X-rays can be divided into two ranges of energy, above and below 11\,keV.  For energies lower than 11\,keV, photoionization, whose cross section is given by equations (\ref{cross_sec_pa}-\ref{cross_sec_pa2}), is the dominant process limiting penetration. 
Note that although our gas-phase metal abundances are very low because of condensation onto grains, we still assume the total metal abundance is in the gas phase for the purposes of this calculation, as if the metals are not locked on grain surfaces. As noted in \citet{1983ApJ...270..119M}, the cross sections do not differ significantly even if most of the heavier elements are condensed in grains, and thus  the estimate of the cross section by \citet{1996ApJ...466..561M} above should still be valid.  However, for higher metallicity than we employ here this approximation fails and the photoionization cross section should be computed more fully.

For energies higher than 11\,keV, Thomson/Compton scattering cross section dominates. Although the ionization rate is still dominated by photoionization for a column $N_{\rm H}\lesssim10^{24}\,$cm$^{-2}$ \citep{1996ApJ...466..561M}, our models have higher midplane column densities. Since the differential cross section of Compton scattering has angular dependence, a precise calculation of the penetration would require a more complicated transport program.  In our calculation, we simply assume that the optical depth for high-energy X-rays is reduced by a factor determined via the assumptions of isotropic scattering and random walk. Until an energy of $\sim$ 100 keV, the cross section given by Klein-Nishina formula does not differ much from Thomson cross section of $\sigma_{T}= 6.7\times10^{-25}$ cm$^{2}$. Once the high-energy X-rays are reduced to 11\,keV energy, the normal one-dimensional treatment is used.

 The total X-ray luminosity from observations is not the best measure of the true luminosity because of obscuration. The intrinsic X-ray luminosity for the AGN core of NGC 1068 is estimated to be $10^{43-44}$\,erg s$^{-1}$ \citep{1997MNRAS.289..443I,2002ApJ...581..182C}. In our calculations, we use total luminosity of $6\times10^{43}$\,erg s$^{-1}$ over 1 keV to 100 keV with a photon number flux with the energy dependence of $E^{-2.0}$.

\subsection{The Relative Ionizing Effects of X-rays and Cosmic Rays}

\begin{figure*}
\centerline{\includegraphics[angle=90,width=8.5cm]{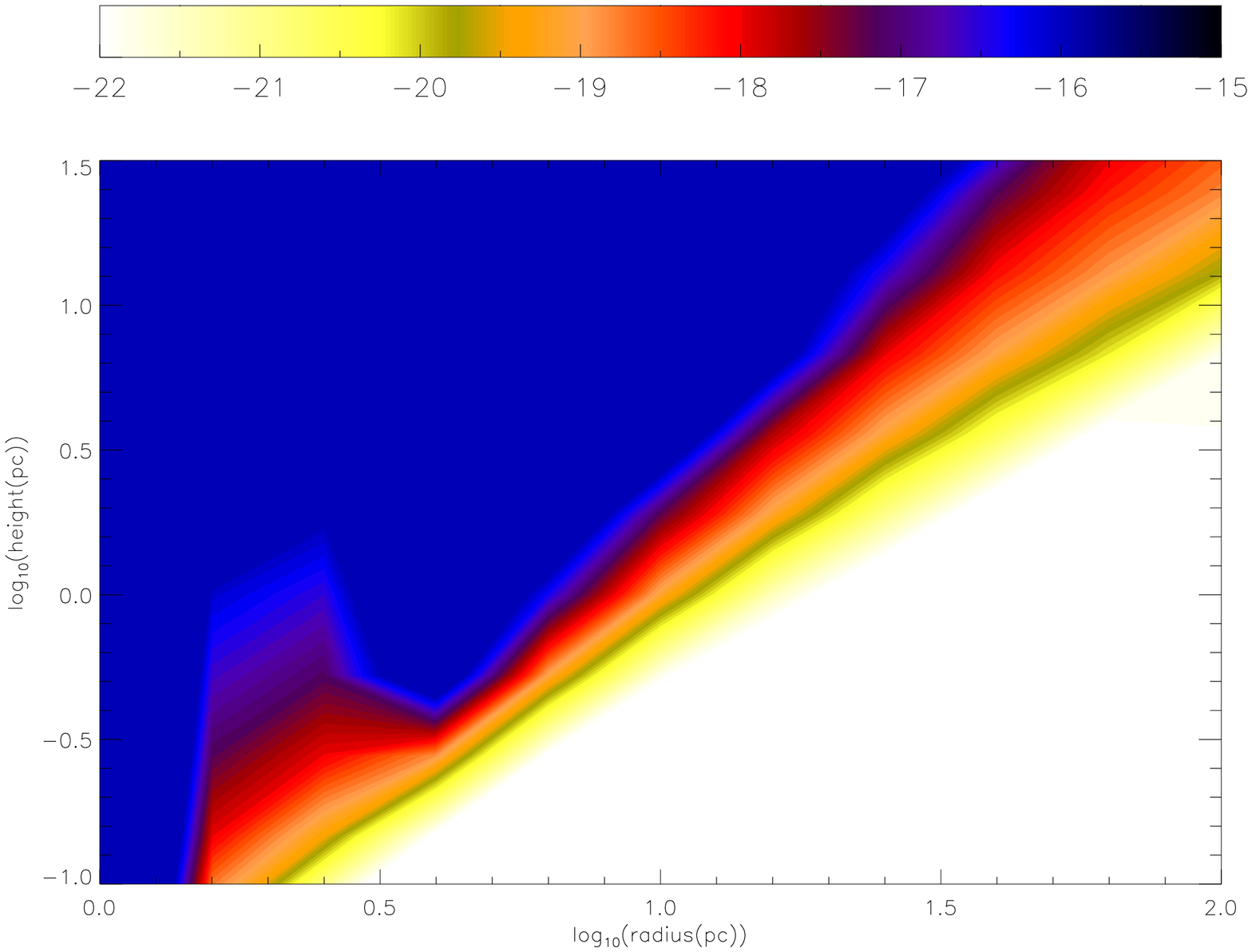}
\includegraphics[angle=90,width=8.5cm]{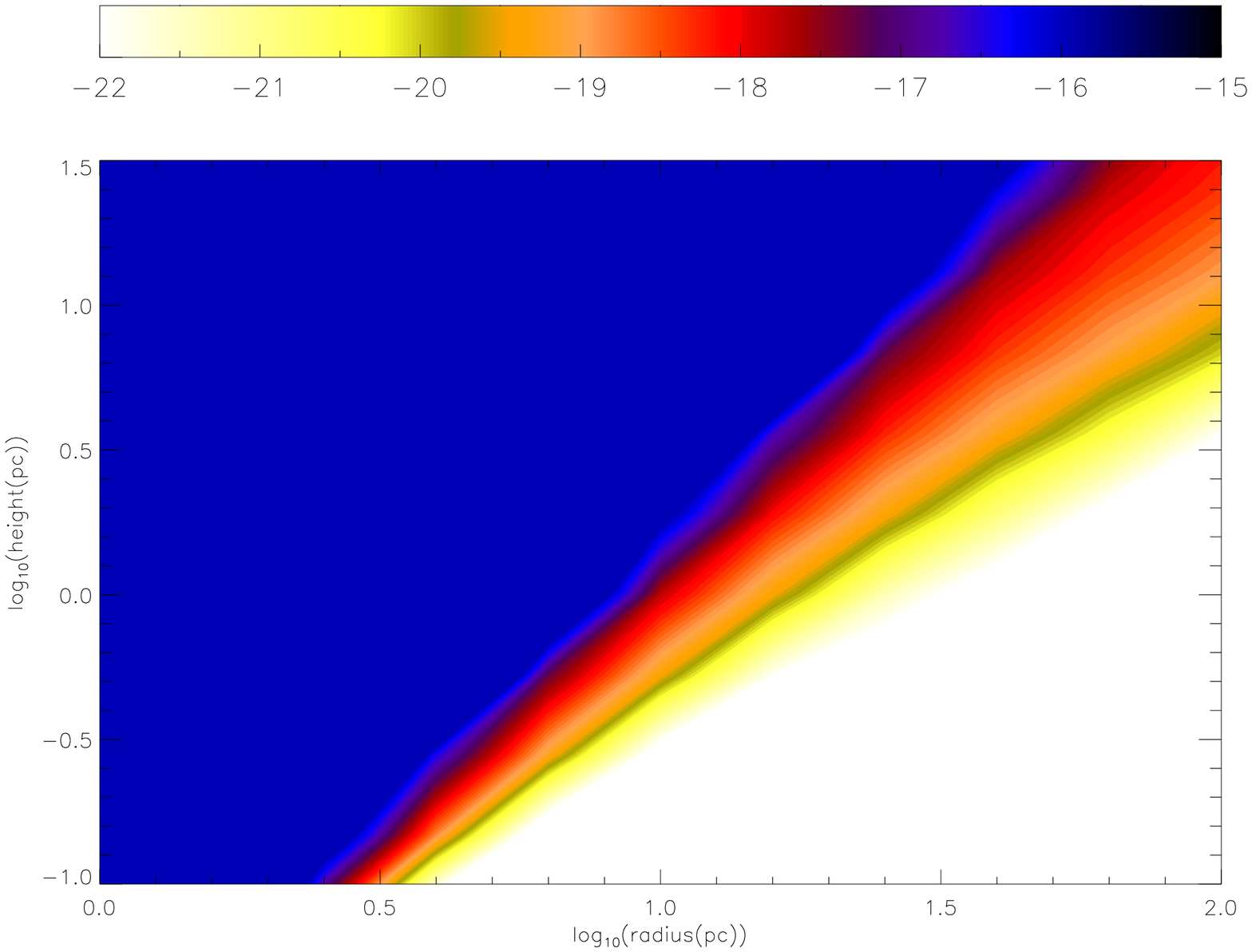}}
\centerline{\includegraphics[angle=90,width=8.5cm]{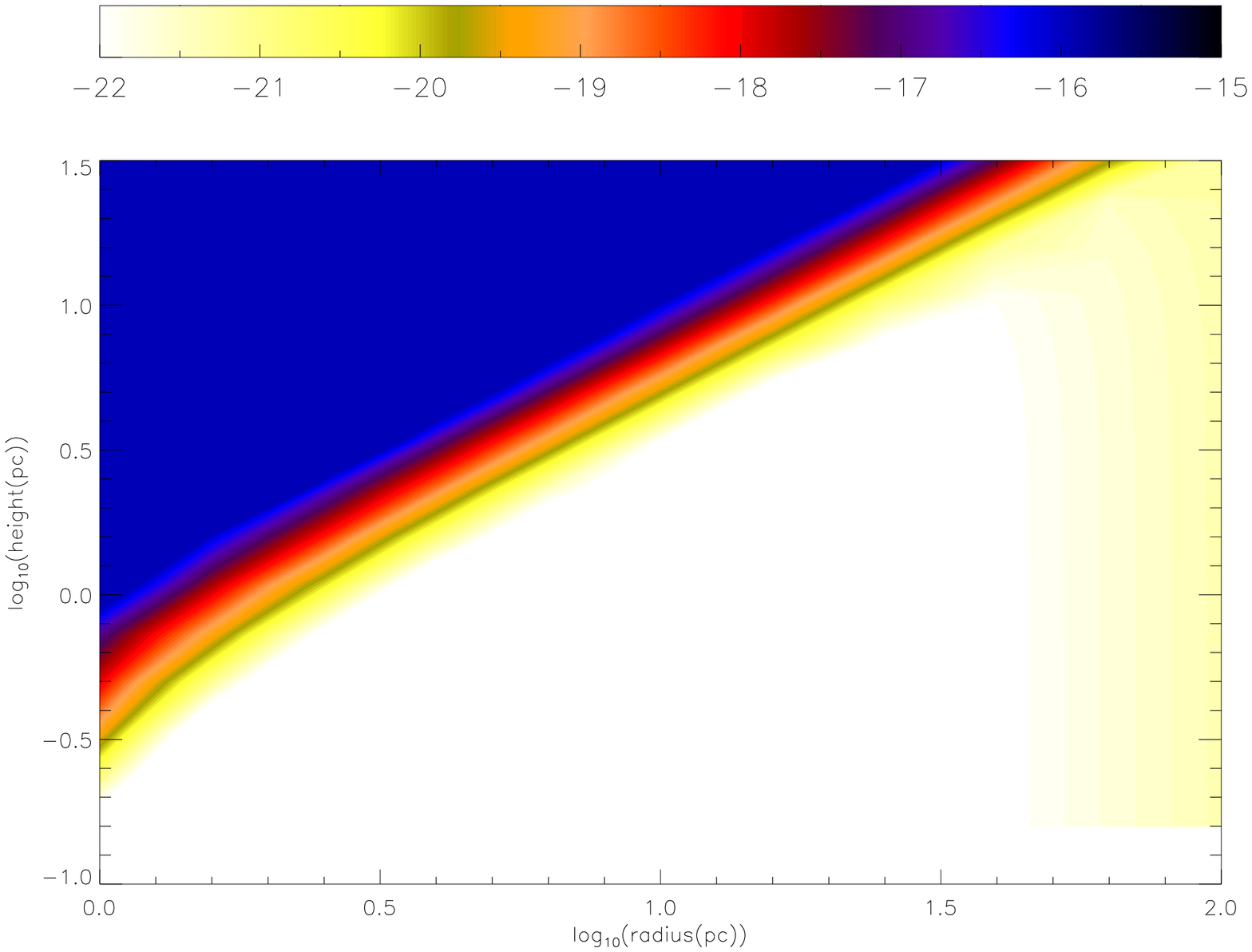}
\includegraphics[angle=90,width=8.5cm]{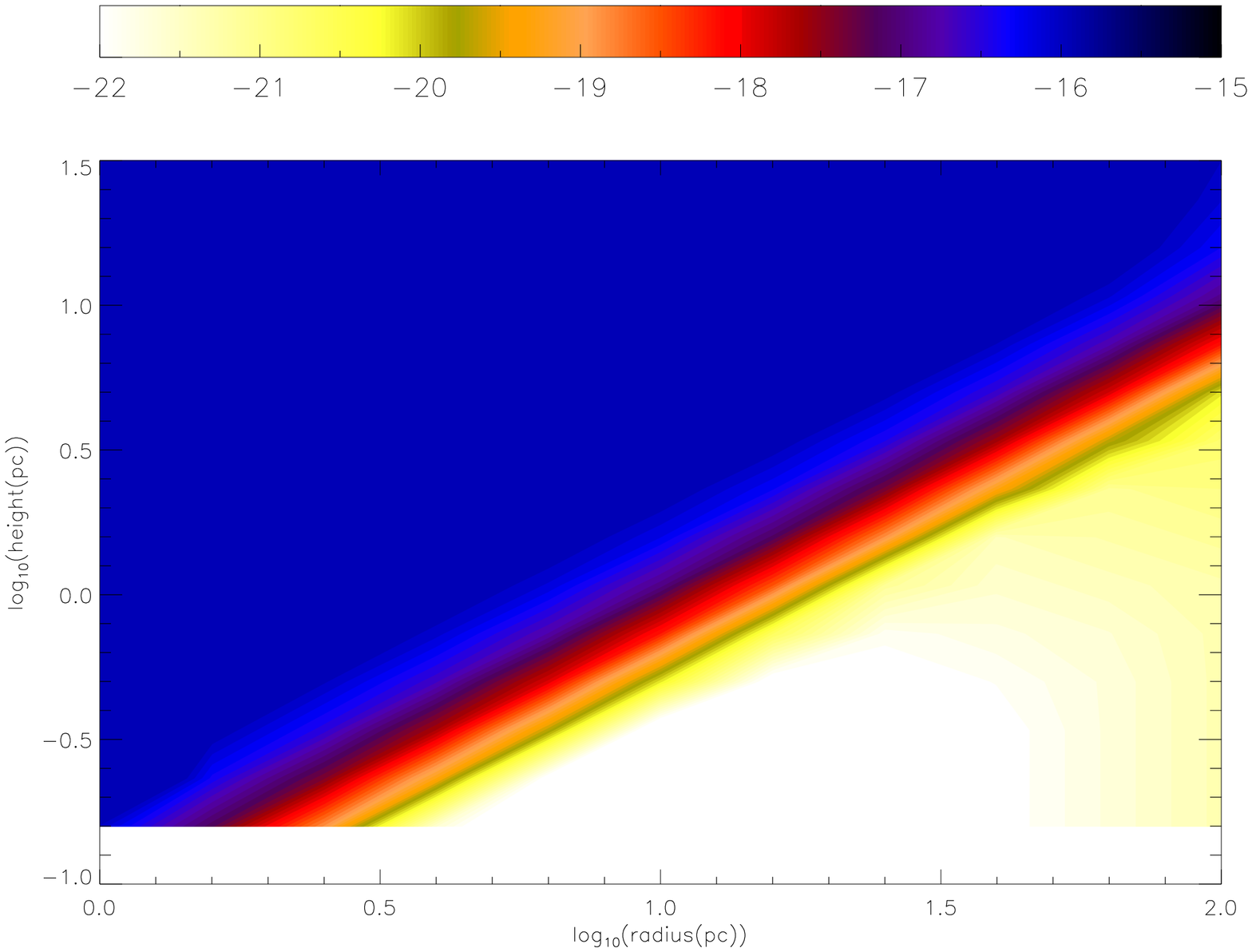}}
\caption{ The radial and height dependence of the ratio of the X-ray ionization rate over total hydrogen density, $\zeta_{\rm X}/n$ (s$^{-1}$cm$^{3}$). The radius, height, and $\zeta/n$ are all shown in logarithmic scale.The XDR referred to in the text is shown in orange, red, and purple regions. The high-temperature region is within $\sim$ 10 pc. {\it Upper left panel:}
Model 1, $Q$ =1; {\it Upper right panel:} Model 2, $Q$ = 1;
{\it Lower left panel:} Model 3, $Q$ = 5; {\it Lower right panel:} Model 4, $Q$ = 5.}
\label{fig:zeta_n}
\end{figure*}

For most regions of the disk, the ionization caused by X-rays exceeds that due to cosmic rays unless $\nu$ is as large as 0.01 (see eq. \ref{eq:ion_sn}).  We therefore first discuss the disk as an X-ray dominated region (XDR).
The ionization due to X-rays depends upon position and the density distribution of the disk model chosen. 
Figure  \ref{fig:zeta_n} shows  the ionization rate throughout our disk models caused by X-rays.  Rather than plotting the ionization rate explicitly, we have chosen to plot $\zeta_{\rm X}/n$, since this parameter determines the steady-state chemical abundances provided that direct dissociation by UV-photons is negligible \citep{1996A&A...306L..21L}.  
For example, \citet{1996A&A...306L..21L}  show that  the fractional abundances of CN and HCN
can be enhanced with moderate values of $\zeta_{\rm X}/n$. In, particular, their peak
fractional abundances are achieved when $\zeta_{\rm X}/n$~$\sim$~several
$\times$ 10$^{-19}\,$cm$^{3}$~s$^{-1}$.  At the lower value of 
$\zeta_{\rm X}/n = 10^{-21}\,$cm$^{3}$~s$^{-1}$, their abundances decrease by two orders of magnitude.
When $\zeta_{\rm X}/n$  $>$
10$^{-18}\,$cm$^{3}$s$^{-1}$, the fractional abundances decrease with
increasing $\zeta_{\rm X}/n$ since the molecules are dissociated rapidly.

The upper left and right panels in Figure~\ref{fig:zeta_n} show $\zeta_{\rm X}/n$ with $Q=1$ for Models 1 and 2, 
respectively.   These are high-density thin-disk models with and without inner tori.  Here it can be seen 
that the ionization caused by X-rays has trouble penetrating to the midplane.  In these models,
X-rays penetrate significantly only into a layer on the surface of the disk and
create a region where $\zeta_{\rm X}/n\sim10^{-21} - 10^{-18}$\,cm$^{3}$ s$^{-1}$ that has typical 
XDR chemistry, but $\zeta_{\rm X}/n<10^{-22}$\,cm$^{3}$~s$^{-1}$  near the midplane.
In the lower left panel in Figure~\ref{fig:zeta_n}, on the other hand, higher ionization 
levels are found; the results are for a lower-density ($Q=5$) version of Model 3, which 
contains a torus out through 100 pc.  Although $\zeta_{\rm X}/n<10^{-22}$\,cm$^{3}$ s$^{-1}$ 
around the midplane, there is a thick layer with $\zeta_{\rm X}/n=10^{-18}-10^{-21}$\,cm$^{3}$ s$^{-1}$.

Model 4, shown in the lower right panel, has a higher X-ray ionization than Model 3 since the 
density falls off faster with height above the disk.  Since X-rays penetrate more because of 
the lower density, there are more regions with $\zeta_{\rm X}/n>10^{-18}$\,cm$^{3}$ s$^{-1}$, where molecules 
cannot exist.
Since Models 1 and 2 have a higher $h/r$ ratio at larger $r$, 
there is a more extended XDR at larger $r$ for these models.  If we compare Models 3 and 4, which have constant $h/r$ for $r>1$\,pc, we find that
Model 3 has a more extended XDR at smaller $r$, while X-rays can reach larger values of $r$ in Model 4.

For those regions where X-ray ionization is relatively small, such as the midplane of  the disk at large radii, 
the ionization rate from cosmic rays can possibly dominate if $\nu$ is relatively large (see eq. \ref{eq:ion_sn}). 
For example, when $Q= 1$ and $\nu= 0.01$, $\zeta_{\rm X}/n$ ranges from
$8\times10^{-21}$\,cm$^{3}$ s$^{-1}$ at 1\,pc to
$8\times10^{-19}$\,cm$^{3}$ s$^{-1}$ at 100\,pc from the AGN core. The effect of
cosmic rays becomes more important for larger $r$ because of the relatively lower density (see Section \ref{section:res_cr} below).

\section{Results}\label{results}

\subsection{Models with X-ray Ionization Only}\label{results_x}

In this section we explore the effects of X-rays without the inclusion of cosmic-ray ionization. The degree to which X-rays can affect the disk chemistry depends on their penetration into the disk, which depends on the disk structure. Far from the midplane, molecular abundances are very low given the very large X-ray ionization and lower density.  Only regions with  considerable molecular abundances are shown in the figures to be discussed below.  In between the ``no-molecule zone" and the midplane, there is an XDR, the extent of which is determined by the ability of X-rays to penetrate.  At closer distances and higher temperatures, the temperature affects the chemistry significantly over a range of X-ray ionization rates, and we refer to these  portions of the disk as ``high temperature-synthesis'' regions. The main results of this high-temperature-synthesis effect are to increase the abundances of some species such as HCN, the family of C$_{\rm n}$H$_{2}$, and cyanopolyynes (e.g., HC$_{3}$N).  The effect is significant for $T\gtrsim 300$\,K at early times before CO fully forms, and can be seen in the results presented at the shock crossing time $t=t_{\rm cross}$ (see Section \ref{section:timescale}).   Yet, in the absence of X-ray heating, this effect still remains at steady-state after CO fully forms for $T\gtrsim500$\,K.  Importantly, though, in our calculations the regions where $T\gtrsim500$\,K are in general strongly heated by X-rays, which tend to dissociate molecules, and therefore the high-temperature-synthesis effect on the final abundances is mitigated in regions of relatively low columns ($\lesssim 10^{24}$\,cm$^{-2}$). Details of this chemistry are discussed in  \citet{2010ApJ...721.1570H}.

We find that 
the results for the abundances in Models 1 and 2 are not qualitatively different except for the values at 1\,pc.  
However, Model 4 is quite different from the others. For these reasons, 
we discuss,  compare,  and contrast Model 1 with $Q = 1$ and Model 4 with $Q = 5$,
with an emphasis on fractional abundances $X$ of important molecules with respect to 
the total hydrogen abundance.  As discussed in Section \ref{section:density}, Model 1 is a high-density thin disk model 
in which X-rays have more difficulty penetrating than in Model 4, which is a relatively low density 
geometrically thick torus.

\subsubsection{ Results for Model 1 with $Q=1$}\label{res_m1}
\begin{figure*}
\centerline{\includegraphics[angle=90,width=8.5cm]{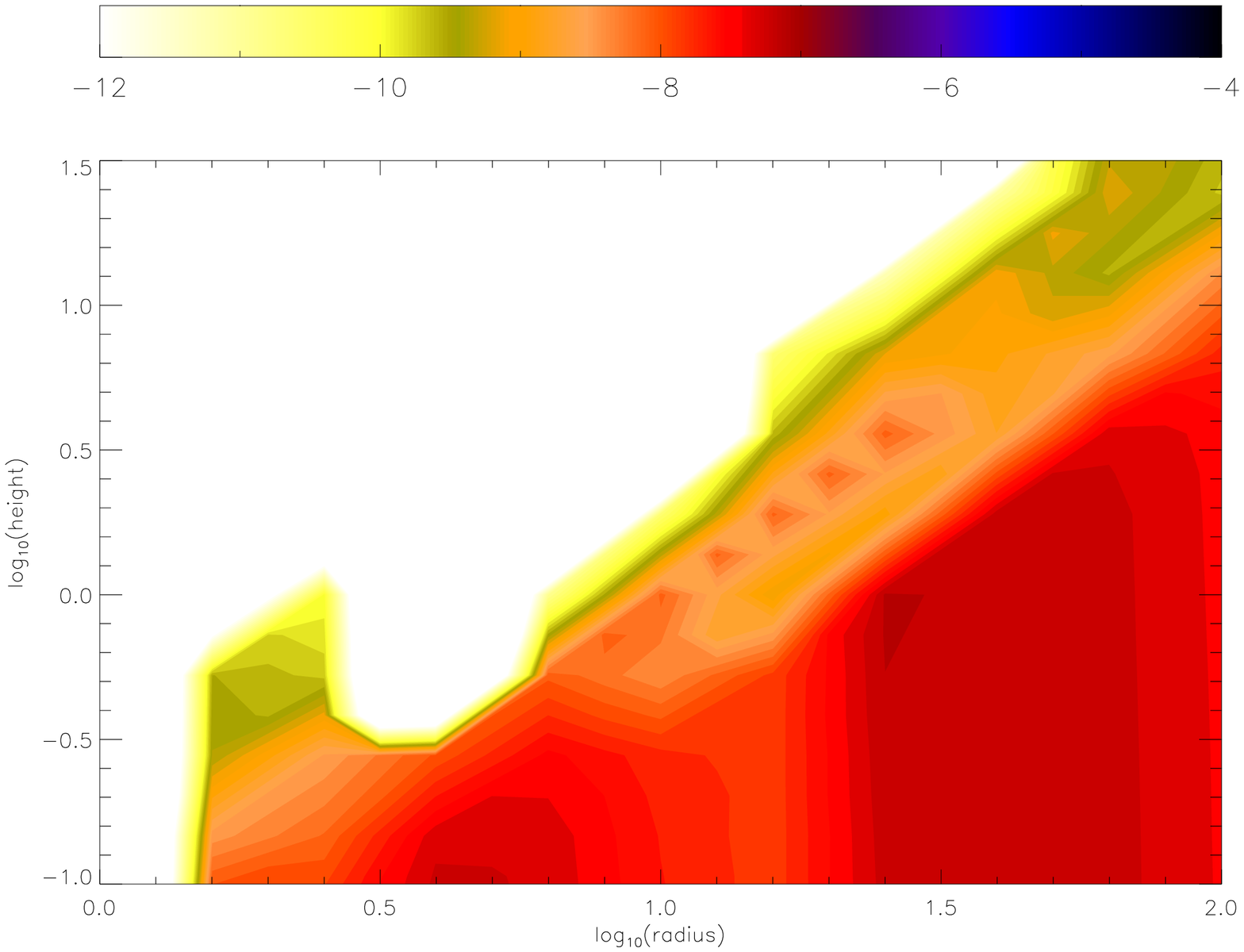}
\includegraphics[angle=90,width=8.5cm]{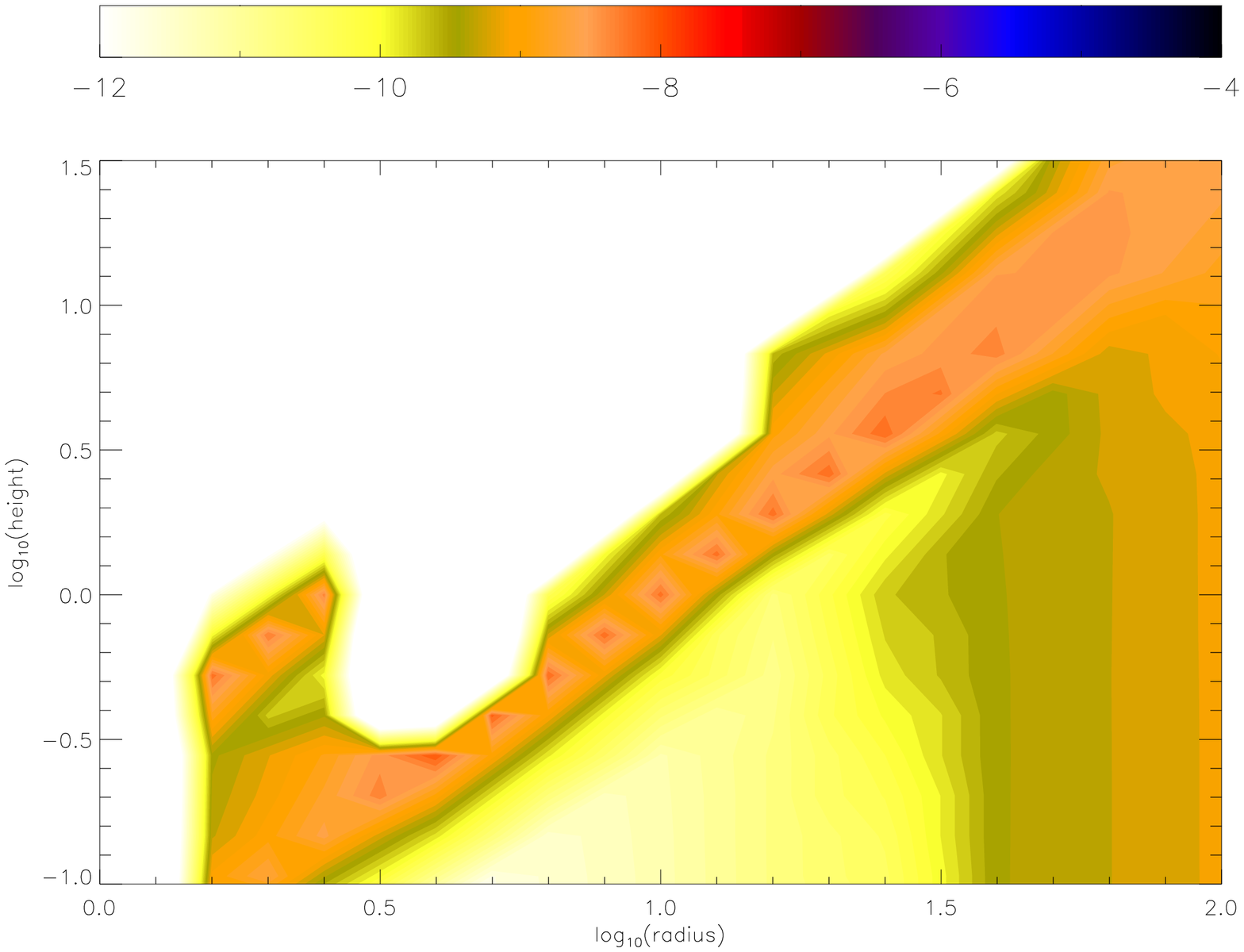}}
\centerline{\includegraphics[angle=90,width=8.5cm]{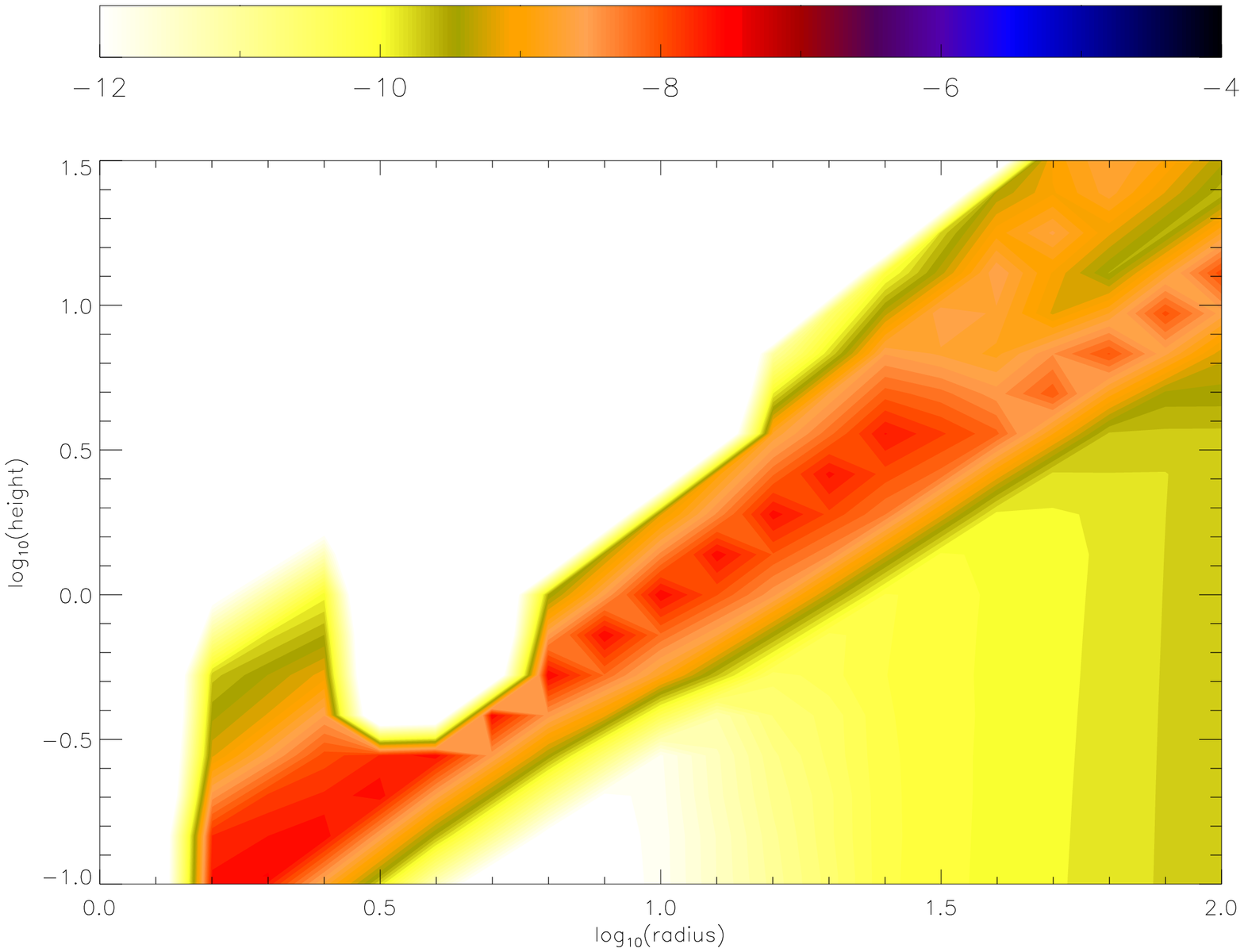}
\includegraphics[angle=90,width=8.5cm]{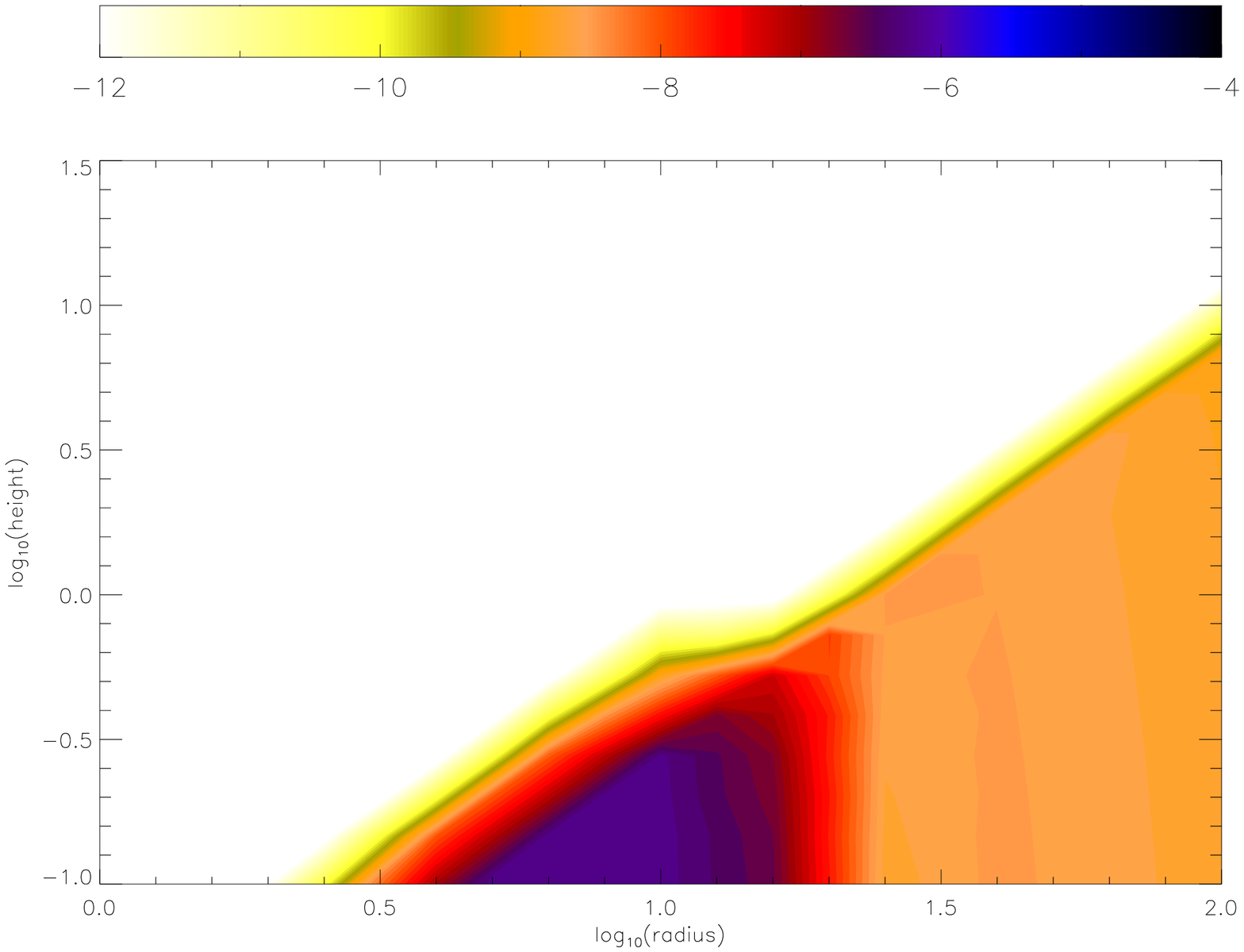}}
\centerline{\includegraphics[angle=90,width=8.5cm]{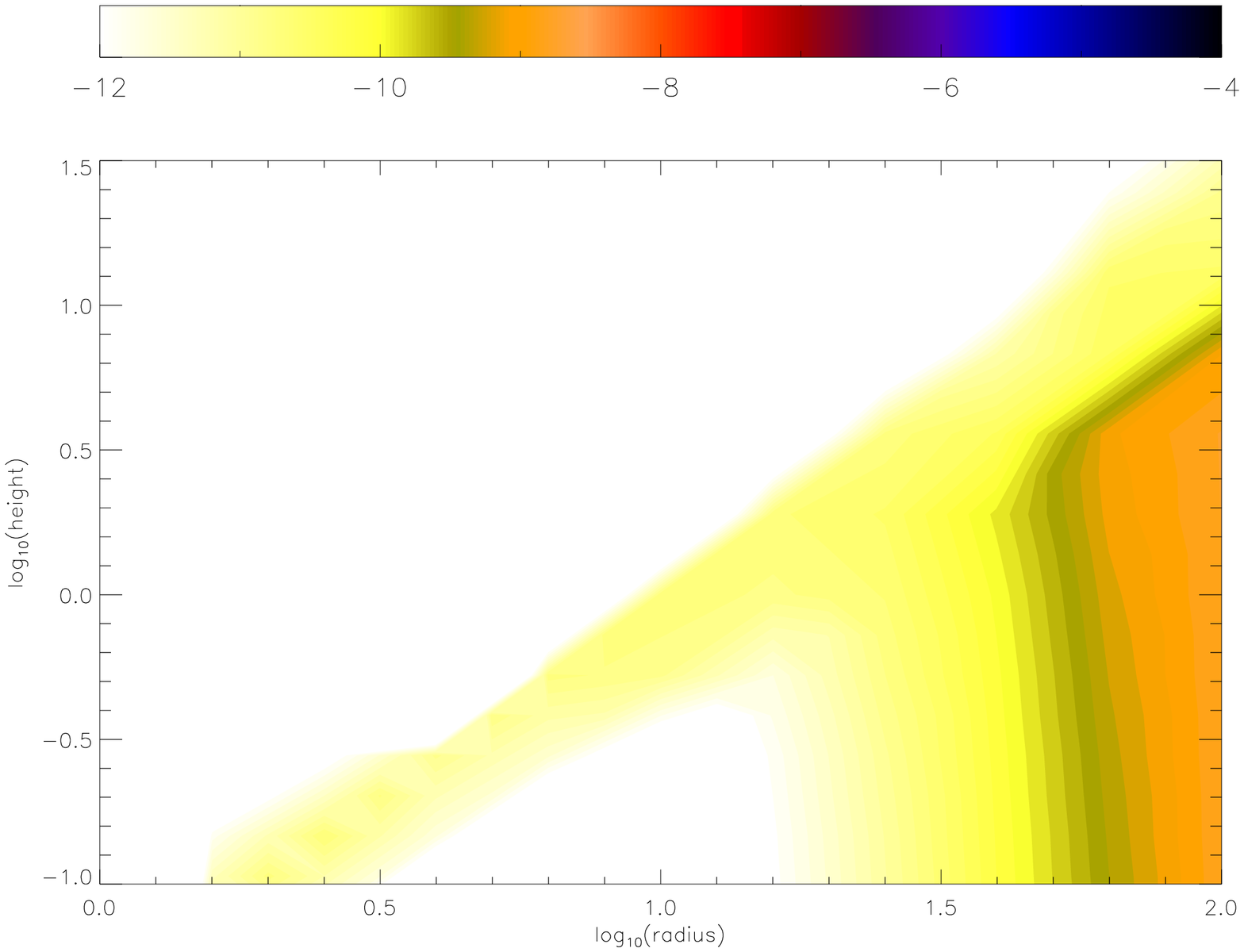}
\includegraphics[angle=90,width=8.5cm]{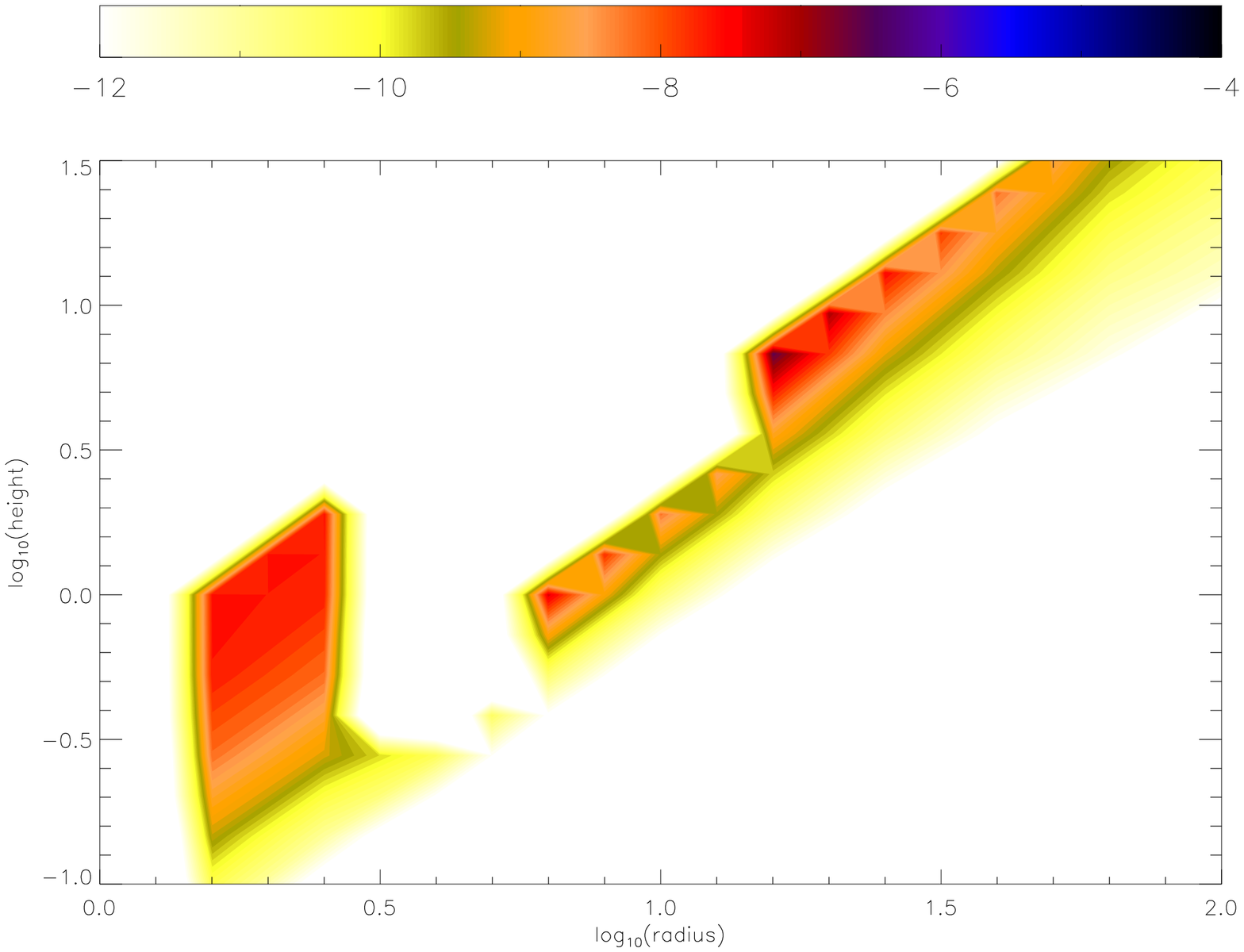}}
\caption{The radial and height dependence of fractional abundances at $t$ = $t_{cross}$ in the Model 1, $Q$ = 1 disk.  {\it Upper left panel:}
HCN; {\it Upper right panel:} CN; {\it Middle left panel:} \cop;  {\it Middle right panel:} HC$_{3}$N;  {\it Lower left panel:} CS;  {\it Lower right panel:} OH$^{+}$.  Note that the color scale is different for CS.}
\label{fig:fab_sfe0_q1_hrv5}
\end{figure*}

\begin{figure*}
\centerline{\includegraphics[angle=90,width=8.5cm]{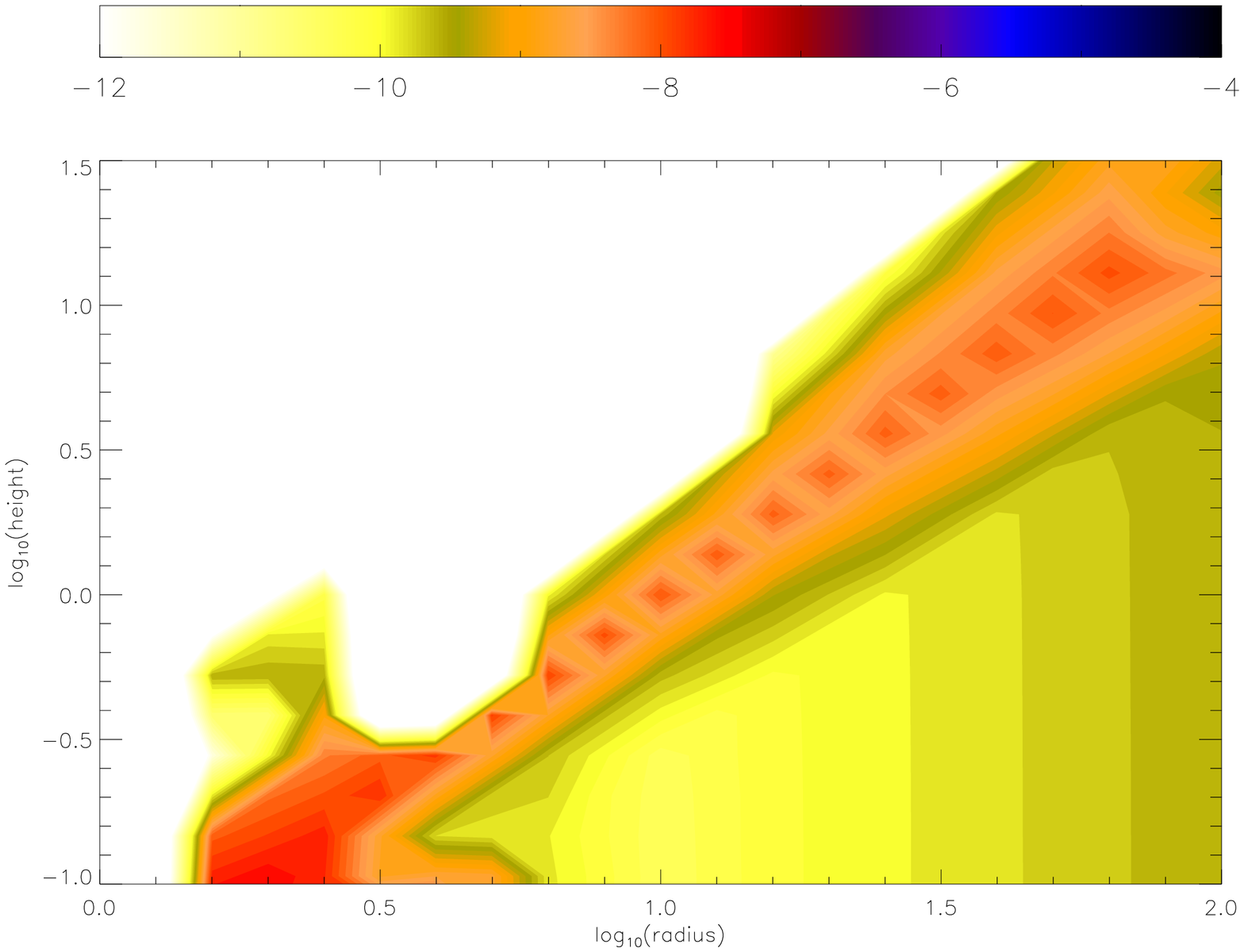}
\includegraphics[angle=90,width=8.5cm]{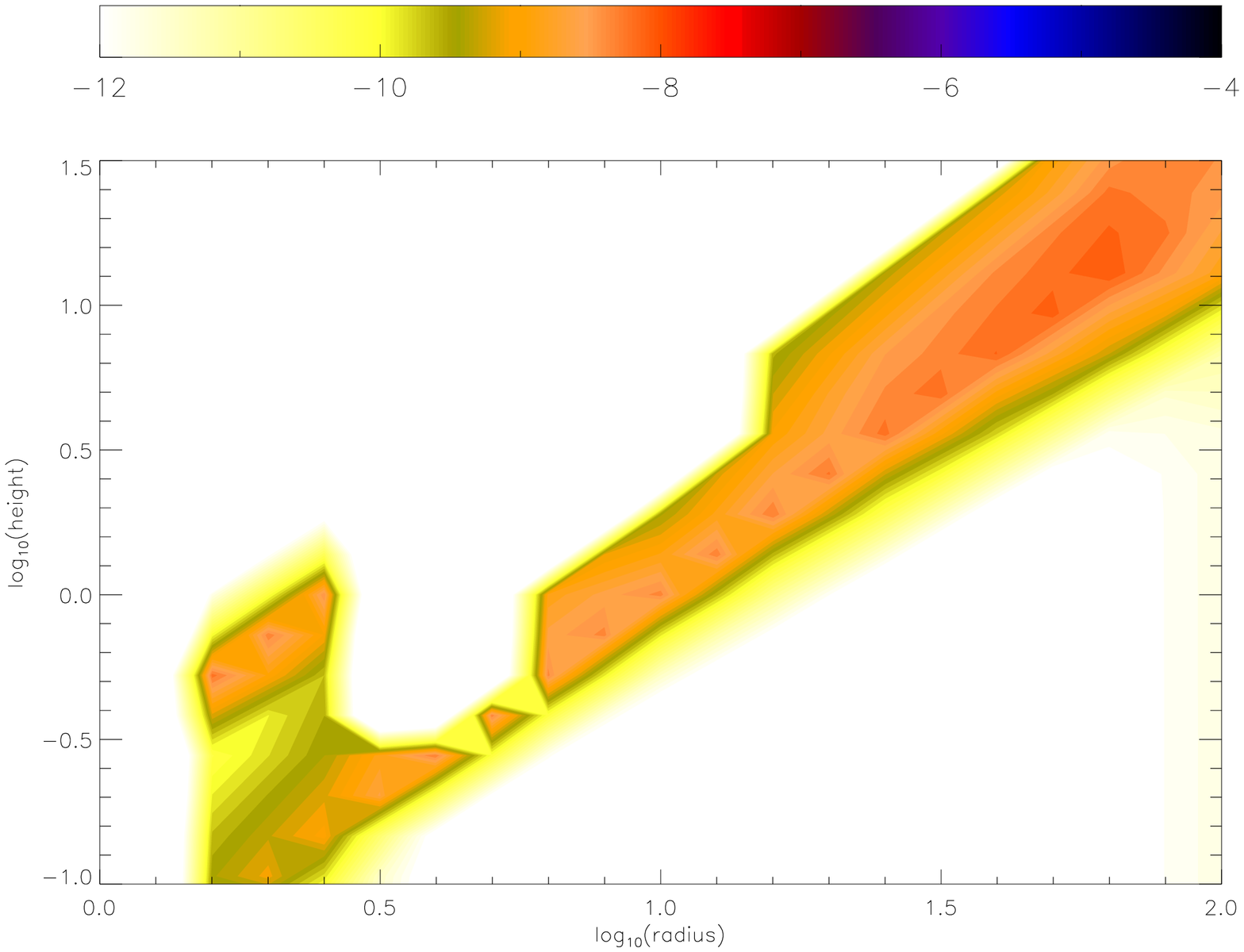}}
\centerline{\includegraphics[angle=90,width=8.5cm]{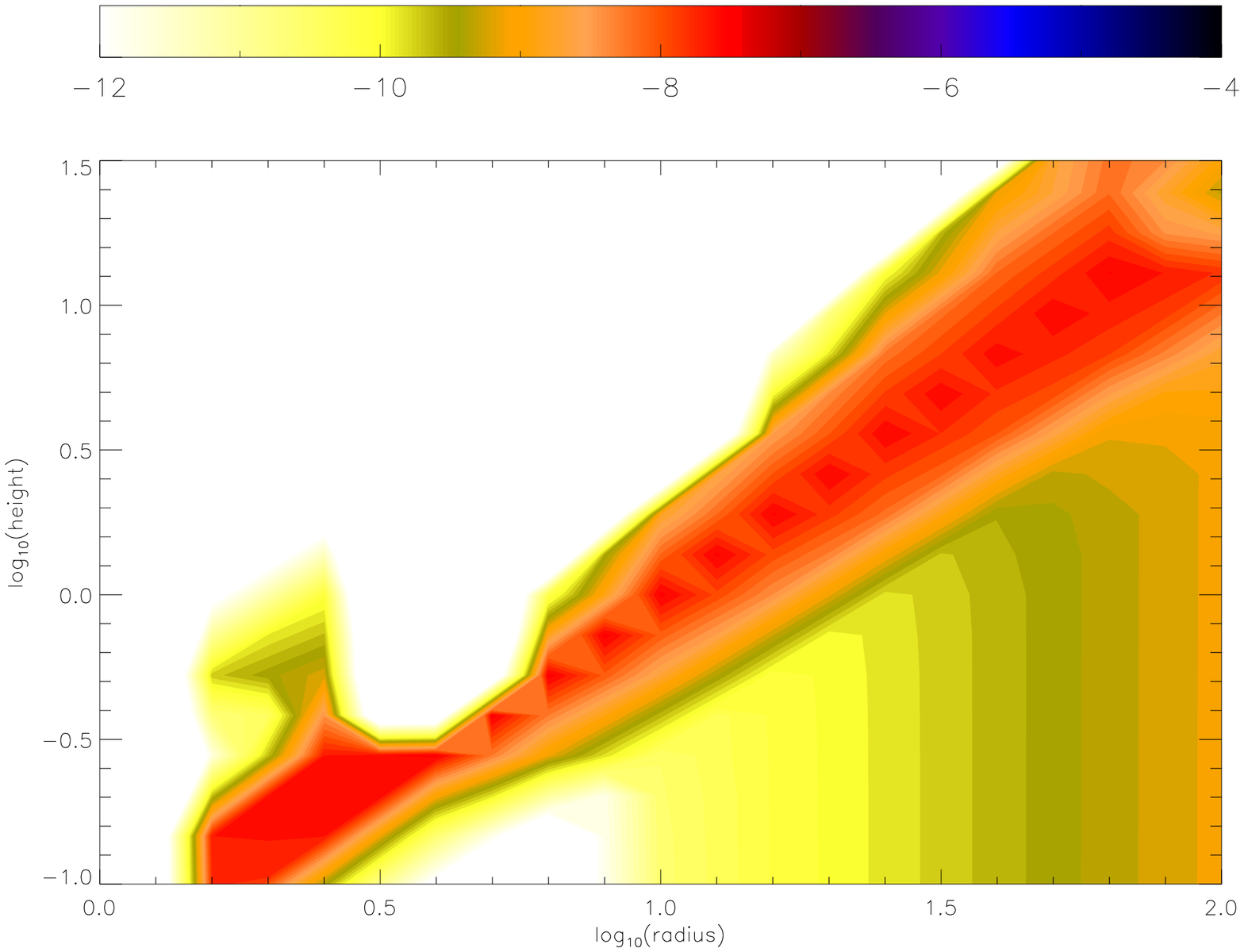}
\includegraphics[angle=90,width=8.5cm]{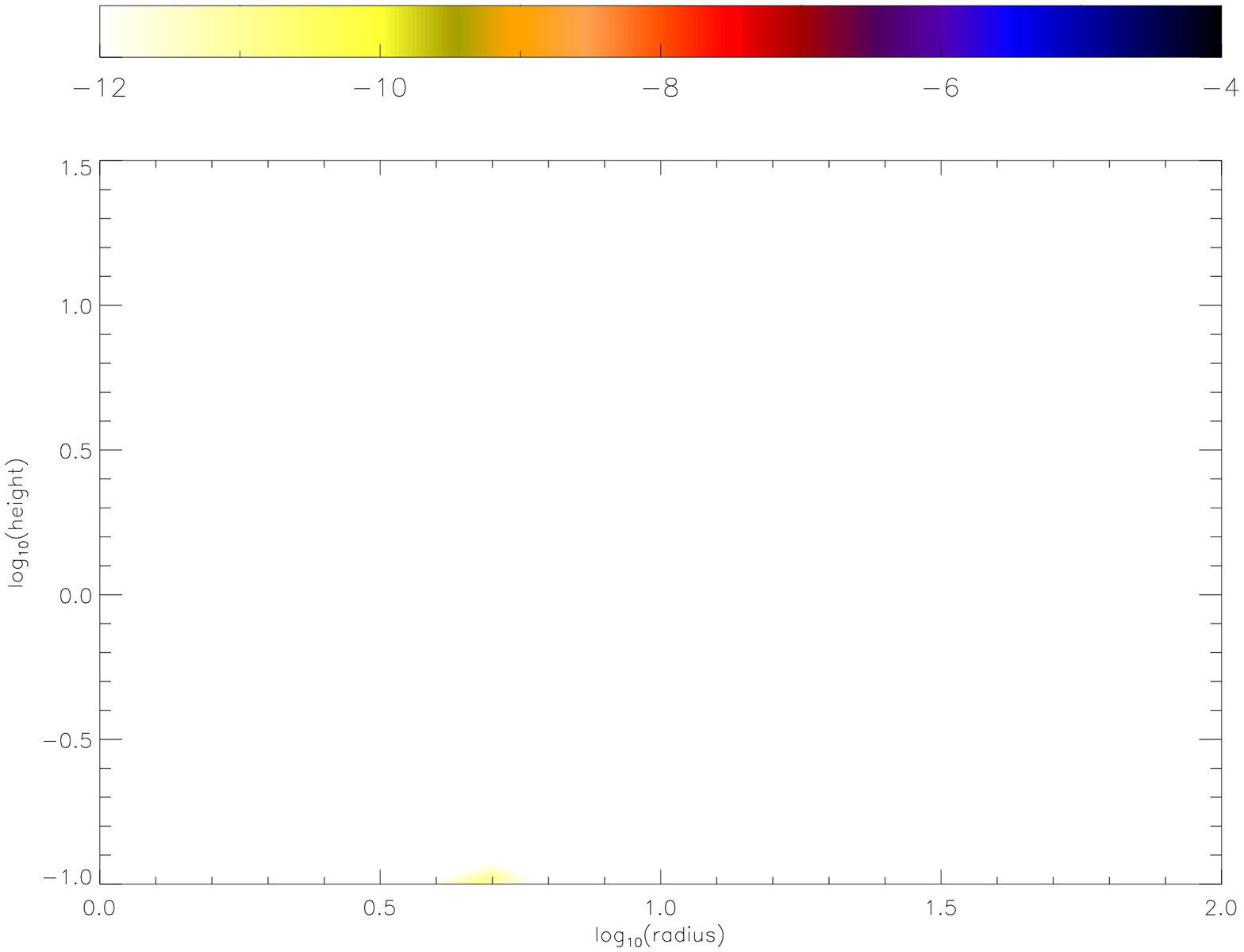}}
\centerline{\includegraphics[angle=90,width=8.5cm]{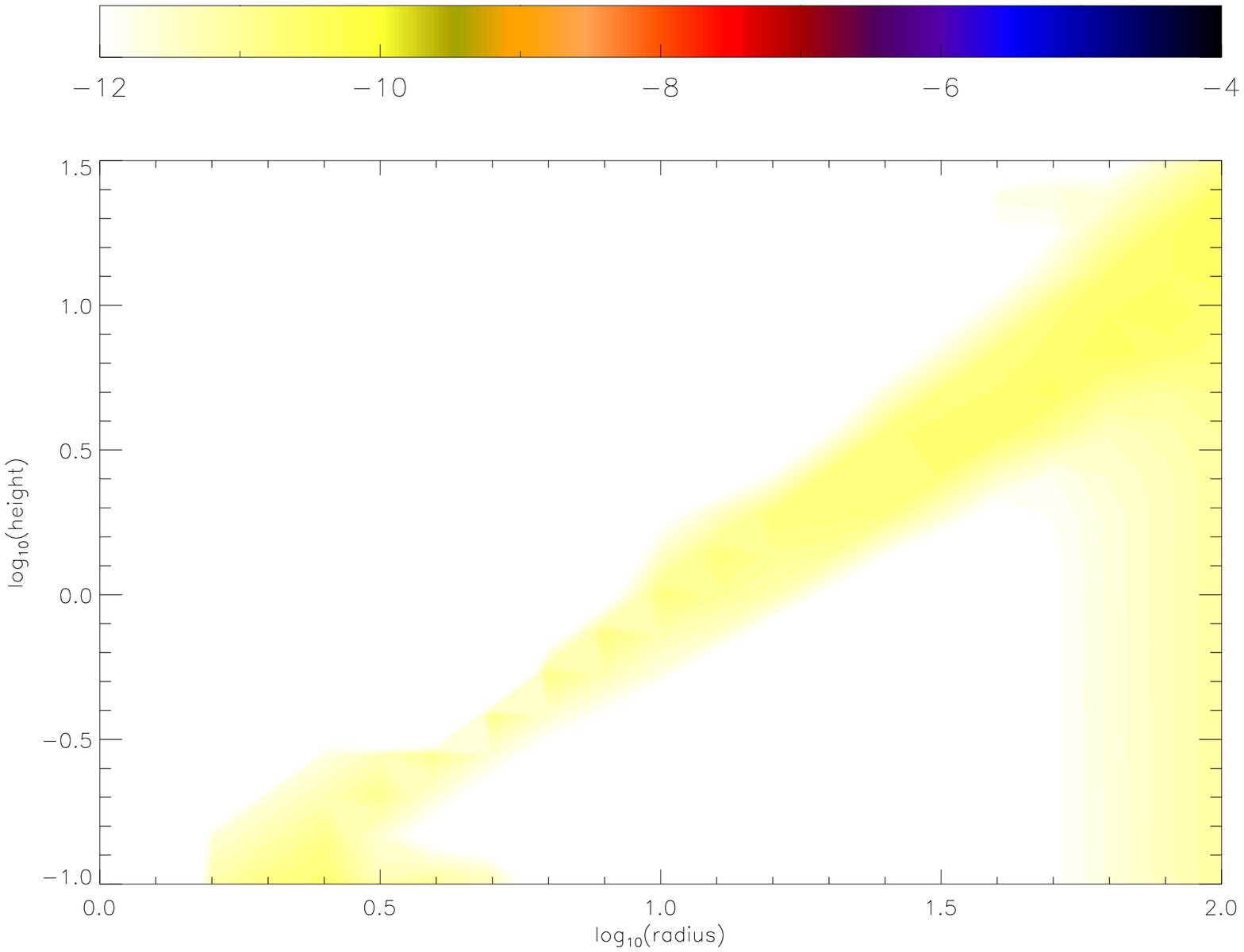}
\includegraphics[angle=90,width=8.5cm]{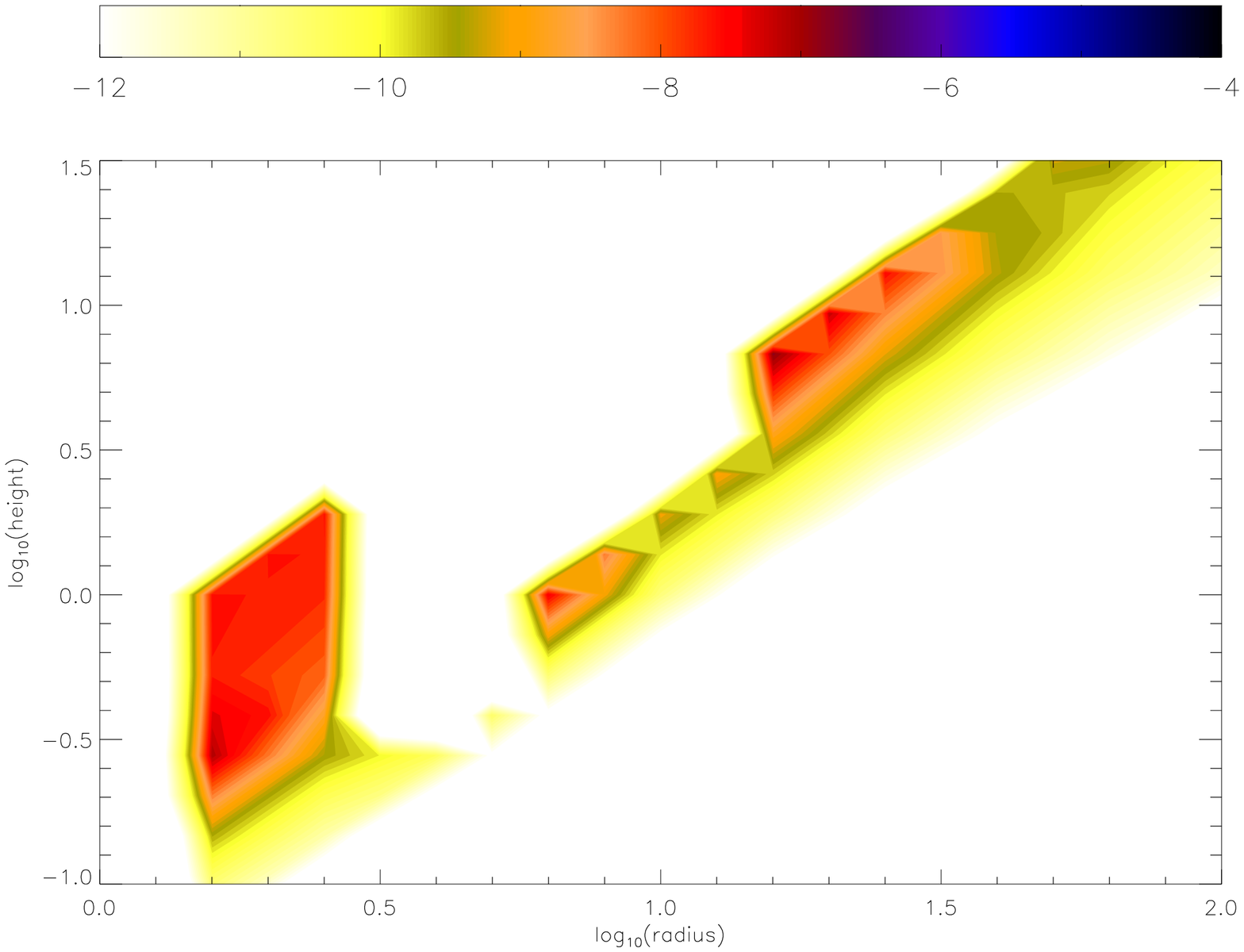}}
\caption{The radial and height dependence of fractional abundances at steady-state in the Model 1, $Q$ = 1 disk.  {\it Upper left panel:}
HCN. {\it Upper right panel:} CN.{\it Middle left panel:} \cop. {\it Middle right panel:} HC$_{3}$N. {\it Lower left panel:} CS. {\it Lower right panel:} OH$^{+}$. Note that the color scale is different for CS.}
\label{fig:fab_ss_sfe0_q1_hrv5}
\end{figure*}

The fractional abundances of selected molecules are shown in Figure \ref{fig:fab_sfe0_q1_hrv5} at $t$ = $t_{\rm cross}$, and Figure \ref{fig:fab_ss_sfe0_q1_hrv5} for steady-state. The fractional abundance of each molecule is discussed below.

{\bf \underline{HCN}:} The fractional abundance of HCN at $t = t_{\rm cross}$ and at steady state is shown in the 
upper left panel of Figures \ref{fig:fab_sfe0_q1_hrv5}~and \ref{fig:fab_ss_sfe0_q1_hrv5}, respectively.  
As mentioned above, we started all elements except for \hmol~from atomic form as in Table \ref{init}. At $t$ = $t_{\rm cross}$, there is a fraction of carbon that is not yet locked up in CO. This free carbon helps to form a high abundance $X({\rm HCN})\gtrsim10^{-7}$ when $T>100\,$K in most regions of the disk. 

 Although the HCN abundance can be enhanced at early times in the calculation (before  steady state is reached) when $T>100\,$K, the temperature in the midplane of AGN disk is in general not high enough ($T\lesssim500\,$K) to keep the HCN abundance high as steady state is approached except for $r\lesssim$\,few pc where the dust temperature is highest. 
When accretion is considered, there might be less of a high-temperature-synthesis effect observed in the HCN abundance, because if accretion is rapid (e.g., $\sim10-100\Omega^{-1}$) $t_{\rm acc}$ can be less than $t_{\rm chem}$, and one does not expect the steady-state abundances to be reached. 

{\bf \underline{CN}:} The time at which the peak in CN abundance is reached is much earlier than that of HCN. Since HCN is more stable, CN eventually becomes HCN either through an indirect process involving ions at cold temperatures or via a more efficient direct reaction with \hmol~at high temperature \citep[see equation (21) in ][]{2010ApJ...721.1570H}. $X({\rm CN})$ peaks at $\sim3\times10^{-8}$ on a timescale $10-100$ times shorter than $t_{\rm cross}$. By $t=t_{\rm cross}$, $X({\rm CN})$ has decreased by orders of magnitude, and is only enhanced in the XDR layer where $X({\rm CN})\sim\,{\rm a \,\,few}\, \times10^{-8}$.  In the high temperature regions, the CN abundance is  lower because of the efficient production of HCN from CN. Within the XDR layer, the CN fractional abundance is much higher in the colder, more distant, part. On longer timescales, the steady-state value of CN near the midplane is orders of magnitude lower than the abundance at $t=t_{\rm cross}$, but a similar degree of enhancement is still seen in the XDR layer (see Figures \ref{fig:fab_sfe0_q1_hrv5} and \ref{fig:fab_ss_sfe0_q1_hrv5}, upper right panel).

{\bf \underline{\cop}:} The fractional abundance of \cop, depicted in  Figures \ref{fig:fab_sfe0_q1_hrv5} and \ref{fig:fab_ss_sfe0_q1_hrv5},  does not show strong time dependence. Once CO is formed, \cop~can be made by receiving a proton from a protonated ion such as H$_{3}^{+}$. Its abundance increases in moderately ionized regions, and we find a higher abundance of \cop~in the XDR layer, similar in location to where the abundance of CN is enhanced.
At both $t=t_{\rm cross}$ and  steady state, the highest abundance is seen at inner disk radii far from the midplane and in the outer XDR region, where the fractional abundance of \cop~is larger than that of CN with a value of $X({\rm HCO^+})\simeq(3-10)\times10^{-8}$.  Starting the simulation with solar elemental abundances, we find that $X$(\cop)~is reduced,
 where $X({\rm HCO^+})<10^{-9}$ even in the XDR layer, since atomic ions such as S$^{+}$, Na$^{+}$, and Mg$^{+}$ hold most of the positive charges instead of \cop.

{\bf \underline{HC$_{3}$N, C$_{2}$H$_{2}$, c-C$_{3}$H$_{2}$, C$_{2}$H}:}
Similar to HCN, there is a high-temperature enhancement for HC$_{3}$N and the family of C$_{n}$H$_{2}$, although the C$_{2}$H abundance decreases with increasing temperature. As seen in Figure \ref{fig:fab_sfe0_q1_hrv5}, 
 the fractional abundance of HC$_{3}$N exceeds 10$^{-6}$ when $r \lesssim5 $~pc at $t= t_{\rm cross}$. In steady-state, however, 
there is high-temperature-synthesis effect only at $r\lesssim$ a few pc where the disk is quite hot, as can be seen in Figure~\ref{fig:fab_ss_sfe0_q1_hrv5}. 
 The fractional abundance of C$_{2}$H$_{2}$ at $t= t_{\rm cross}$ also peaks at high temperatures at a value of  about 10$^{-6}$ for $r\sim10\,$pc, 
 and the steady-state abundances decrease by orders of magnitude as they do in HC$_{3}$N.
 The abundance of c-C$_{3}$H$_{2}$ follows a similar trend as C$_{2}$H$_{2}$.   
 
The fractional abundance of C$_{2}$H is orders of magnitude lower than that of C$_{2}$H$_{2}$ in the high-temperature-synthesis regions. For $r\sim100$\,pc, the fractional abundance of C$_{2}$H is around $1\times10^{-9}$ at $t=t_{\rm cross}$, and decreases by orders of magnitude on longer timescales as  steady-state is reached.   
When calculated using solar elemental abundances, the abundances of these carbon-chain molecules peak at slightly higher values in high-temperature-synthesis regions simply because of the higher elemental carbon abundance, but are orders of magnitude lower for other regions.

{\bf \underline{CS}:}
In general, the CS abundance peaks at times before $t=t_{\rm cross}$, after which most of the sulfur goes into SO or SO$_{2}$ in an oxygen-rich (C/O$<$1) environment. By the time $t=t_{\rm cross}$, $X({\rm CS})<10^{-9}$ for almost all  regions,  and SO/CS$>$10 for $r\lesssim20\,$pc. For a larger $r$, due to the lower density and lower temperature, the chemistry evolves more slowly, yet there is still a high abundance of CS.

{  \bf \underline{OH$^{+}$}:}
Since OH$^{+}$ can survive in regions that have higher X-ray ionization rates than other molecules, it is abundant in the XDR layers over a wide range of $r$. In Model 1, there is only a small region where OH$^{+}$ is enhanced because of the adopted density structure and the attenuation of X-rays from the torus.
 
\subsubsection{Results for Model 4 with $Q=5$}\label{res_m4}

\begin{figure*}
\centerline{\includegraphics[angle=90,width=8.5cm]{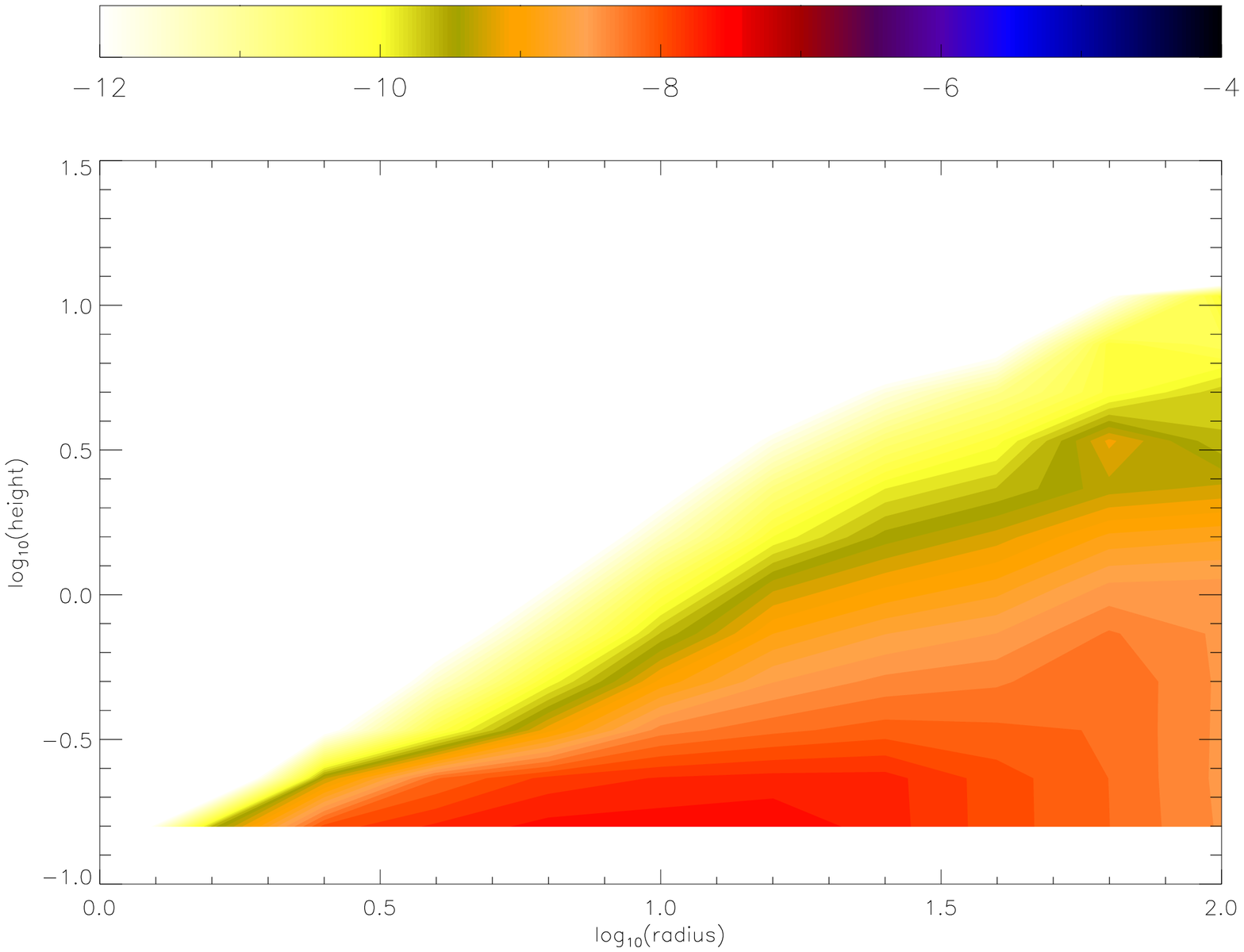}
\includegraphics[angle=90,width=8.5cm]{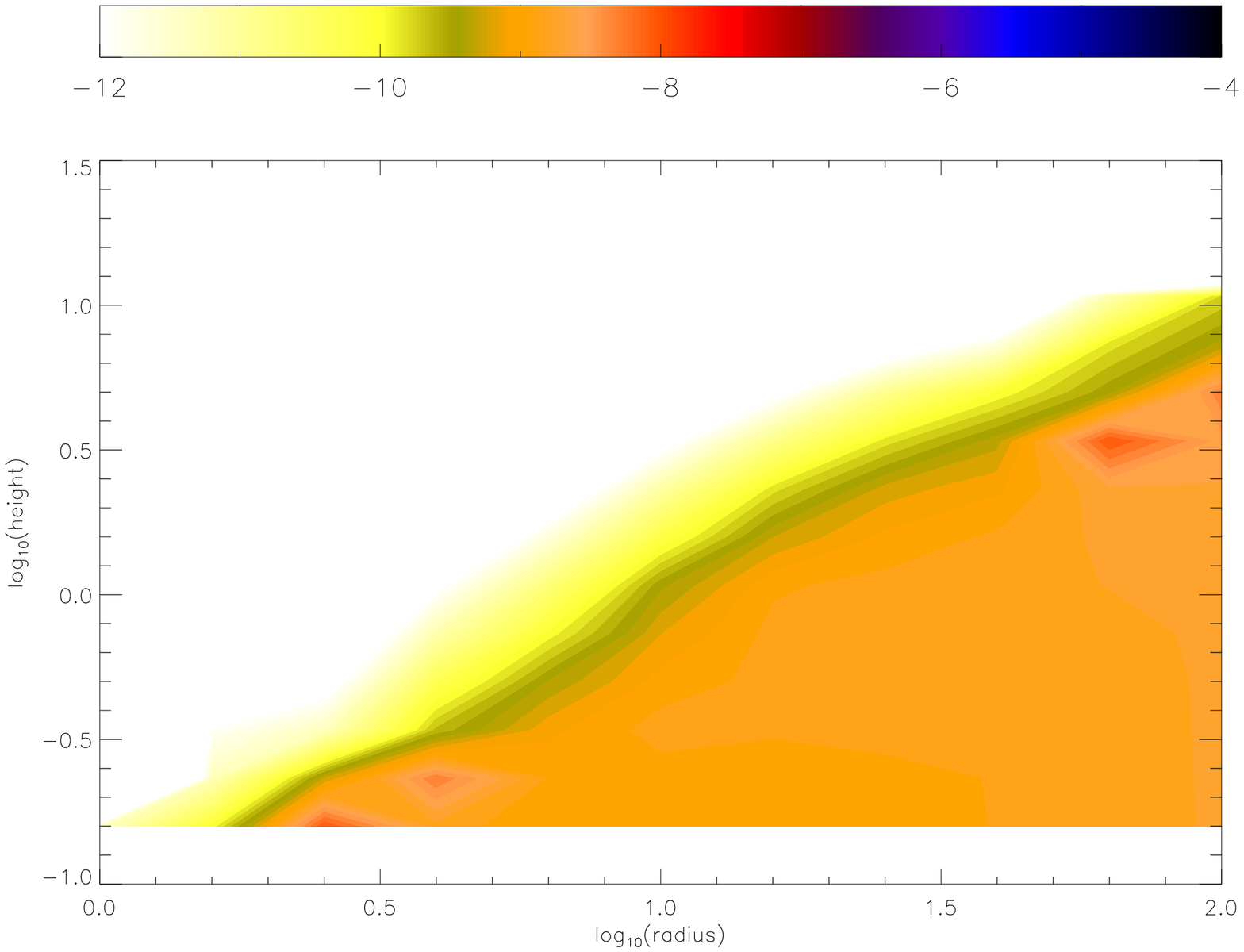}}
\centerline{\includegraphics[angle=90,width=8.5cm]{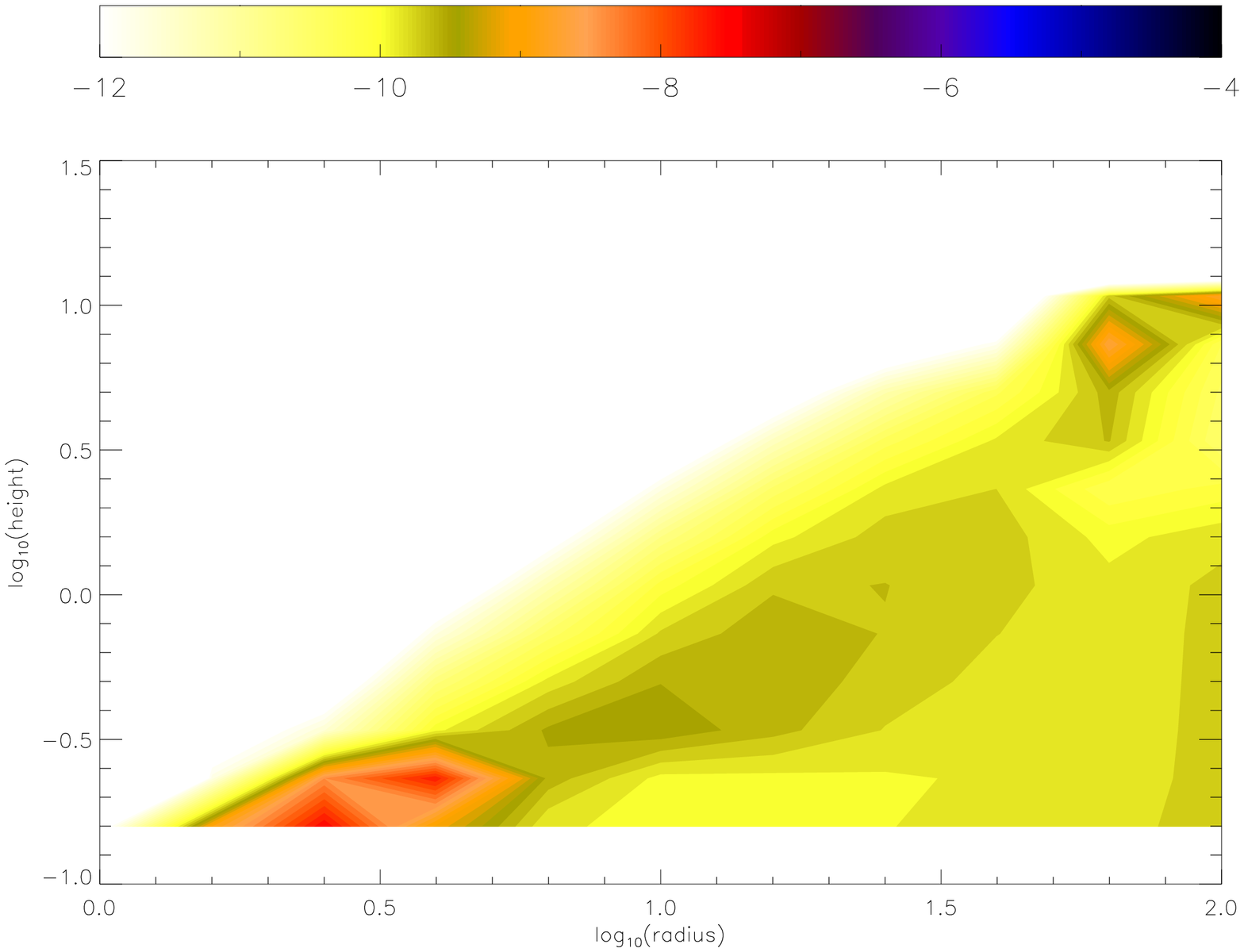}
\includegraphics[angle=90,width=8.5cm]{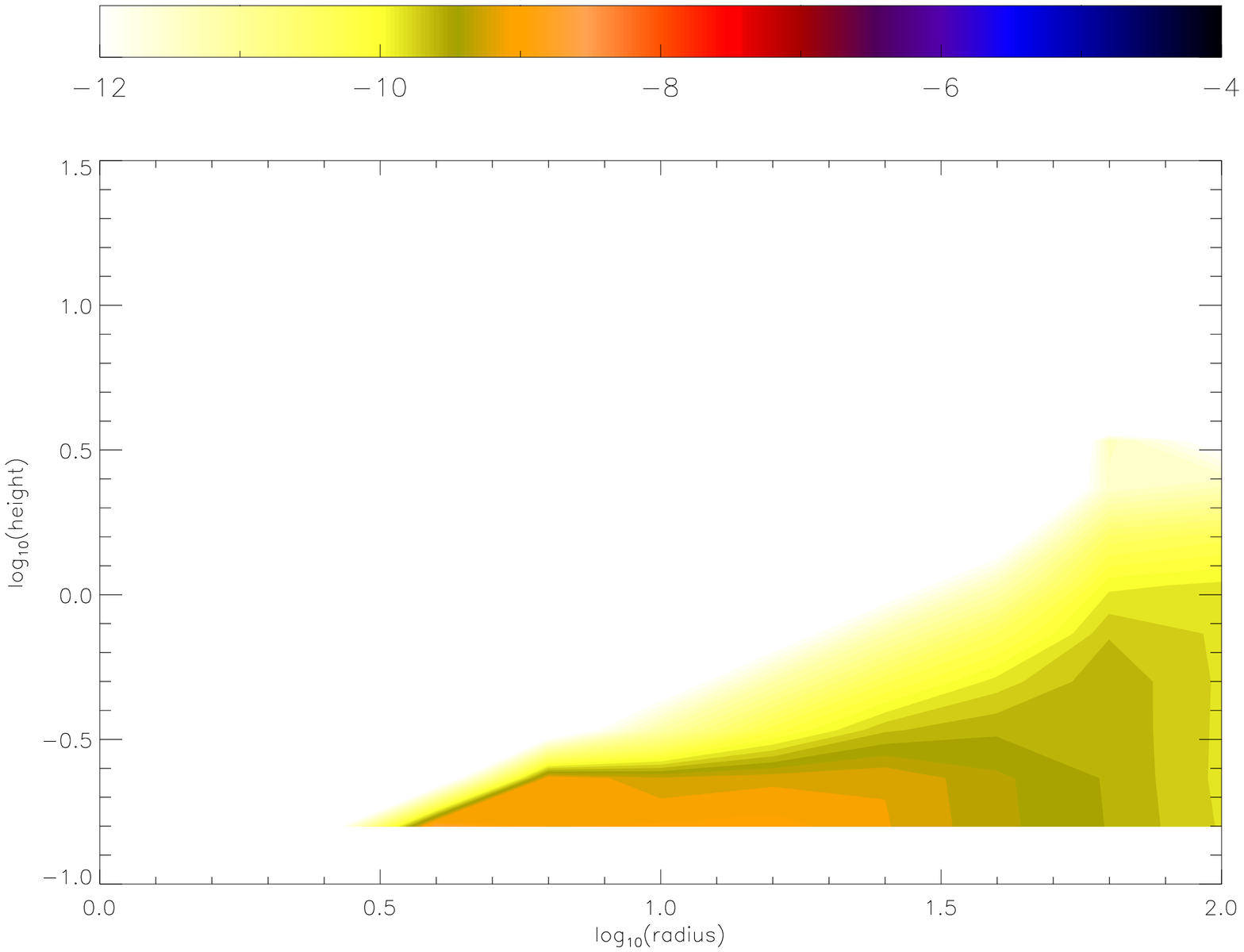}}
\centerline{\includegraphics[angle=90,width=8.5cm]{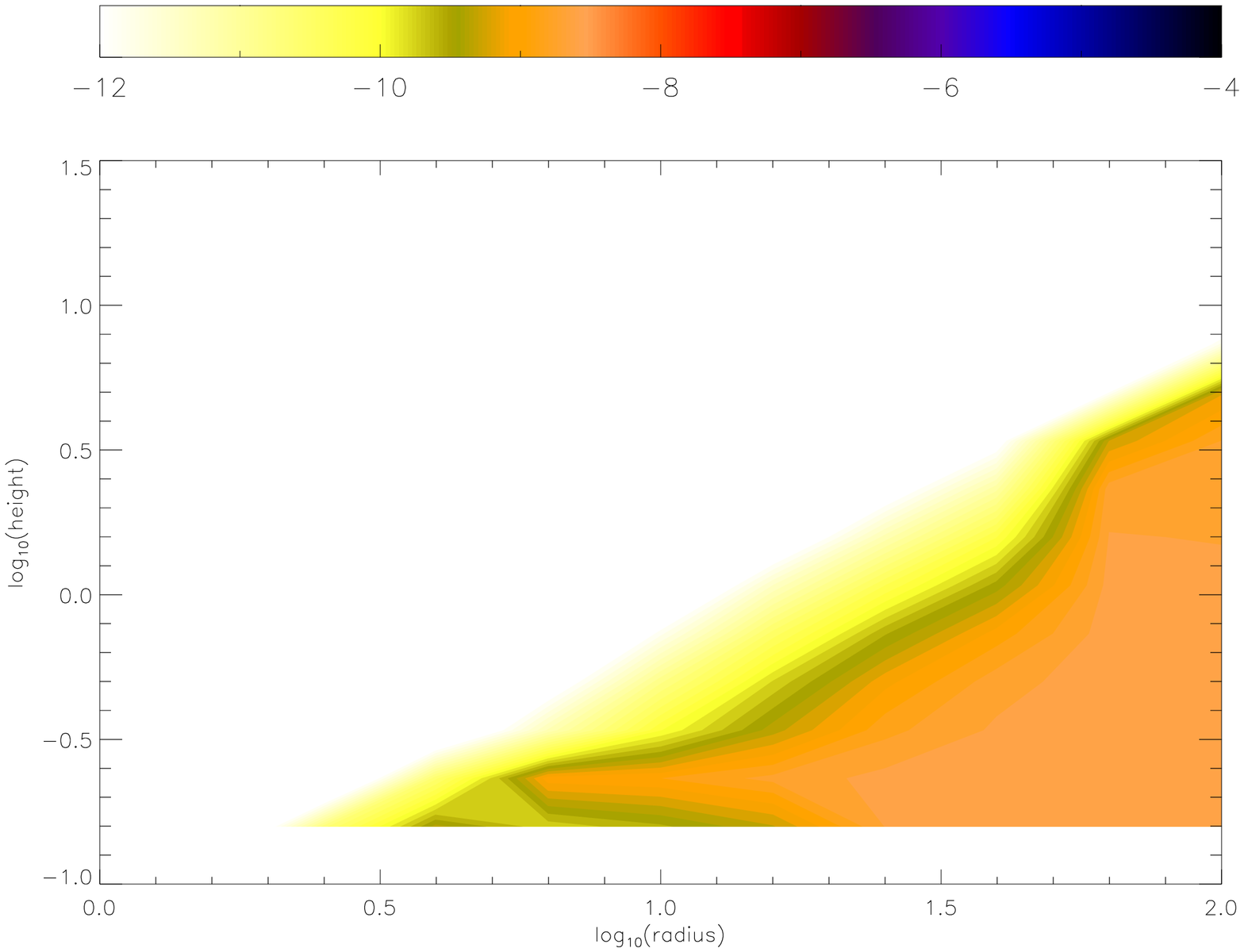}
\includegraphics[angle=90,width=8.5cm]{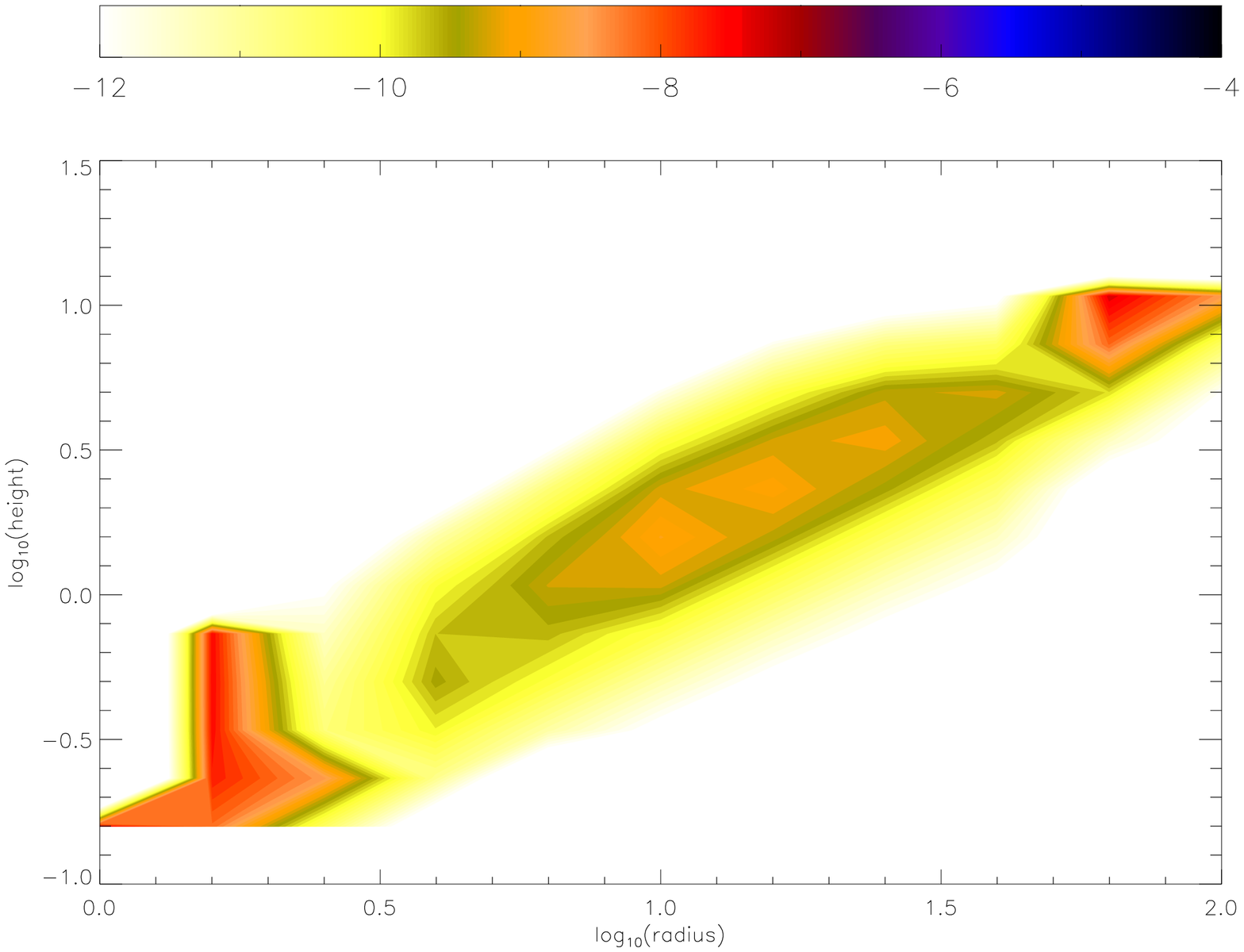}}
\caption{The radial and height dependence of fractional abundances at $t$ = $t_{cross}$ in the Model 4, $Q$ = 5 disk.  {\it Upper left panel:}
HCN; {\it Upper right panel:} CN; {\it Middle left panel:} \cop;  {\it Middle right panel:} HC$_{3}$N;  {\it Lower left panel:} CS;  {\it Lower right panel:} OH$^{+}$.  Note that the color scale is different for CS.}
\label{fig:fab_sfe0_q5_hrv22}
\end{figure*}

The biggest difference between Models 1 and 4 is the larger fraction of the XDR layer over the entire disk in the latter. Its presence changes the abundance pattern of a number of species.  As depicted in Figure~\ref{fig:fab_sfe0_q5_hrv22}~an enhancement of CN compared with Model 1 (see Figure~\ref{fig:fab_sfe0_q1_hrv5}) corresponds generally to the expanded location of the XDR layer. Here, the CN abundance  at $t= t_{\rm cross}$ is $\sim(1-5)\times10^{-8}$, which slightly exceeds the steady-state value of $\sim(1-2)\times10^{-8}$.  Other abundances and abundance ratios are affected by the lowered densities in Model 4.  
For $r\gtrsim5$\,pc, the CS/SO abundance ratio becomes higher than unity  at $t$ = $t_{\rm cross}$. The enhancement does not correspond to the XDR region, and it is instead due to the slow chemical evolution because of the relatively low density.  Finally, the additional X-ray flux can 
dissociate the carbon-chain molecules in regions that correspond to high-temperature-synthesis regions in Model 1.  These species only have moderate abundances at $t=t_{cross}$, and their abundances decrease by orders of magnitude at steady state. 

\subsubsection{Calculated Column Densities and Ratios}
\label{column_ratio}

Our models lead to the prediction of vertical column densities by suitable integration of the calculated abundances.  We list the calculated vertical columns at six radii ranging from $1 - 100$\,pc for 22 atomic and molecular species as a function of radius and temperature.  Column densities at $t= t_{\rm cross}$ are listed in Tables \ref{model1}, \ref{model2}, \ref{model3}, and \ref{model4} for Model 1 ($Q=1$), Model 2 ($Q=1$), Model 3 ($Q=5$), and Model 4 ($Q=5$), respectively, while Tables \ref{model1s}, \ref{model2s}, \ref{model3s}, and \ref{model4s} contain results for the column densities at steady state.  

The calculated column densities are highly dependent on the disk models. 
The molecular column densities of Model 4 are much lower than in  Models 1, 2, and 3 partly because of the larger "no-molecule zone" due to high penetration of X-rays, and also because the $z$-dependence of the density  gives a  lower total column density. For Model 1 at 1 pc, Model 3 at $r<3$\,pc, and  Model 4 at 1 pc, the molecular column densities are smaller than the column densities at larger radii because the molecules are dissociated by X-ray irradiation.

\begin{figure*}
\centerline{\includegraphics[angle=270,width=8.5cm]{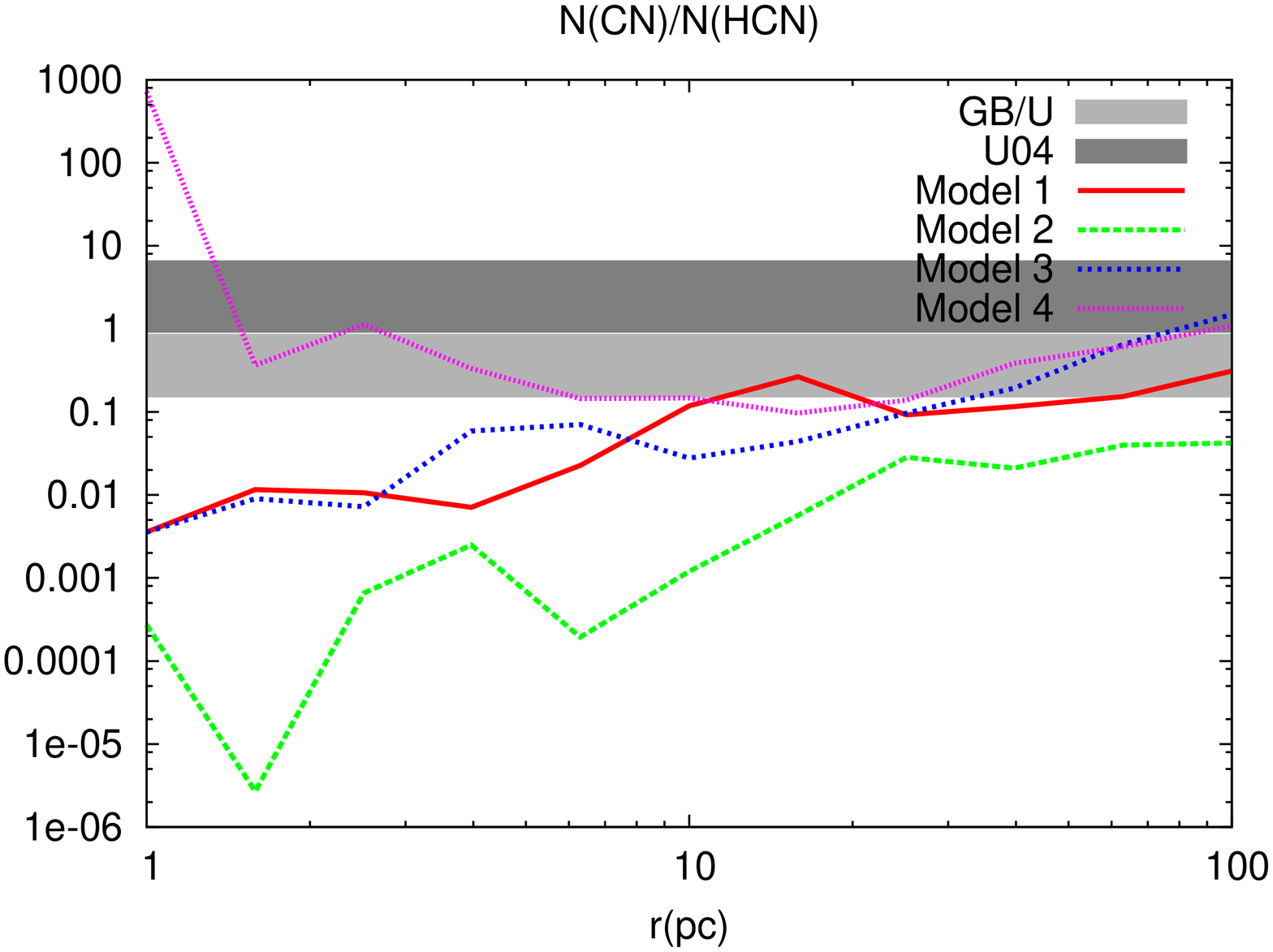}
\includegraphics[angle=270,width=8.5cm]{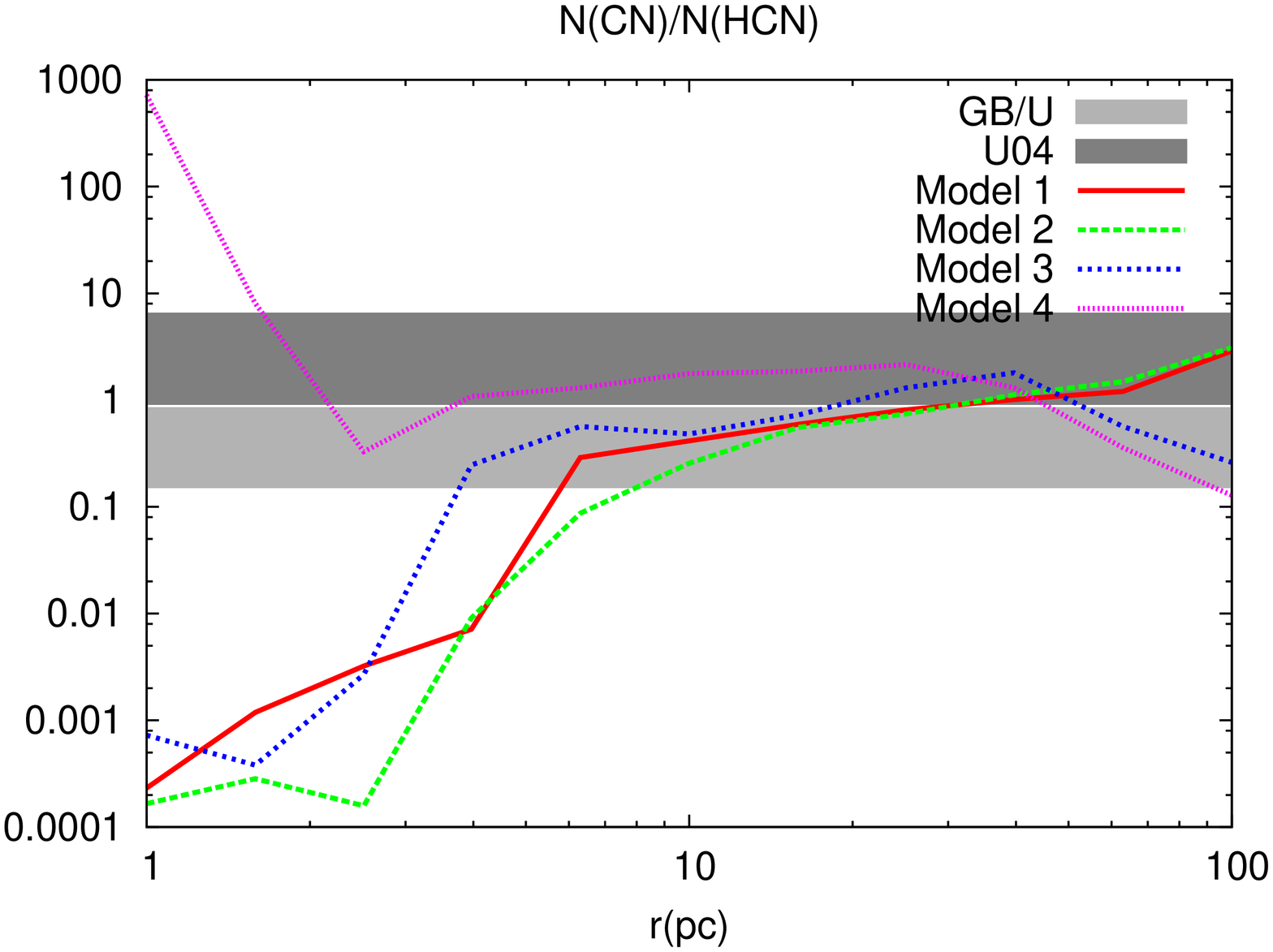}}
\caption{The ratio of CN/HCN vertical column densities vs radius.  {\it Left panel:} at $t$ = $t_{cross}$; 
 {\it Right panel:} at steady-state.  GB/U refers to the ratio of the CN abundance by \citet{2010A&A...519A...2G} over the HCN abundance by \citet{2004A&A...419..897U}. U04 refers to the abundance ratio inferred by \citet{2004A&A...419..897U}.}
\label{fig:cn_hcn_column_q}
\end{figure*}

\begin{figure*}
\centerline{\includegraphics[angle=270,width=8.5cm]{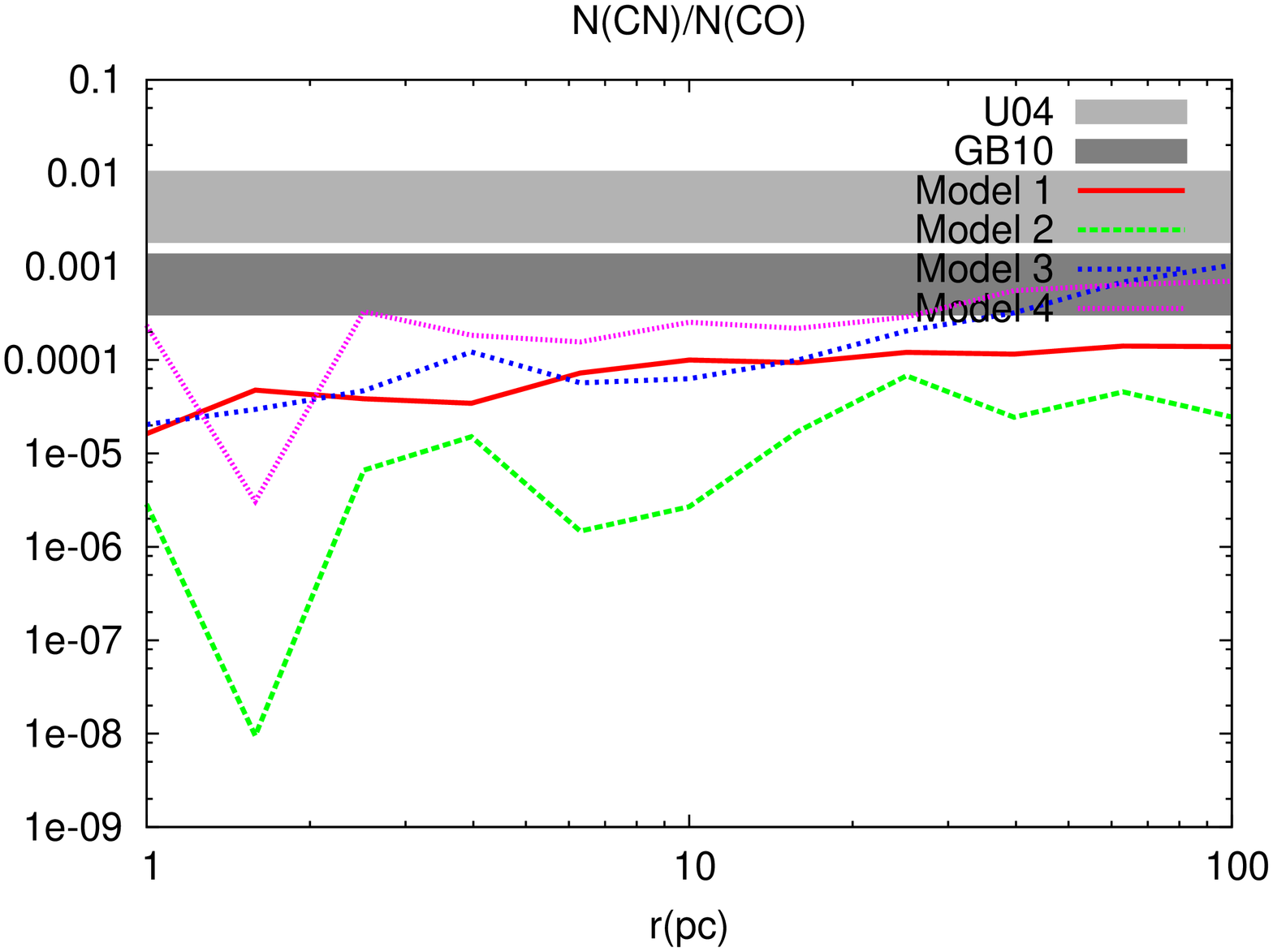}
\includegraphics[angle=270,width=8.5cm]{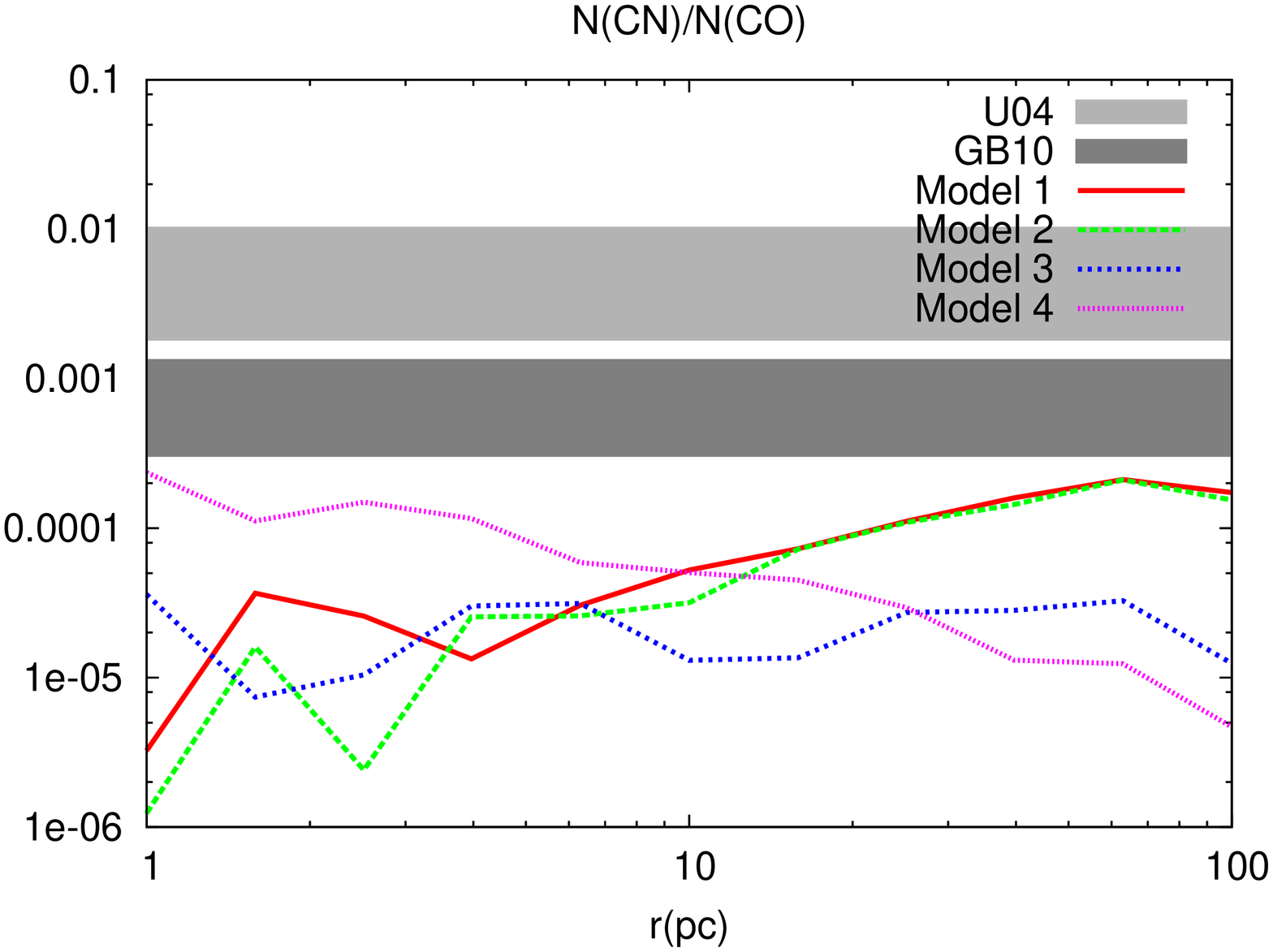}}
\caption{The ratio of CN/CO vertical column densities vs radius.  {\it Left panel:} at $t$ = $t_{cross}$;
 {\it Right panel:} at steady-state. GB10 refers to the observed abundance by \citet{2010A&A...519A...2G}. U04 refers to the abundance ratio inferred by \citet{2004A&A...419..897U}.}
\label{fig:cn_co_column_q}
\end{figure*}

From the calculated columns, we can obtain ratios relevant to observations for each of our models. Below we discuss the specific ratios CN/HCN, CN/CO, HCN/CO, and \cop/CO.   Although strictly ratios of column densities, we often refer to them as abundance ratios below.
The ratio of vertical column density at $t$ = $t_{\rm cross}$ and at steady-state is  plotted vs $r$ for CN/HCN in Figure \ref{fig:cn_hcn_column_q}. In general, this ratio decreases with decreasing $r$ due to the higher temperatures at smaller radii, which favor HCN over CN except for the very inner part where the regions contain very few  molecules. These regions do not contribute to the total molecular column densities.   Because of their larger XDR layers, Models 3 and  4 have the highest ratio of CN/HCN, approaching unity at 100 pc.  
The column density ratio for CN/CO is the highest in Model 4 because of the higher portion of XDR layers, ranging from 10$^{-4} - 10^{-3}$ at $t=t_{cross}$ and 10$^{-5} - 10^{-4}$ at steady state for most of the disk. There is no strong radial dependence for $r>3\,$pc (Fig. \ref{fig:cn_co_column_q}).
 
\begin{figure*}
\centerline{\includegraphics[angle=270,width=8.5cm]{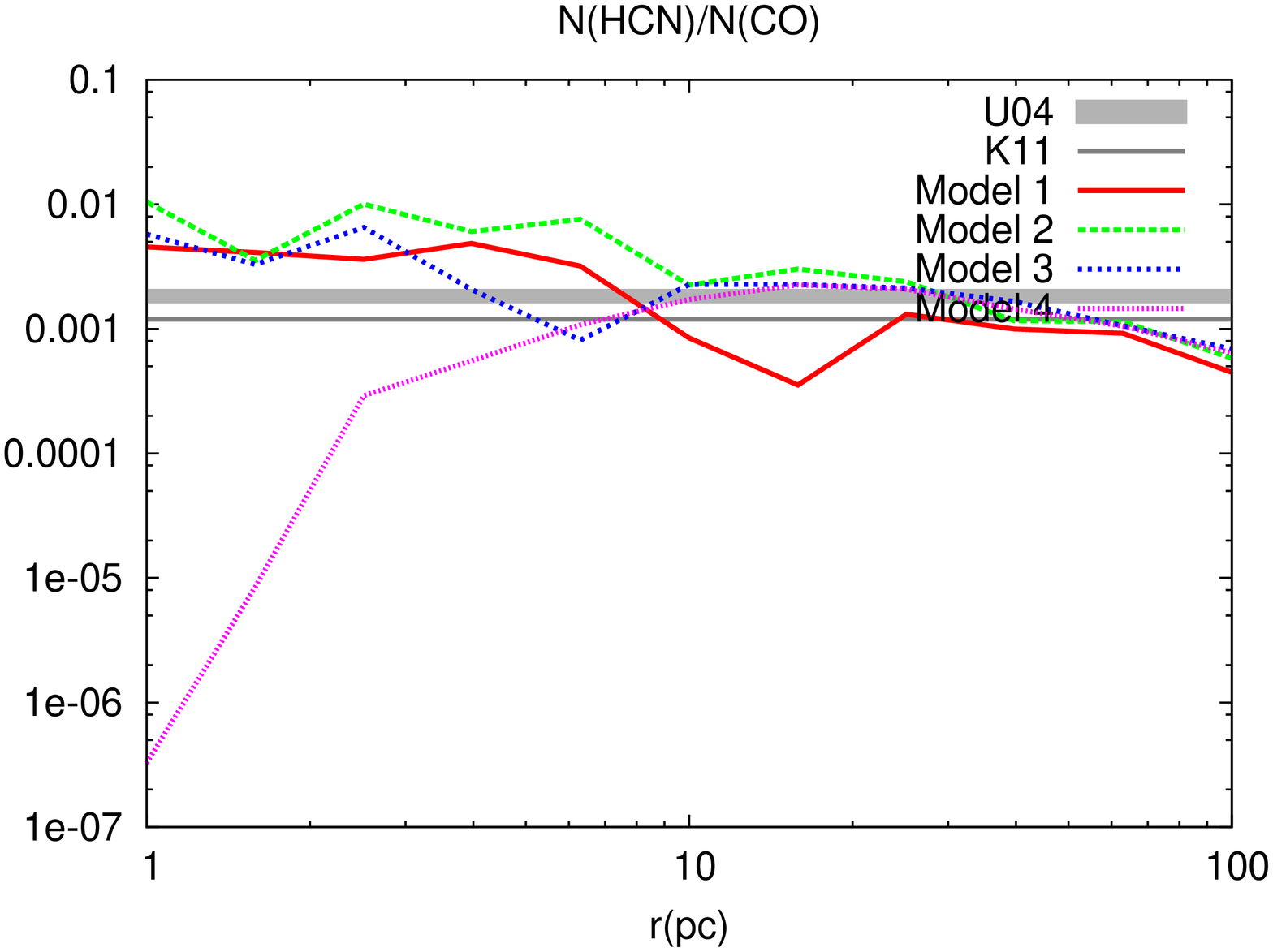}
\includegraphics[angle=270,width=8.5cm]{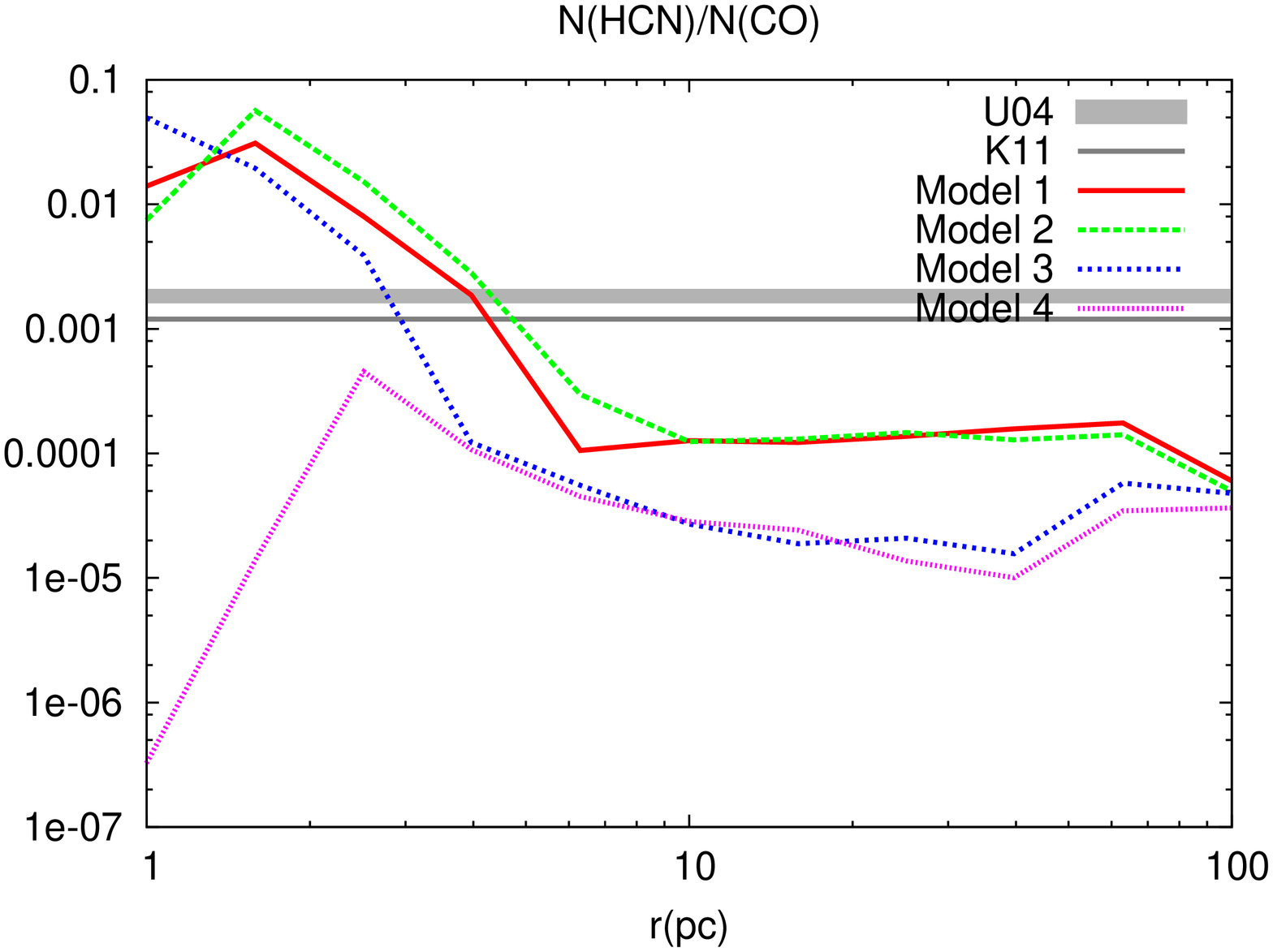}}
\caption{The ratio of HCN/CO vertical column densities vs radius.  {\it Left panel:} at $t$ = $t_{cross}$; 
 {\it Right panel:} at steady-state. K11 refers to the observed abundance of HCN by \citet{2011ApJ...731...83K}.  U04 refers to the abundance ratio inferred by \citet{2004A&A...419..897U}.}
\label{fig:hcn_co_column_q}
\end{figure*}

\begin{figure*}
\centerline{\includegraphics[angle=270,width=8.5cm]{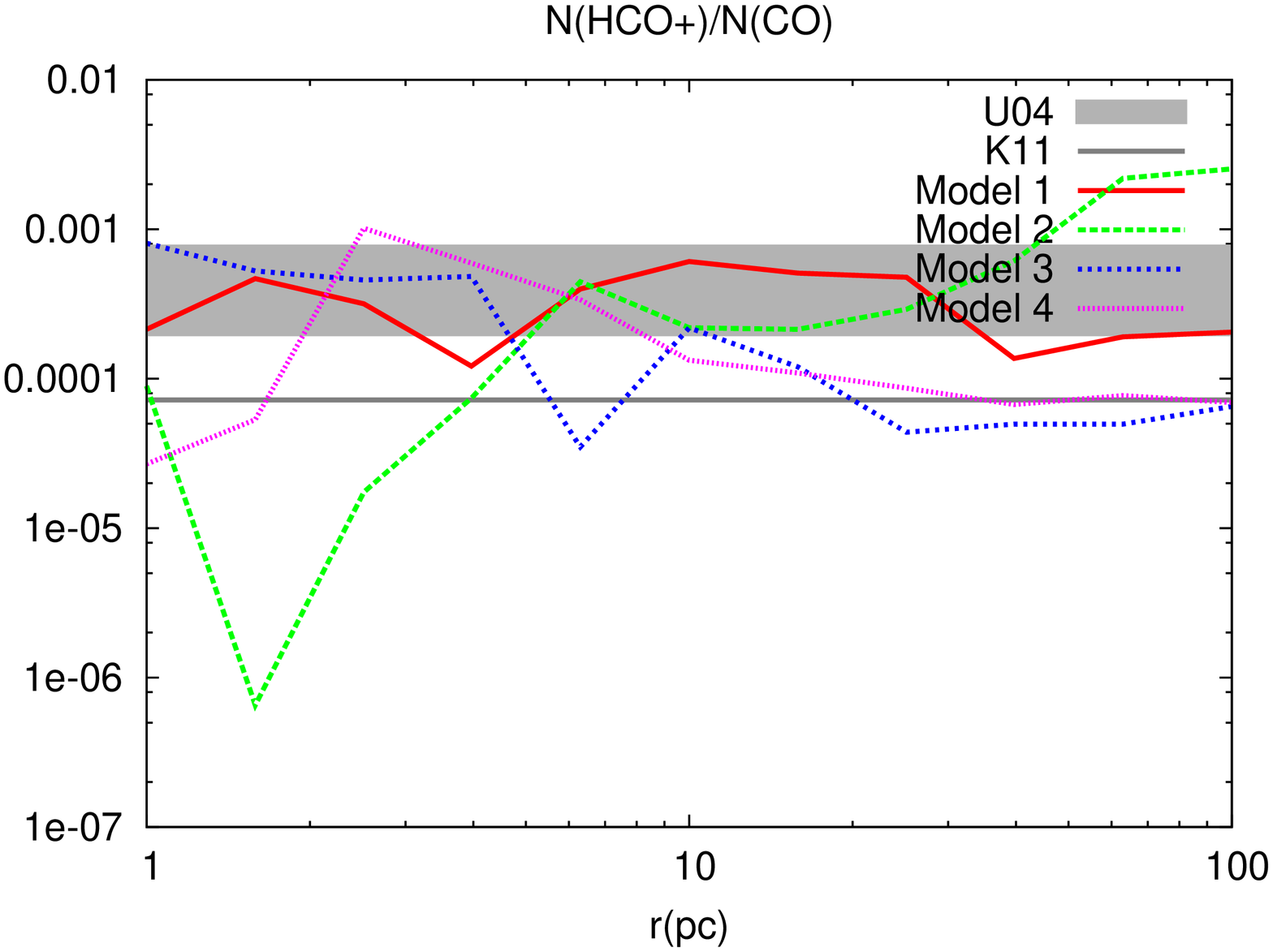}
\includegraphics[angle=270,width=8.5cm]{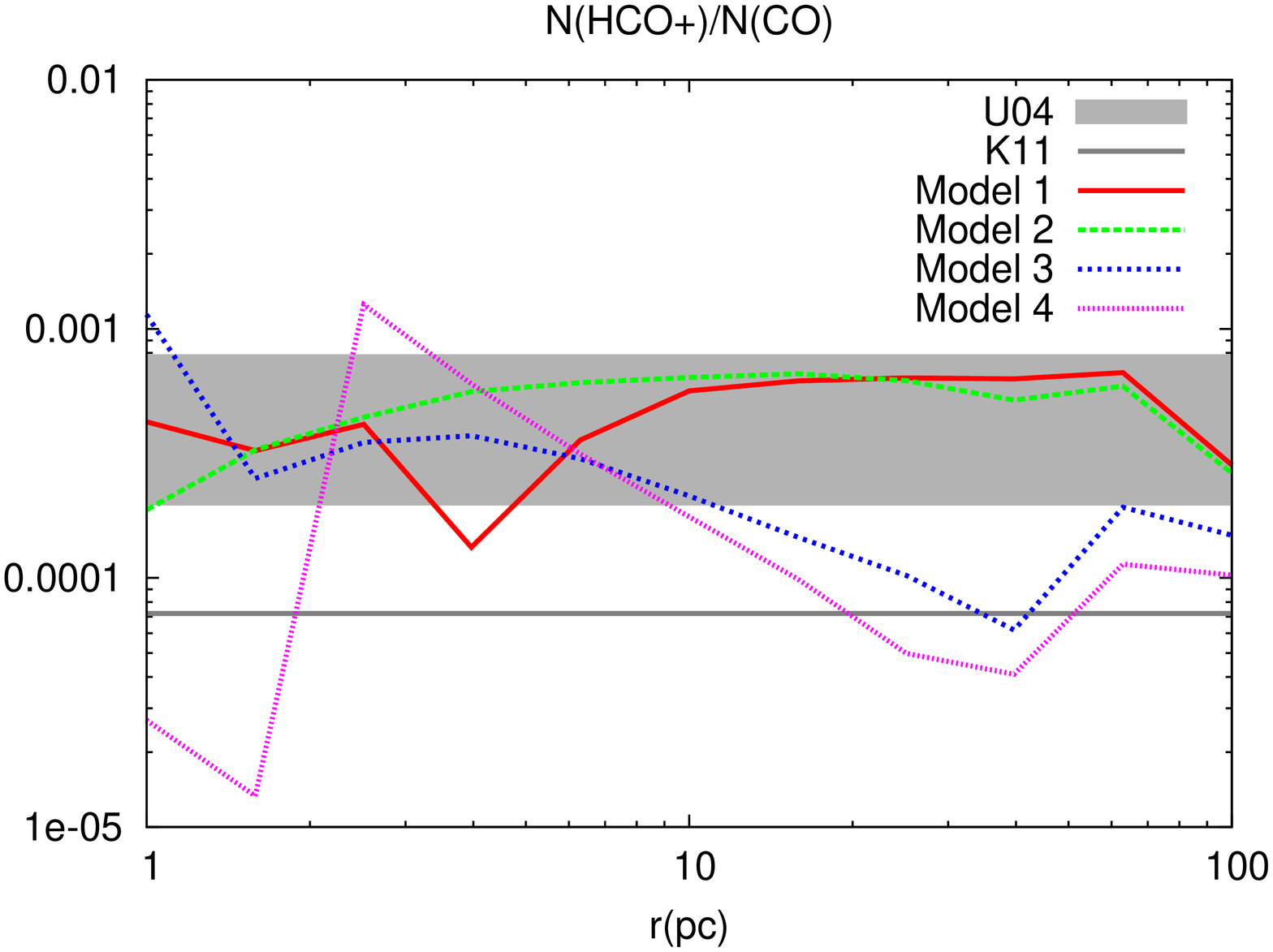}}
\caption{The ratio of \cop/CO vertical column densities vs radius.  {\it Left panel:} at $t$ = $t_{cross}$; 
 {\it Right panel:} at steady-state.  U04 refers to the abundance ratio inferred by \citet{2004A&A...419..897U}.  K11 refers to the observed abundance of \cop~by \citet{2011ApJ...731...83K}.}
\label{fig:hco+_co_column_q}
\end{figure*}

\begin{figure*}
\centerline{\includegraphics[angle=270,width=8.5cm]{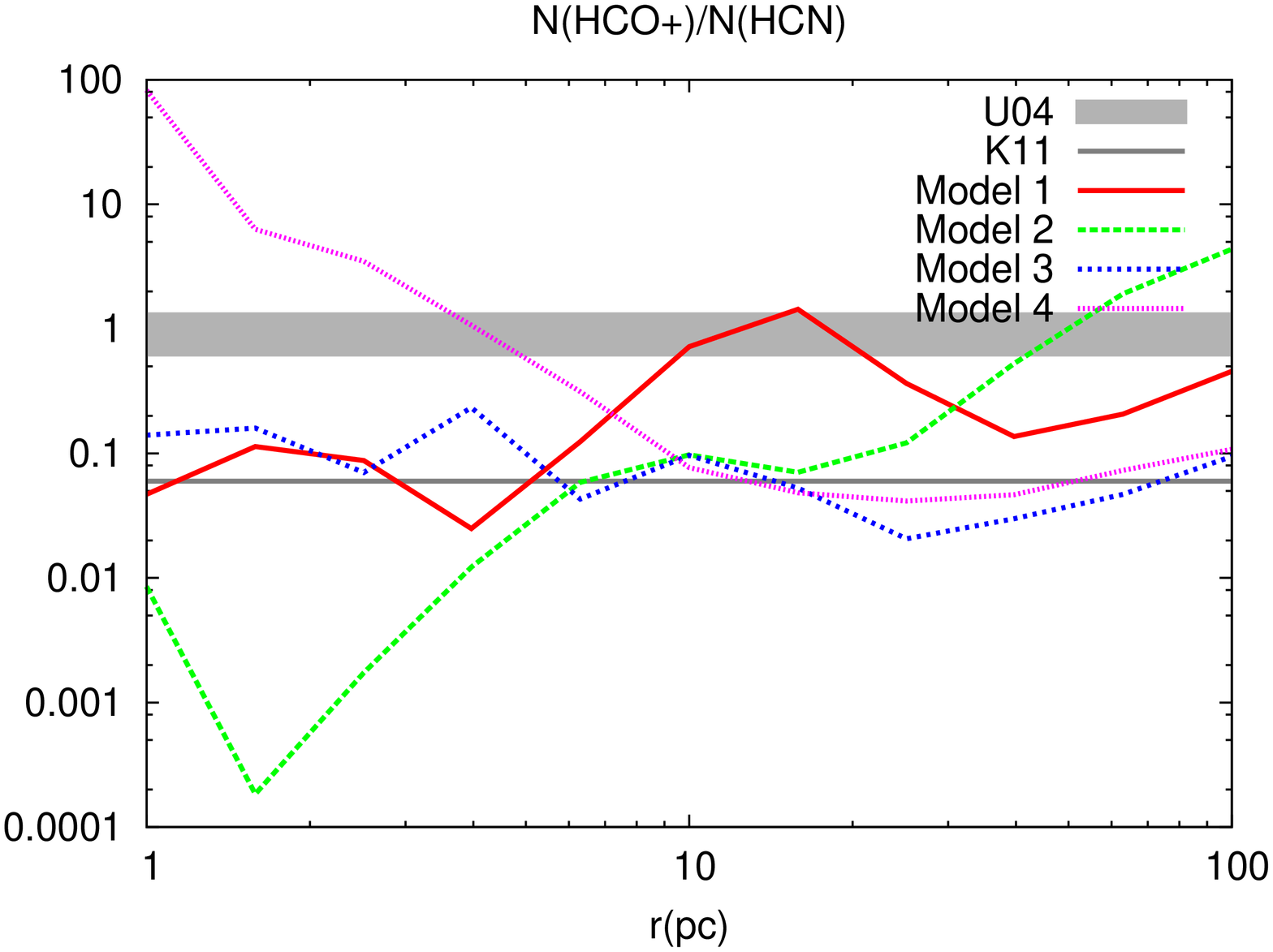}
\includegraphics[angle=270,width=8.5cm]{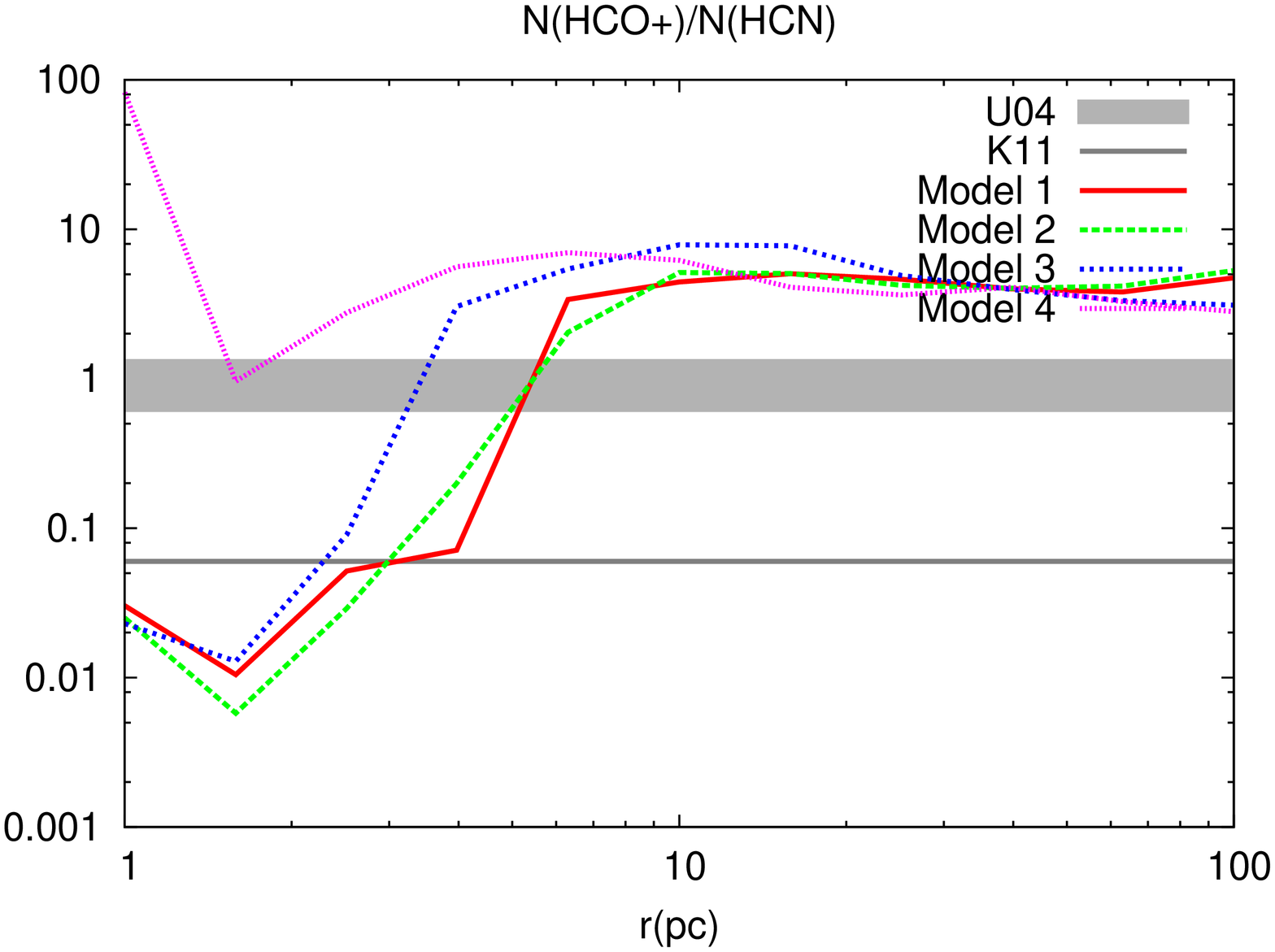}}
\caption{The ratio of \cop/HCN vertical column densities vs radius.  {\it Left panel:} at $t$ = $t_{cross}$;
 {\it Right panel:} at steady-state.  U04 refers to the abundance ratio inferred by \citet{2004A&A...419..897U}.  K11 refers to the observed abundance ratio of \cop/HCN by \citet{2011ApJ...731...83K}.}
\label{fig:hco+_hcn_column_q}
\end{figure*}

 Figure \ref{fig:hcn_co_column_q} shows the HCN/CO column ratio at $t=t_{\rm cross}$ and at steady state. The abundance ratio directly reflects the high-temperature enhancement of HCN at $t=t_{cross}$, except for the inner regions of Model 4, where the strong X-ray dissociation causes HCN/CO to decrease. The HCN/CO ratio is  10$^{-3}$-0.01 for $r<10$\,pc, and decreases for larger $r$ because of the decreasing temperature to several $\times10^{-4}$ for $t= t_{\rm cross}$.  At steady state, the HCN/CO ratio drops down to $\lesssim$10$^{-4}$.  
As shown in Figure \ref{fig:hco+_co_column_q}, the~\cop/CO ratio is around 10$^{-4}-10^{3}$ for most  of the disk.
 Finally, Figure \ref{fig:hco+_hcn_column_q} shows the ratio of \cop/HCN. When the \cop/HCN ratio is too low, it is influenced by the high HCN/CO abundance ratio.

\subsection{Effects of Inclusion of Cosmic Ray Ionization Rate}\label{section:res_cr}
\begin{figure*}
\centerline{\includegraphics[angle=270,width=8.5cm]{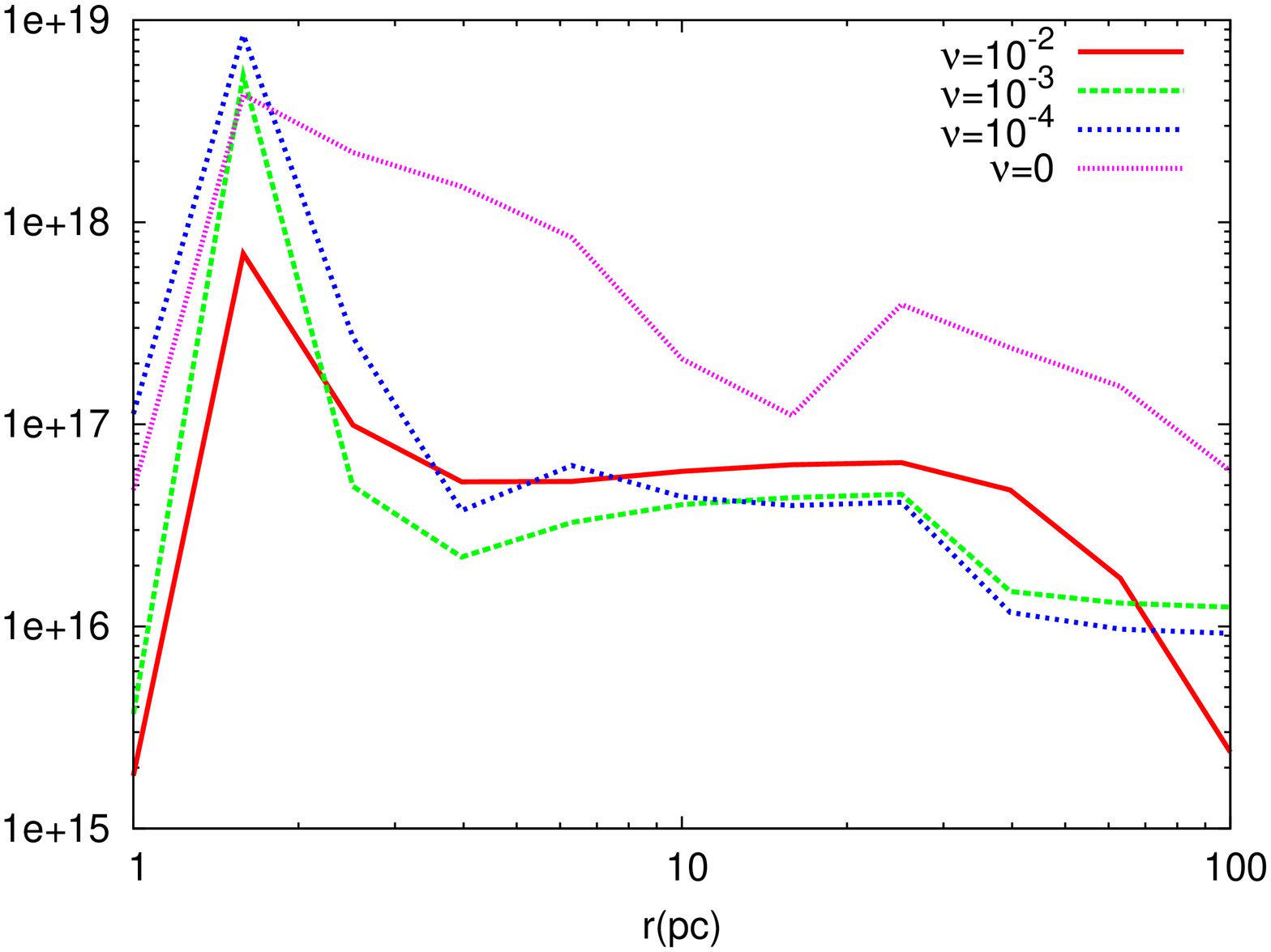}
\includegraphics[angle=270,width=8.5cm]{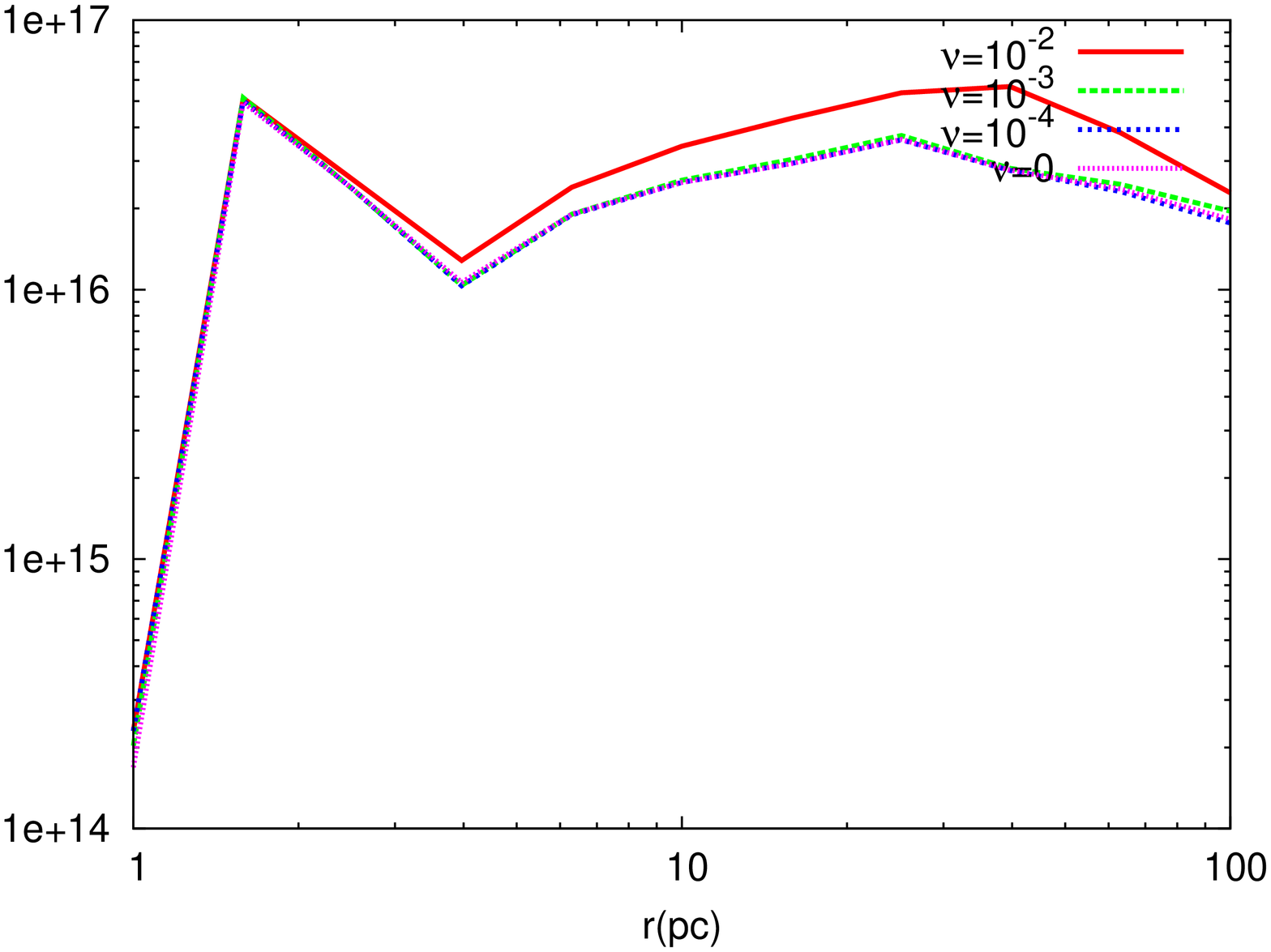}}
\centerline{\includegraphics[angle=270,width=8.5cm]{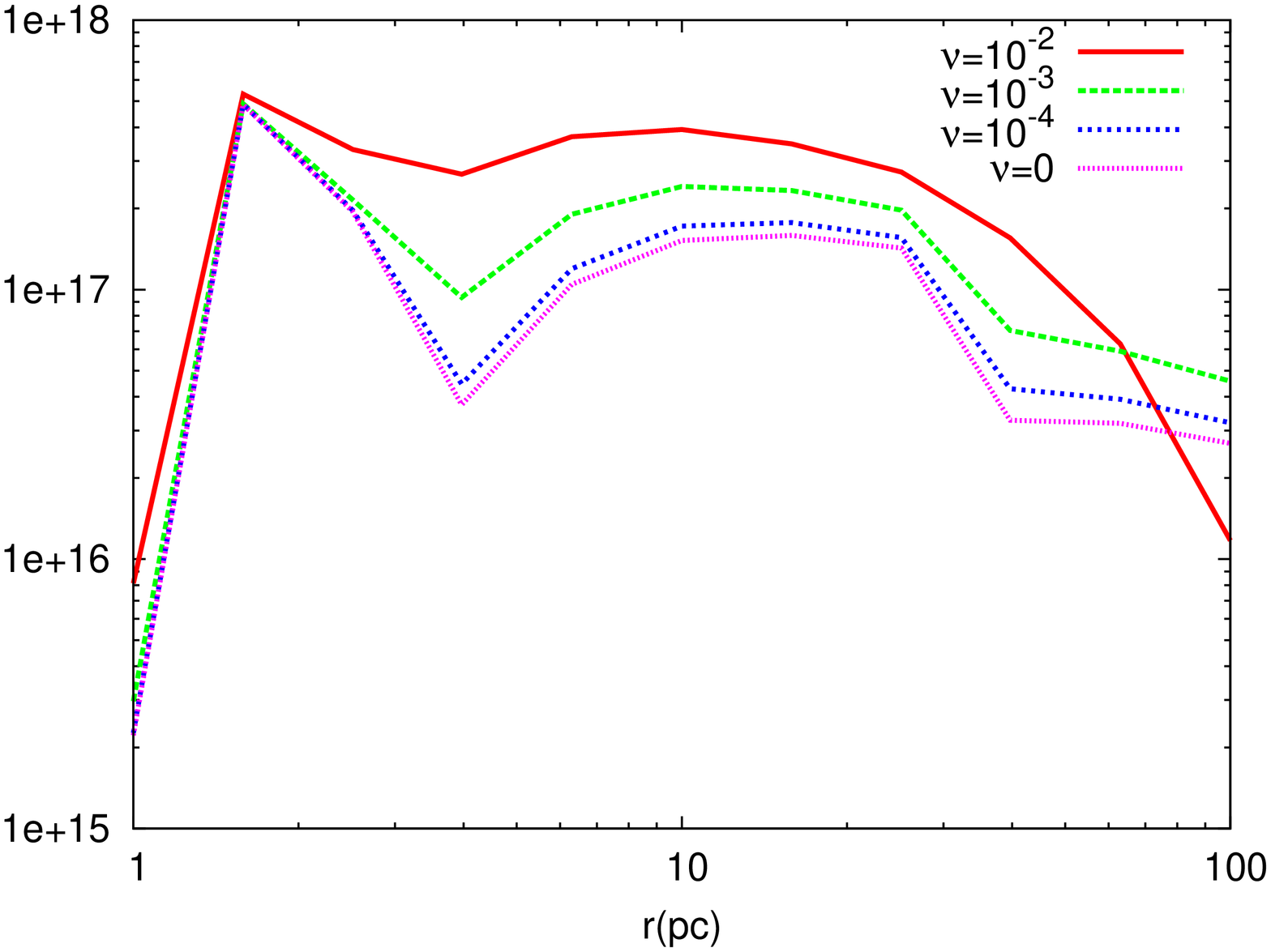}
\includegraphics[angle=270,width=8.5cm]{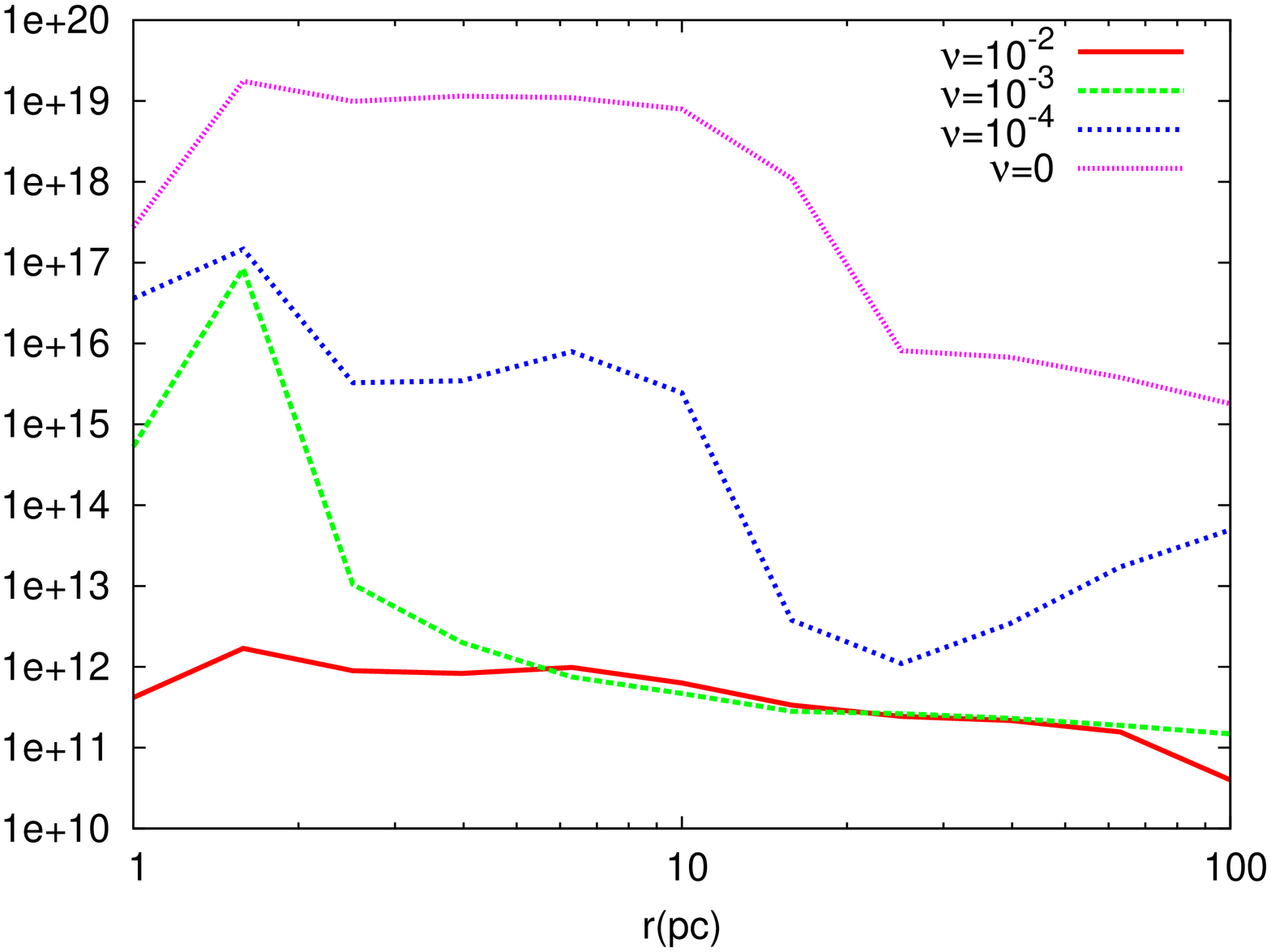}}
\caption{The variation of  assorted abundances vs radius with different cosmic-ray ionization rates caused by differing star formation efficiencies $\nu$ for the Model 1, $Q$ = 1 disk.  {\it Upper left panel:} HCN;
 {\it Upper right panel:} CN;  {\it Lower left panel} \cop; {\it Lower right panel:} HC$_{3}$N.}
\label{fig:column_sfe}
\end{figure*}

Our results presented so far have been obtained without the inclusion of cosmic ray ionization.  We have also run models that include a range of cosmic ray ionization rates by varying the star formation efficiency $\nu$ (see Section \ref{section:cr}).  The effects of cosmic rays from star formation become dominant  over X-ray ionization only when $\nu\gtrsim0.01$.  
Figure \ref{fig:column_sfe}~shows the variation of the total vertical column densities of HCN, CN, and HCO$^{+}$~respectively as a function of radius for values of $\nu$ equal to 0, 10$^{-4}$, 10$^{-3}$, and 0.01 at $t$ = $t_{\rm cross}$ for Model 1, $Q$ = 1. The HCO$^{+}$ abundance can be enhanced by a factor of a few when the star formation efficiency is higher, although HCN decreases when cosmic-ray ionization is included. The column of CN is enhanced only slightly when $\nu=0.01$.

 Inclusion of cosmic-rays can also dissociate complex molecules such as C$_{3}$H$_{2}$ and HC$_{3}$N for $\nu>10^{-3}$
(see eq.~\ref{sfrvol}). Although cosmic-rays can heat the gas to a few $\times100\,$K, the high ionization rate can cause dissociation of molecules preventing more complex molecules from forming.

\subsection{Line intensities: Preliminary Results}\label{section:intensity}

We have computed observable intensities of selected molecules using the publicly available three-dimentional radiative transfer code $LIME$ \citep{2010A&A...523A..25B}.  Since our models span a wide range of physical conditions, including those that are highly optically thick, we are unable to achieve complete convergence in some regions, for some species and transitions.   In particular, in $LIME$ the signal-to-noise ratio of the population convergence is a ratio of the current population over the geometric mean of the fluctuations in the population over the last 5 iterations. 
Out of 80,000 grid points used, the lowest the signal-to-noise ratio is below 1. The average value of signal-to-noise ratio for CO reaches more than 100 although these ratios are lower for HCN and HCO$^{+}$, which is somewhere above 40 for model 2 and 20 for model 4. Lower levels of convergence come mostly from less populated high $J$ levels with weaker emission, and should not affect the overall results for the low-$J$ lines ($<4$) discussed in this paper.  Even so, the results presented here are thus preliminary, and should only be used to understand general trends. With this precaution, we present our preliminary results on the line transfer here with selected models at $t=t_{cross}$.

The $LIME$ code requires density, temperature, molecular abundance, velocity, and the doppler width for the input parameters. The velocity of the rotation and doppler width of $h \Omega$ are used and images are produced with a viewing angle of 45 degrees. Results presented here are convolved with a beamsize of 0.25 arcsecond ($\sim$ 18\,pc at the distance of NGC 1068) using {\it MIRIAD} \citep{1995ASPC...77..433S}. Finally, figures are produced with the GREG program in the GILDAS package \footnote{http://www.iram.fr/IRAMFR/GILDAS}. When intensity ratios are presented, we use units of K km s$^{-1}$ instead of Jy beam$^{-1}$ km s$^{-1}$. Here we present the results for Model 2 and 4 since Model 2 is simpler to interpret than Model 1, which has a "torus," and Model 4 has significantly different results from other models.

\begin{figure*}
\centerline{\includegraphics[angle=0,width=8.5cm]{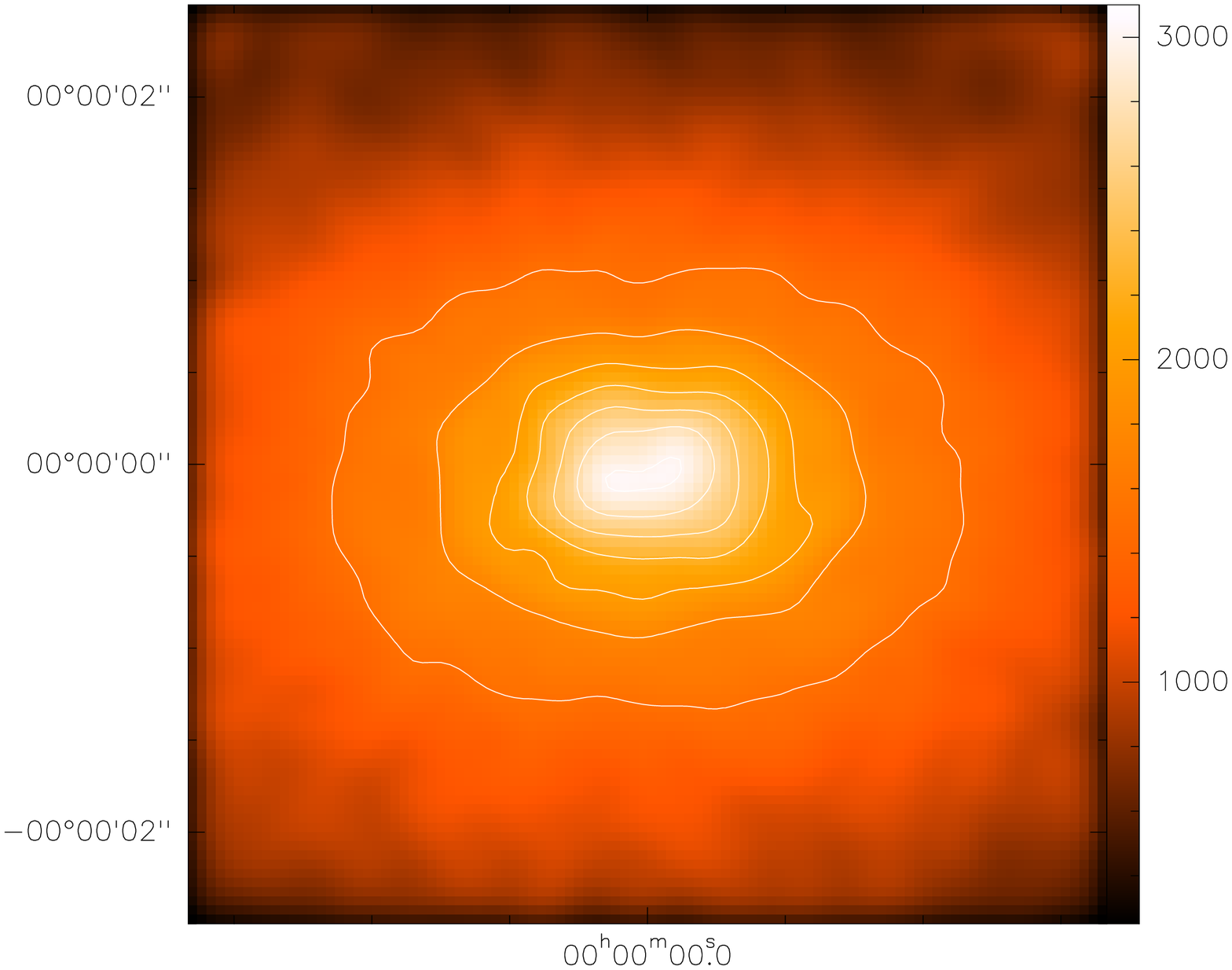}
\includegraphics[angle=0,width=8.5cm]{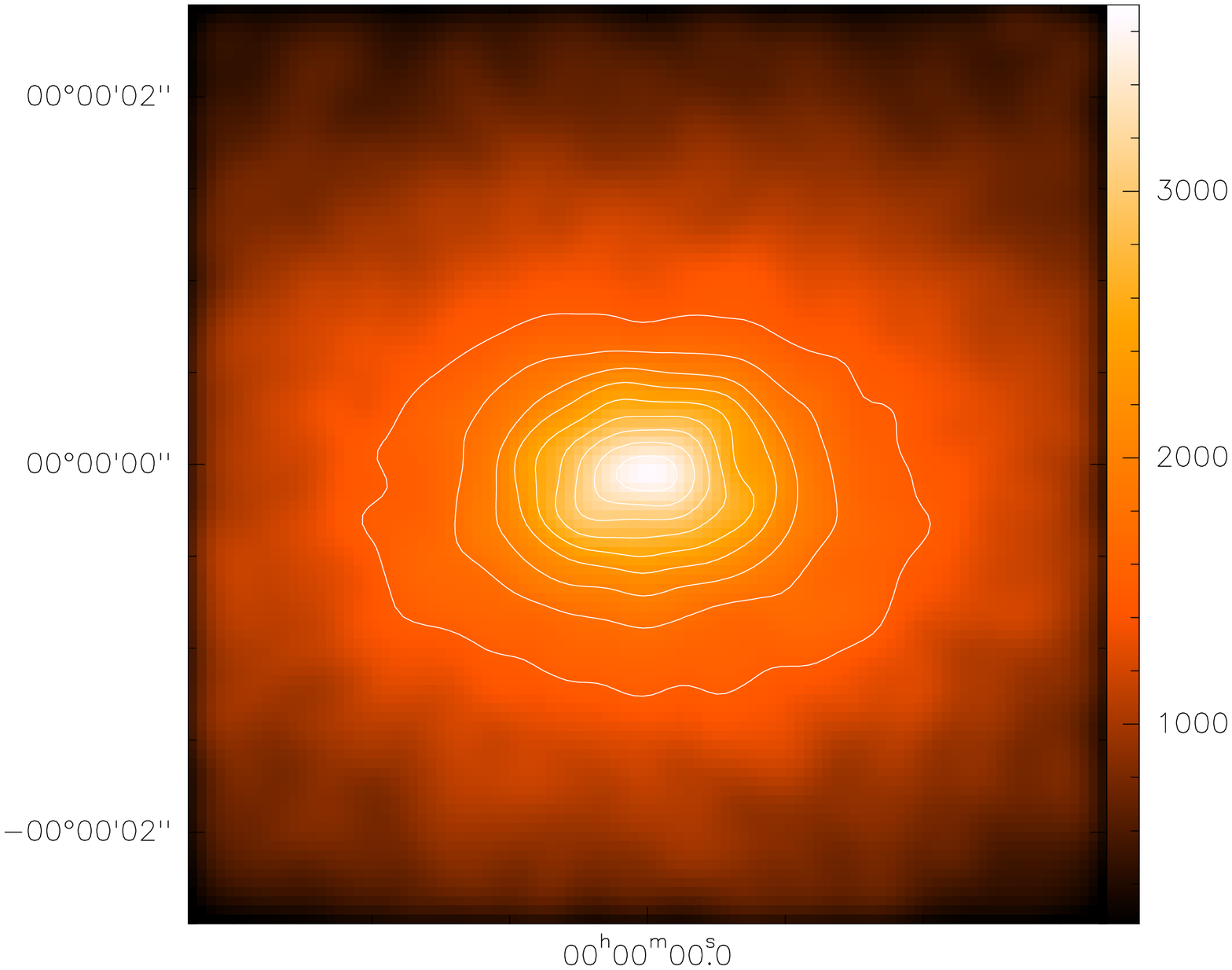}}
\centerline{\includegraphics[angle=0,width=8.5cm]{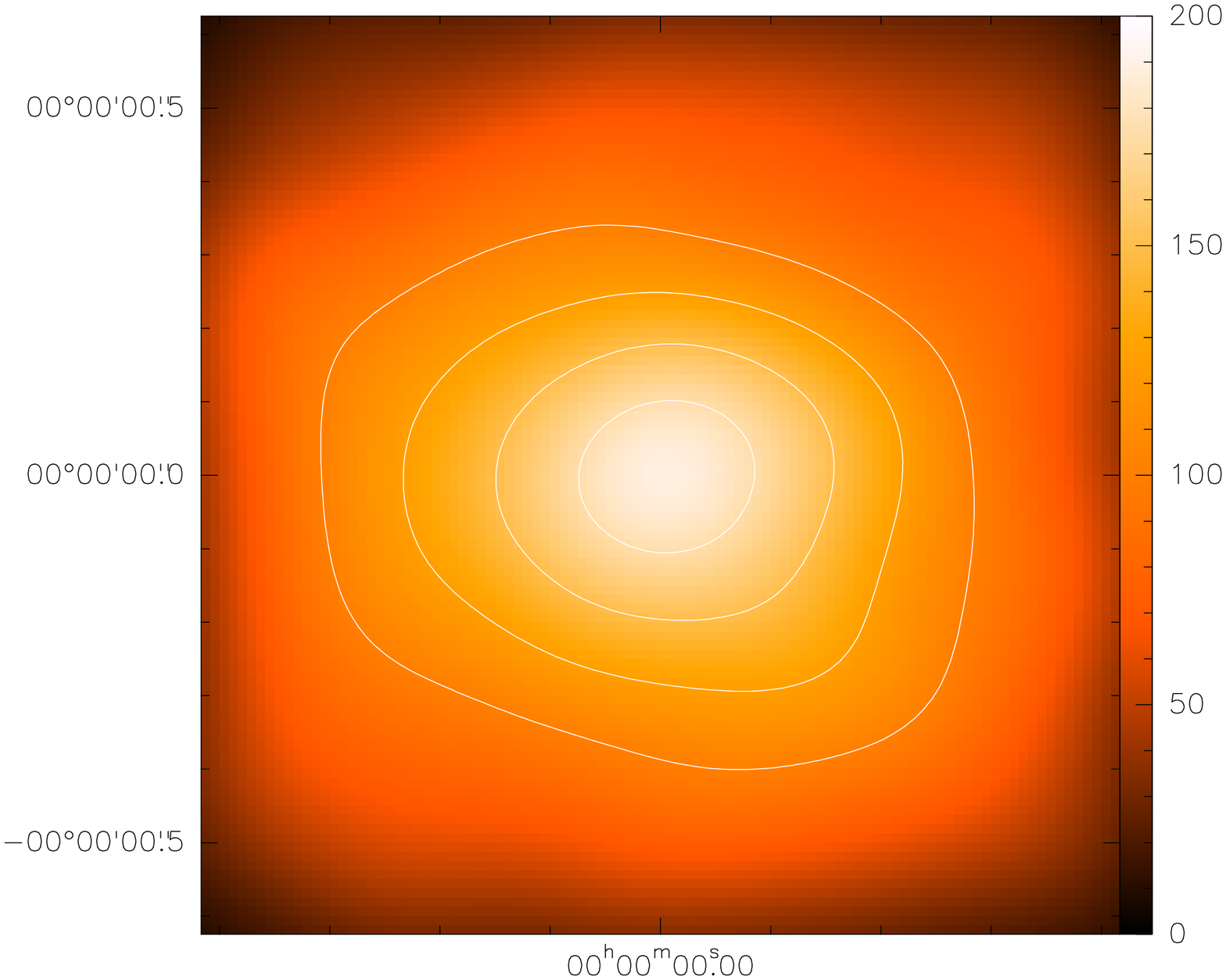}
\includegraphics[angle=0,width=8.5cm]{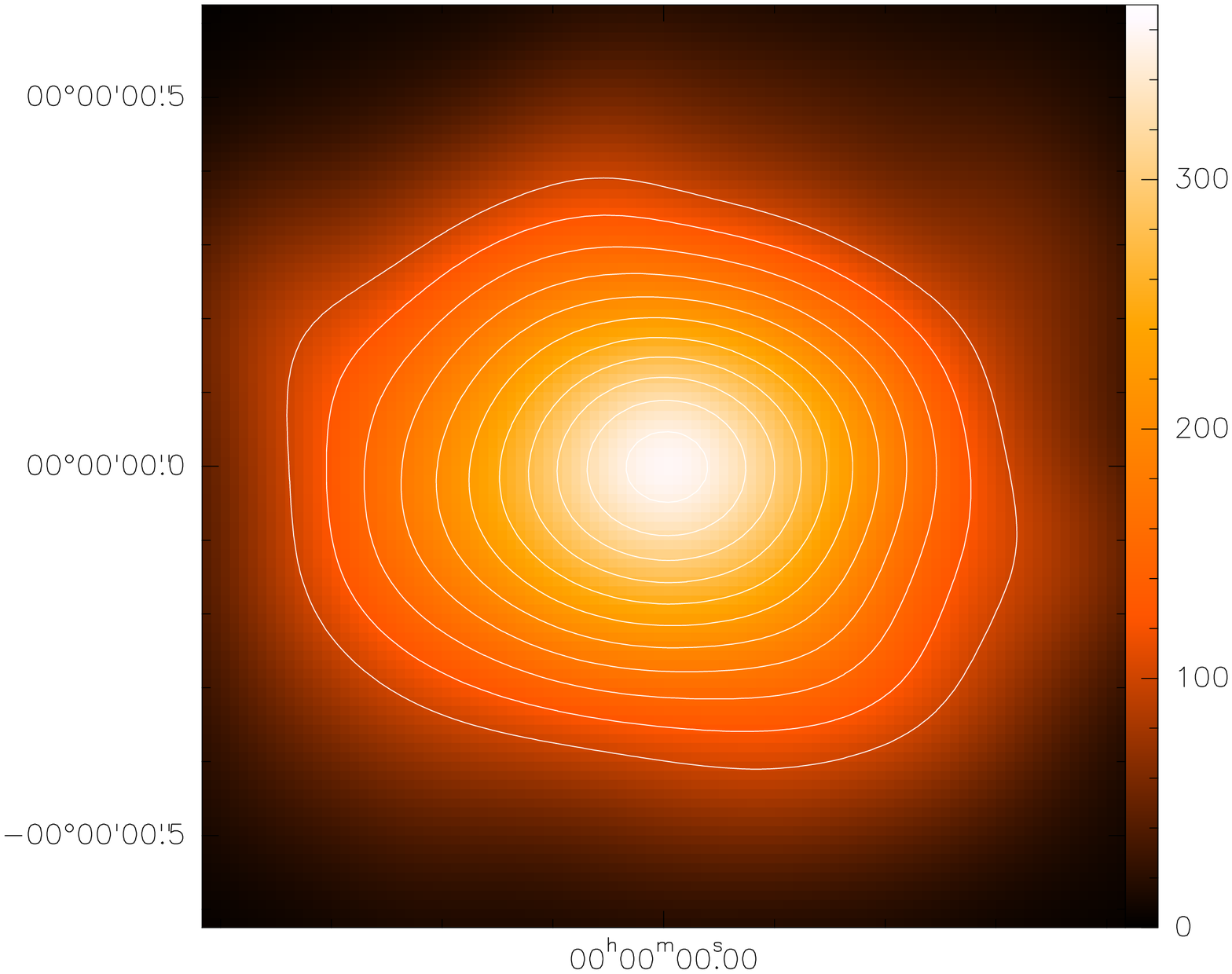}}
\caption{Velocity-integrated line intensities of CO emission are shown.  {\it Upper left panel:} Model2 CO(1-0). Contour levels are from 1500 to 3000 K km s$^{-1}$ in steps of 250 K km s$^{-1}$.;
 {\it Upper right panel:} Model 2 CO (3-2). Contour levels are from 1500 to 3500 K km s$^{-1}$ in steps of 250 K km s$^{-1}$.;  {\it Lower left panel} Model 4 CO (1-0). Contour levels are from 100 to 175 K km s$^{-1}$ with 25 K km s$^{-1}$ increments.; {\it Lower right panel:} Model 4 CO (3-2). Contour levels are from 100 to 350 K km s$^{-1}$ in steps of 25 K km s$^{-1}$.}
\label{fig:co_int}
\end{figure*}

%
{\bf \underline{Model 2:}}
Because of the radial differences in abundances and excitation conditions, the apparent peaks in the intensity of molecular emission can change 
depending on the species and the line considered. 
Transitions of CO $J=1-0$, $J=2-1$, and $J=3-2$ peak around 10\,pc from the AGN core with velocity-integrated intensities of about $\sim 3000 - 3500\,$K km s$^{-1}$ (see Fig \ref{fig:co_int}~upper panels). In general, since higher transition lines have higher critical densities, their intensities decrease with increasing radius faster than lower transition lines. For example, we find that 
the CO($3-2$)/($1-0$) ratio is slightly higher than 1 ---  around 1.2 --- within 30\,pc of the AGN, but falls below unity farther away.
The HCN $J=1-0$ and $J=3-2$ emission peaks have intensities of 2500 and 2200 K km s$^{-1}$ as it can be compared in the upper panels of Fig \ref{fig:hcn_int}. The ratio of HCN($J=3-2$)/($J=1-0$) is highest at 100\,pc away from the AGN with a value of 1.4. This ratio is smaller ($\sim$0.8) closer to the AGN.
Since the CN abundance is lower near the AGN, the peak of CN (2$_{5/2}$-1$_{3/2}$) is about 100\,pc away from the AGN with an intensity of 90\,K km s$^{-1}$ (see Fig \ref{fig:hcn_int}~left panel). 
The HCO$^{+}$($1-0$) emission is about a half of HCN($1-0$) on scales of a few to 10\,pc, but the ratio HCO$^{+}$/HCN$(2-1)$ becomes close to unity with increasing radius as shown in Fig \ref{fig:hcop_int}.
%
%
The intensity of HC$_{3}$N$(10-9)$ is on the order of thousands of K km s$^{-1}$, equivalent to the HCN emission intensity.  Intensity ratios of HC$_{3}$N(10-9) over HCN(1-0) can be only slightly lower than the order of unity in (U)LIRGs \citep[e.g, $\sim 0.78$ in Arp 220 and 0.40 in NGC 4418][]{2011A&A...527A.150L}, but not in typical AGNs. The abundance may instead be suppressed because of the grain depletion. Alternatively, the bottom right panel of Fig \ref{fig:column_sfe} shows that even a moderate amount of cosmic-ray ionization can dissociate HC$_{3}$N and also suppress its abundance.

{\bf \underline{Model 4:}}
Since the overall column density of CO is much lower than that of Model 2, the peak velocity-integrated intensity is also much lower. The peak intensity of CO($1-0$), CO($2-1$), and CO($3-2$) are 200, 300, and 350\,K km s$^{-1}$. The intensity ratios of CO($3-2$)/($1-0$) and CO($2-1$)/($1-0$) are 1.5 and 2 respectively around the AGN. 
The velocity-integrated intensity of HCN($1-0$) is $\sim150\,$K km s$^{-1}$ within 30\,pc, but it falls rapidly with increasing distance from the AGN, and is around 50\,K km s$^{-1}$ near 50\,pc. The intensity ratio of HCN($3-2$)/($1-0$) is about 2 around 10\,pc from the AGN, but it decreases with increasing radius, and it is below unity when $r\sim 50\,$pc.
Much lower intensities than those of CO and HCN are predicted for CN. The intensity of CN($1_{3/2}-0_{1/2}$) is 20\,K km s$^{-1}$ at 10\,pc from the AGN, and drops to 5 K km s$^{-1}$ near 20 pc. CN($2_{5/2}-1_{3/2}$) has about 1.5 times more intensity than CN($1_{3/2}-0_{1/2}$) in around 10\,pc, although it drops faster further out from the AGN. The higher level transition CN($3_{7/2}-2_{5/2}$) has a higher intensity of 80 K km s$^{-1}$, but decreases to less than 1 K km s$^{-1}$ farther out.
Similar intensities to HCN are found for \cop($3-2$) and ($1-0$) with the same $J$-number transition near 10pc from the AGN core, but their intensity drops to about a tenth of the HCN lines around 30\,pc away from the AGN.
 Although the abundance of HC$_{3}$N is on the order of 10$^{-9}$, which is much smaller than that in Model 2, there is a small region of several pc in Model 4 where HC$_{3}$N emission is seen with HC$_{3}$N(10$-$9)/HCN(1$-$0)$\sim$0.1.

In general, almost the entire circumnuclear disk (CND) of Model 4 is XDR-dominated, and has higher intensities for the higher $J$-transition lines compared with Model 2. Just as the abundance ratio of CN/HCN is higher in Model 4 than other models, the intensity ratio of CN/HCN is also higher.

\section{Discussion}\label{discussion}

\subsection{Comparison with Observations: A Case of NGC 1068 - a Prototypical AGN}

Although our models have not been tuned to represent any individual galaxy or AGN disk, a  general comparison of models with observations helps to understand the physics and chemistry, and can serve as a guide both to future modeling and observations.  Here, we compare our models with observations of a prototypical AGN containing galaxy NGC 1068. As a reference, observed intensities in NGC 1068 are listed in Table \ref{table:obs}. It must be noted that some of the observations in this table have different beamsizes and thus cannot be compared directly.

\subsubsection{CO}

The CO observations by \citet{1995ApJ...450...90H} show spiral arms on kpc scales, and a nuclear "ring" at radius $\sim$ 200 pc. A higher resolution image by \citet{2000ApJ...533..850S} reveals two peaks, each $\sim$ 100 pc away from the central black hole, the so-called ``Eastern" and ``Western" knots, with Eastern knot having higher peak intensities by a factor of 2 than the Eastern knot for $J=1-0, 2-1$, and $3-2$ lines of CO as summarized in Table \ref{table:obs}.  If the disk is a regular Keplerian disk and the CO abundances do not vary strongly, then the high velocity components must come from regions close to the center of the CND, and the two emission peaks should be symmetric. However, the CO emission appears asymmetric, and off from the expectations of a simple disk.  \citet{2000ApJ...533..850S}~suggest  a warp in the disk as a possible cause for the observed double knots  with a complicated structure. \citet{2003A&A...412..615G} constructed a disk model without a warp, which is irradiated by X-rays with an angle offset from the perpendicular axis of the disk. According to \citet{1996ApJ...466..561M}, the CO intensity peaks where the X-ray intensity per density is moderately high, 
with $H_{\rm X}/n\sim10^{-26}-10^{-25}\,$erg cm$^{3}$ s$^{-1}$ (See Eq. \ref{eq:hx}~for the definition of $H_{\rm X}$). 
If the peaks are indeed created at regions illuminated by the X-ray ionization cones, one of them should appear more obscured than the other, as illustrated in Figure 9 of \citet{2003A&A...412..615G}. 
 A hydrodynamic simulation by \citet{2005ApJ...619...93W} without X-ray irradiation and with a constant value for the so-called X-factor, the conversion factor from CO intensity to \hmol~column density, also explains this asymmetric double peak with sufficiently large CO beams of $\sim1$'', although smaller beams at a fraction of an arcsecond are predicted to reveal ring-shaped emission in their model.  
 
 In our models, we find that the CO fractional abundance is mostly constant with $(5-7)\times 10^{-5}$ throughout the entire molecular region, and higher molecular emission also seem to come from the regions with larger column densities.  
Our models 2 and 4 both show centrally-peaked emission, which does not match the observations of NGC 1068. Although some of the emission near the AGN core might be absorbed by the surrounding colder gas, it is likely that there is simply a lack of molecular gas in this region, perhaps caused by feedback from the AGN. The total gas mass presented in \citet{2009ApJ...696..448H} from their warm H$_{2}$ observation within 60 pc from the AGN is 10$^{7}\,M_{\odot}$, which is about an order of magnitude lower than the case in Model 2.

The ratio between high-$J$ and low-$J$ transitions --- e.g., CO($3-2$)/CO($1-0$) --- is higher in the observations than in Models 2 and 4, which could also be explained if the density is lower than our models since there will be smaller amount of cold gas with the lower density. In Model 2, this ratio does not exceed unity in units of K km s$^{-1}$ and in Model 4 these ratios are only $\simeq1.5$ around their peak location. As shown in Table \ref{table:obs}, the observed ratio of CO$(3-2)/(1-0)$ in NGC 1068 is 7. 
Although Model 4 is closer to the observations in terms of the ratio of high-$J$ to low-$J$ transitions, the intensities fall more rapidly than 
the observations with increasing radius. Model 2, on the other hand, has too high an intensity near the AGN core, but matches the observations fairly well on larger scales. Thus the actual CND of NGC 1068 must have lower density than our models on pc scales, but similar structure to Model 2 on 10 pc scales.  Of course, because we have not attempted to fit the observations of NGC 1068, only a qualitative comparison should be made.  We discuss other molecules below, 
highlighting qualitative and interesting differences between the models  and the observations as a guide to future work.

\subsubsection{HCN \& HCN/CO}
Similar to CO, the observed HCN emission also has Western and Eastern knots, with stronger emission from the former.
Although detailed spatial information for HCN within the CND is not known, \citet{1994ApJ...426L..77T} observed the HCN($1-0$) at a  similar level of intensity to CO ($1-0$). For $J=3-2$, the HCN/CO intensity ratio in units of K km s$^{-1}$ is around $\sim 0.2$, and increases to $\sim 0.3$ around the AGN core. 
\citet{2004A&A...419..897U} estimated the column density ratio of $N$(HCN)/$N$(CO) to be $(1.6-2.0)\times10^{-3}$, which is close to the value from a more recent observation by \citet{2011ApJ...731...83K}, who estimated that $N$(HCN)/$N$(CO)\,$=1.2\times10^{-3}$. \citet{2011ApJ...736...37K} estimated slightly lower column density ratio of $N$(HCN)/$N$(CO)=3$\times$10$^{-4}$.

These estimates assume a single component of homogeneous density and temperature, and different physical conditions within the CND cannot be taken into consideration.

Our results in Figure \ref{fig:hcn_co_column_q} show that the calculated $N$(HCN)/$N$(CO) column density ratio can be more than one order of magnitude higher than the observed value for $r < 20- 30$~ pc,  but it is within an order of magnitude to the observed ratio  for larger $r$ at $t$ = $t_{\rm cross}$.  The steady-state column density ratio of $N$(HCN)/$N$(CO) is at least an order of magnitude lower than the observed value for $r>10\,$pc in Model 3 and 4. In Models 1 and 2, the ratio stays around the observed value because of the flared structures, as mentioned in Section \ref{column_ratio}.

In our models 2 and 4, and for transitions  $J=3-2$ and $J=1-0$, there are regions where the intensity ratio of HCN/CO is larger than unity. Those are the regions where the X-ray irradiation is stronger than other regions.  For $J=3-2$, the region where the intensity ratio HCN/CO$>$1 is physically smaller, and this will lead to lower HCN/CO$(3-2)$ with respect to HCN/CO$(1-0)$ if observed with a large beamsize. The peaks of both molecules reside within 10\,pc in our models, which is again different from NGC 1068.

\subsubsection{HCO$^{+}$/HCN}
Similar to HCN and CO, both the Western and Eastern knots are seen for \cop($J=3-2$ and $4-3$) \citep[see Fig. 2][]{2011ApJ...736...37K}. Unlike HCN or CO, \cop~emission does not seem to have a high intensity red-shifted component that traces a jet \citep{2011ApJ...736...37K}. The emission is likely to be originating in the XDR regions since it peaks slightly closer to the AGN core than the CO and HCN peaks by a few tens of pc. \citet{2004A&A...419..897U} suggested that the column density ratio of $N$(HCO$^{+}$)/$N$(HCN) $\sim0.6 - 1.3$ while analysis by \citet{2011ApJ...731...83K} shows that $N$(HCO$^{+}$)/$N$(HCN) = 0.06, which is an order of magnitude lower. \citet{2011ApJ...736...37K} also obtain a value of column density ratio \cop/HCN that is lower than unity, i.e.,  \cop/HCN\,$\sim2\times10^{-3}-0.1.$  In most of our models,  \cop/HCN abundance ratios   at $t= t_{\rm cross}$ are somewhere between the two observed abundance ratios, but the steady-state abundance ratios at $r> 10$\,pc are greater than indicated by these observations (see Fig.~\ref{fig:hco+_hcn_column_q}). Besides the non-steady-state abundances at $t= t_{\rm cross}$, higher metal abundances can also suppress \cop, causing a lower \cop/HCN column density ratio.  If some fraction of the HCN emission is coming from the jet, as discussed above, the estimated \cop/HCN ratio in the XDR layer could in principle be higher than so far observed, and a high abundance ratio of \cop/HCN in our results in some models cannot be excluded.

In the nuclear region ($r\lesssim 90\,$pc) of the disk in Model 2, \cop/HCN ($1-0$) $<$1, and its value becomes close to unity for a larger radius.  The value in \citet{2005AIPC..783..203K}, which is the average of the entire CND, is 0.5 for NGC 1068. When our results are seen with a larger beamsize, the ratio becomes close to unity.
 
\subsubsection{CN, CN/CO, CN/HCN}

Based on the maps of 
\citet{2008JPhCS.131a2031G} and \citet{2010A&A...519A...2G}, CN also has two peaks, Western and Eastern knots. 
The location of the CN/CO intensity peak is seen at about 0.5 arcsec ($\approx 36$~pc) away from the AGN core while the CN/HCN velocity-integrated intensity peak is  about 0.1 at the AGN core and increases to 0.3 at a peak seen at about 1 arcsec ($\approx 72$~pc).   \citet{2004A&A...419..897U} estimated the $N$(CN)/$N$(CO) column density ratio to be $(1.8-10)\times10^{-3}$ from their CN(2-1) observation, but more recent observations estimate a lower overall column density ratio of $N$(CN)/$N$(CO)\,$=(0.3-1.3)\times10^{-3}$  
\citep{2010A&A...519A...2G}.\footnote{The CN fractional abundance is listed in terms of CN/\hmol,  so we use CO/\hmol = 8$\times$10$^{-5}$, which was used in \citet{2004A&A...419..897U}, to obtain CN/CO.}  
In our models, the column density ratio $N$(CN)/$N$(HCN) is in general higher at larger $r$ in Models 1 and 2 at both times, while for $N$(CN)/$N$(CO) the dependence is far less pronounced, as can be seen in Figures~\ref{fig:cn_hcn_column_q} through \ref{fig:cn_co_column_q}.   The higher intensity of CN with respect to HCN in the colder and outer part of the disk may come from the higher $N$(CN)/$N$(HCN) column density ratio due to the combination of non-steady-state chemistry ($t\sim t_{dyn}$) and the effect of X-rays.  
Our Model 2 calculation also has a higher CN/HCN intensity ratio with increasing distance from the AGN while Model 4 has a higher ratio near the AGN core. When the region is an XDR, it must have a higher CN/HCN intensity ratio. A lower intensity ratio of CN/HCN may indicate that the X-rays may not be the only source of heating, but there might also be thermal heating (i.e., from coupling of gas with warm dust, or heating from shocks). 

\subsubsection{CS}
The fractional abundance of CS that \citet{2004A&A...419..897U} estimated is 1$\times$10$^{-8}$. This is an order of magnitude higher than the value in \citet{2009ApJ...694..610M}. Both values are relatively high with respect to our models,  except for Models 3 and 4  on 100\,pc scales  if we utilize the degree of sulfur depletion onto the grains used standardly in the chemical models of dense clouds in our Galaxy \citep{1982ApJS...48..321G}. In galactic PDRs, there is  evidence that sulfur depletion is less than in regular cold dense clouds, making S/H$\sim (2-5)\times10^{-6}$ \citep{2006A&A...456..565G}. In diffuse clouds, there are claims that the sulfur abundance is even higher than the solar value. For example, \citet{2002A&A...384.1054L} show that the ratio of elemental abundance S/O is four times higher in a diffuse cloud towards $\zeta$Oph than that in a solar neighborhood. The observed abundance of CS cannot be achieved in our models unless there are shock waves to cause early-time chemistry (i.e., abundances at $t = t_{\rm cross}$),  
there is less depletion of sulfur onto the grains than in our models, or the metallicity is even higher than the solar 
value. 
Super-solar metallicities are commonly inferred in other AGNs, and it is likely that the nuclear environment in NGC 1068 
is similarly metal-rich \citep{2007ApJ...666..828F}.

\subsubsection{Carbon-Chain Molecules}
\begin{sloppypar}
Carbon-chain molecules  HC$_{3}$N \citep{2011A&A...528A..30C}, C$_{2}$H \citep{2011A&A...528A..30C,2011ApJ...728L..38N}, and c-C$_{3}$H$_{2}$ \citep{2011ApJ...728L..38N} have been detected in NGC 1068. Emissions from carbon-chain molecules are in general weaker than the aforementioned molecules, and there is little spatial information within the CND. Although HC$_{3}$N was detected in NGC 1068, the line ratios of HC$_{3}$N/HCN are around $0.2-0.4$ in (U)LIRGs, which are much higher than the other types of galaxies, which have a ratio less than 0.05 \citep{2011A&A...528A..30C}.  The reason for the enhancement of HC$_{3}$N is unknown, but concentration of its emission in galactic nuclei is seen for the case of Mrk 231 \citep{2011arXiv1111.6762A}~and NGC 4418 (Costagliola et al. A\&A, submitted). In addition, the ALMA early science observation by Takano et al. shows HC$_{3}$N in the CND, not in the starburst ring.
Our results show the enhancement of HC$_{3}$N at high temperature; nevertheless, it seems that HC$_{3}$N is more susceptible to X-ray irradiation than 
c-C$_{3}$H$_{2}$, as can be seen by comparing results in Tables \ref{model1}, \ref{model2}, and \ref{model3}. 
 Although the central concentration of HC$_{3}$N in the CND of NGC 1068 is seen with the ALMA early science observation, the emission is much weaker than the our Model 2 predicts according to Takano et al. (in prep). 
 \end{sloppypar}

 \subsection{A Case for (U)LIRGs}
 Luminous infrared galaxies (LIRGs) are star-forming galaxies with infrared luminosities of $L_{\rm IR}>10^{11}L_{\odot}$ ($L_{\rm IR}>10^{12}L_{\odot}$ for ultra-luminous infrared galaxies (ULIRGs)). It is believed that high infrared luminosities are due to  star formation, and that these galaxies are potentially an evolutionary phase in galaxies that precedes the AGN phase \citep{1988ApJ...325...74S,1996ARA&A..34..749S,2006ApJS..163....1H}. 
In general, they have a higher gas fraction, a lower metallicity, and a higher star formation efficiency \citep{1998ApJ...507..615D,2008ApJ...674..172R,2004ApJ...606..271G}. 
In addition to an enhancement of HC$_{3}$N, elevated abundances of C$_{2}$H$_{2}$ are seen in (U)LIRGs by \citet{2007ApJ...659..296L}.
In our models with $Q = 1$, some complex molecules such as \acet~and HC$_{3}$N are predicted to have elevated abundances in particular regions of the disk discussed in Section \ref{results_x}. These molecules indicate the presence of a high-temperature medium with a relatively low ionization rate. At warm temperatures around $200-300$\,K, the steady-state abundances of these molecules are low, but the peak abundances at earlier times are still enhanced.  Observational studies of c-C$_{3}$H$_{2}$ in (U)LIRGs might be useful and may give further information on the condition of the interstellar medium in these galaxies. A detailed discussion of the relation between molecular abundances and  physical conditions in (U)LIRGs will be presented in a future paper. 

\subsection{Variation in Physical Conditions}
There are many models for the accretion disks of galactic nuclei, and there are still many unknowns in some of the physical processes, such as the mass accretion rate onto the supermassive black hole, the cosmic-ray energy density, the star formation rate,  and the disk thickness (i.e. the scale height $h(r)$).  The scale height of the dust torus in NGC 1068 is found to be surprisingly high 
( its thickness is 2.1\,pc over 3.4\,pc radius \citep{2004Natur.429...47J}) in the central few pc with no conclusively identified mechanism to support the disk
vertically against its gravity. 

The model by \citet{2005ApJ...630..167T}, on which our calculation for midplane density is based, assumes a Toomre $Q$ parameter of close to unity, while the disk height is supported by  radiation pressure on the dust from  OB-stars. 
On the other hand, the model by \citet{2009ApJ...702...63W} is based on  feedback from the supernovae and stellar winds described in \citet{2001ApJ...547..172W}. This model does not assume $Q \sim 1$, and the midplane density within $r<30$\,pc is at least an order of magnitude lower than in the model of \citet{2005ApJ...630..167T}, 
although it is difficult to compare a full hydrodynamic model containing strong time-dependent density inhomogeneities to our simplified and smooth disks. In the model by \citet{2009ApJ...702...63W}, the mean density at a few tens of pc is around 10$^{4}$ cm$^{-3}$ as opposed to the model by \citet{2005ApJ...630..167T} where the mean midplane density varies from 10$^{6}-10^{8}$~cm$^{-3}$. At the lower densities, the total gaseous column density through the midplane is averaged to be a little less than 10$^{25}$ cm $^{-2}$, and a significant amount of higher energy X-rays can still penetrate into the midplane. The mean fraction of \hmol~in the \citet{2009ApJ...702...63W} model is around 0.3, while our results have molecular fractions close to unity. In the \citet{2009ApJ...702...63W} model, the medium is inhomogeneous, and the molecules can exist only in higher-density "clumps". Another model, by \citet{2009MNRAS.393..759S,2010MNRAS.403.1801S},  also contains an inhomogeneous medium with a lower midplane density of 10$^{4}$~cm$^{-3}$ within 10 pc, and even lower farther out from the AGN core. The chemistry in this third model would also be strongly affected by the larger XDR.

The physical conditions change depending on the evolutionary stage and the dominant processes at work inside the CND. Comparing our models with the observations of NGC 1068, the molecular density of our models on scales smaller than 10\,pc appears much higher.  The CND of 1068 may thus simply have a lower gas fraction in this region, perhaps as a result of preceding AGN feedback and/or stellar feedback. If the density near the nucleus is low enough to produce a "hole" in the molecular disk, the location of the XDR will change, and the peak locations of each molecular line change. 
Our models with $Q=1$ (Models 1 or 2) fail to capture these time-dependent effects, and thus probably have much higher density than the CND of NGC 1068. The assumption that $Q=1$ may not hold in parts of the disk where it is, or has been, strongly disturbed by AGN activity or recent stellar feedback, as shown in \citet{2012MNRAS.424.1963S}. However, the assumption of $Q\sim1$ appears to hold true in many of galactic disks \citep[e.g.,][]{2001ApJ...555..301M,1998ApJ...507..615D}. Effects of a systematic and time-dependent variation in the Toomre $Q$ paramter in more models must be studied in order to understand the chemical abundances and star formation condition in active galaxies.

\section{Summary and Future Work}\label{conclusion}
 The chemistry in the molecular disk around active galactic nuclei can be affected by many physical components of the system, including X-rays from the AGN core, the disk density structure, star formation activity and cosmic rays, and dynamical processes such as shocks, turbulence, rapid accretion, and jets. In this paper, we considered simple density structures for the molecular disk, with X-rays, cosmic-rays, and gas-grain collisions as the primary heating mechanisms for the gas.   We calculated the time-dependent abundances in the gaseous disk, reporting the abundances at both the characteristic shock-crossing time at each disk annulus and in steady state, and we coupled this abundance determination to a self-consistent calculation of the gas temperature.
Although shocks are considered in the sense that they dissociate molecules, we did not include shock heating and the interdependence between this heating and the chemistry in this paper, which involves more details.

 Our models of AGN disks show that there is an XDR layer, where the molecules are irradiated by X-rays, and a midplane, where the gas is colder, and more fully shielded, and it is this reservoir of mass that may be accreted into the black hole. Depending on the disk density structure and X-ray illumination, the location of the peak abundances, emission, and line ratios --- both between different rotational levels and different molecular species --- change. For example, the abundance ratio of CN/CO can be enhanced in the XDR layer. In addition, the abundance of CN in the crossing time is enhanced with respect to the steady-state abundance, and the abundance ratio of HCO$^{+}$/CO can also be enhanced in XDR layer with little time dependence. Time dependence is also important for HCN; the abundance ratio of HCN/CO can be enhanced in a warm part of the midplane at $t=t_{\rm cross}$. We also provided preliminary radiative transfer 
calculations in our model disks and abundance profiles in order to make a first comparison with observations. The more the fraction of XDR layer the model has, the higher the ratio of higher-$J$ line intensity over lower-$J$ line intensities.  The enhancement of abundance in CN and HCO$^{+}$ in the XDR layer also appear as higher intensities.
Our goal in the future is to present the full modeled intensities, and to apply the calculations in this paper to many different disk models in order to fit the observations in specific galaxies. There are also parameters that need more careful consideration such as the clumpiness, and variation in metallicity. Although existing observations can provide high-resolution intensity maps of CN and HCN \citep{2008JPhCS.131a2031G,2010A&A...519A...2G}~in 10-pc scale, even higher resolution of pc-/subpc-scale images are possible with ALMA.  In addition, the ALMA early science also demonstrated its power to observe more complex molecules. Strong lines of HC$_{3}$N(11-10) and (12-11) were observed. It will be interesting to observe higher-J lines of HC$_{3}$N such as $J=25-24$ that was observed with high excitation temperature in NGC 4418 by \citet{francesco_thesis}.
  Radial dependences of abundances for CN, HCN, \cop, sulfur-containing species, and carbon chain molecules can be used to constrain high/low density physical models, metallicity, and the effects of X-rays. In particular, with the radial abundance gradient, one could constrain the disk structure including density and temperature from the molecular observations.
\acknowledgments
N. H. thanks Yuri Aikawa, Nozomu Kawakatu, Phillip Maloney, Rowin Meijerink, Taku Nakajima, Hideko Nomura, Paul Rimmer, Marco Spaans, Shuro Takano, Kohji Tomisaka, and Anton Vasyunin for helpful discussions, and especially David Neufeld for sending his code to calculate molecular line cooling rates. E. H. acknowledges the support 
of the National Science Foundation for his astrochemistry program, and his program in chemical kinetics through the Center for the 
Chemistry of the Universe. He also acknowledges support from the NASA Exobiology and Evolutionary Biology program through a subcontract from Rensselaer Polytechnic Institute, and from NASA/JPL for science in support of the Herschel telescope.  TAT is supported in part by an Alfred P. Sloan Foundation Fellowship and by NASA grant number NNX10AD01G.

\appendix
\section{Temperature Calculation}
\label{ap_temp}
\subsection{Heating}
\subsubsection{X-ray heating}
When atoms and molecules are ionized by X-rays, fast electrons are produced, which can interact with ambient electrons. \citet{1996ApJ...466..561M} have shown that this Coulomb heating dominates heating when the electron fraction is high. The Coulomb rate is expressed as \citep{2005A&A...436..397M}:
\begin{equation}
\Gamma=\eta \,n\, H_{\rm X}
\end{equation}
where n is the total hydrogen density, $\eta$ is the heating efficiency as in \citet{2005A&A...436..397M} based on the results by \citet{1999ApJS..125..237D}, and $H_{\rm X}$ is defined as 
\citep{1996ApJ...466..561M}
\begin{equation}\label{eq:hx}
H_{\rm X}=\int^{E_{max}}_{E_{min}} \sigma_{pa}(E)F(E)dE
\end{equation}
where $\sigma_{pa}$(E)F(E) is the photoabsorption cross section and F(E) is the local photon energy flux at energy E. 
 The gas can also be heated when X-rays ionize molecular hydrogen by a series of reactions. As described in \citet{1973ApJ...186..859G}, the ionization heating rate is:
\begin{equation}\label{eq:kh2}
{\rm \gamma + H_2 \rightarrow H_2^+ + e^-}
\end{equation}
   \begin{equation}
{\rm H_2 + H_2^+ \rightarrow H_3^+ + H}
\label{eq:kh}
\end{equation}
\begin{equation}
{\rm H_3^+ + e^- \rightarrow H_2 + H}.
\label{eq:ke}
\end{equation}
The last two reactions are exothermic, and a large fraction of the energy goes into heating. We follow the formula for the heating rate in \citet{2005A&A...436..397M}:
\begin{equation}
\Gamma = \frac{17.5k_ex_e+1.51k_{\rm H}x_{\rm H}+13.7k_{\rm H_2}x_{\rm H_2}}{k_ex_e+k_{\rm H}x_{\rm H}+k_{\rm H_2}x_{\rm H_2}}\times 10^{-12}\zeta_{\rm H_2}x_{\rm H_2}n\, {\rm erg~cm^{-3}~ s^{-1}},
\end{equation}
where $k_{e}$, $k_{H}$, and $k_{H2}$ are the total reaction rates of Equations \ref{eq:ke}, \ref{eq:kh}, and \ref{eq:kh2}, while $x_{\rm e}$, $x_{\rm H}$, and $x_{\rm H2}$ are the ionization fraction, and the fractional abundances of H and H$_{2}$, respectively. $\zeta_{\rm H_2}$ is the ionization rate per hydrogen molecule, and $n$ is the density. 

\subsubsection{Cosmic-ray heating}
As noted earlier, cosmic-rays can ionize atoms and molecules, which can produce secondary electrons. These secondary electrons can excite molecular hydrogen, which can be used in heating.  With the amount of energy deposited per primary ionization by \citet{1978ApJ...219..750C} and \citet{1973ApJ...186..859G}, the heating rate by cosmic rays is given by \citet{1985ApJ...291..722T} to be 
\begin{equation}
\Gamma=1.5\times10^{-11} \zeta_{H2} n_{H_{2}} {\rm \,erg~cm^{-3}~s^{-1}},
\end{equation}
 where $\zeta_{H2}$ is the cosmic-ray ionization rate per molecular hydrogen and $n_{\rm H2}$ is the H$_{2}$ density.
 
\subsubsection{Dust-gas interaction}
When the dust temperature and the gas temperature are different, they can be heated or cooled by each other through collisions. We utilize the heating/cooling rate given by \citet{1989ApJ...342..306H}:
\begin{equation}
\Gamma = 1.2\times 10^{-31}n^{2} \left( \frac{T}{1000\,K}\right)^{1/2}\left( \frac{100\,{\mbox{\normalfont \AA}}}{a_{min}}\right)^{1/2}\times[1-0.8{\rm exp}(-75\,K/T)](T-T_{dust}){\rm \,erg~cm^{-3}~s^{-1}},
\end{equation}
 where $n$ is again the gas density, $T$ is the kinetic gas temperature, $T_{\rm dust}$ is the dust temperature, and $a_{\rm min}$ is the minimum grain size.

\subsection{Cooling}
\begin{sloppypar}
The vibrational and rotational cooling rates of \hmol, CO, and H$_{2}$O are taken from \citet{1993ApJ...418..263N} and \citet{1995ApJS..100..132N}. In addition, atomic line cooling by CII 158 $\,\mu$m, OI 63$\,\mu$m and 6300 \AA~is included. These cooling rates are calculated using the method described in \citet{2005pcim.book.....T}:
\begin{equation}
n^2 \Lambda = \frac{g_u/g_l exp(-E_{ul}/kT)}{1+n_{cr}\beta(\tau)+g_u/g_l exp(-E_{ul}/kT)}\mathcal{A}_{j}n A_{ul} h\nu_{ul} \beta(\tau),  
\end{equation}
where $n$ is the total number density, $g_{\rm u}$ and $g_{\rm l}$ are the statistical weights  of the upper and lower levels, $E_{\rm ul}$ is the energy difference between the upper and lower levels, $k$ is the Boltzman constant, $\mathcal{A}_{j}$ is the abundance of species $j$, $A_{\rm ul}$ is the Einstein coefficient, and $\nu_{ul}$ is the frequency of the line of the transition from the upper state to the lower state. The critical density $n_{\rm cr}$ is defined as 
\begin{equation}
n_{\rm cr}\equiv \frac{\beta(\tau)A_{ul}}{\gamma_{ul}},
\end{equation}
where $\beta(\tau)$ is the escape probability at an optical depth $\tau$. We used the form of escape probability as follows \citep{2005pcim.book.....T}
\begin{equation}
\beta(\tau)=\frac{1-{\rm exp}(-2.34 \tau)}{4.68 \tau},~(\tau<7)
\end{equation}
\begin{equation}
=\frac{1}{4 \tau \left[{\rm ln}\left(\frac{\tau}{\sqrt{\pi}}\right)\right]^{0.5}},~(\tau \geq 7).
\end{equation}
\end{sloppypar}

The optical depth is defined by
\begin{equation}
\tau_{ul}=\frac{A_{ul}c^3}{8\pi \nu^3} \frac{n_u}{b/\Delta z}\left[\frac{n_l g_u}{n_u g_l}-1\right].
\label{eq:tau}
\end{equation}
In Equation~(\ref{eq:tau}), $b$ is the Doppler broadening, and $\Delta z$ is the distance from the surface. This quantity of $b/\Delta z$ can be considered as a velocity gradient, and the value of $\Delta z \Omega$ was used.


\begin{deluxetable}{ccc}
\tablewidth{0pc}
\tabletypesize{\scriptsize}
\tablehead{
\colhead{Species} &\colhead{Low Metal} &\colhead{Solar} }
\tablecaption{Initial Fractional Abundances With Respect to Total Hydrogen.\label{init}}
\startdata
\hmol        &0.5 &0.5\\
He        &0.14 &0.09\\
O        &1.76(-4) &2.56(-4)\\
N        &2.14(-5) &7.6(-5)\\
F        &2.0(-8) &1.8(-8)\\
Cl        &3.0(-9) &1.8(-7)\\
C$^{+}$        &7.3(-5) &1.2(-4)\\
Fe$^{+}$        &3.0(-9) &2.0(-7)\\
Mg$^{+}$        &3.0(-9) &2.4(-6)\\
Na$^{+}$        &3.0(-9) &2.0(-7)\\
P$^{+}$        &3.0(-9) &1.17(-7)\\
S$^{+}$        &2.0(-8) &1.5(-5)\\
Si$^{+}$        &3.0(-9) &1.7(-6)\\
\enddata
\end{deluxetable}

\begin{deluxetable}{cl}
\tablewidth{0pc}
\tabletypesize{\scriptsize}
\tablehead{
\colhead{Symbols} &\colhead{Meaning} }
\tablecaption{List of Symbols.\label{symbols}}
\startdata
$\rho$ & gas mass density\\
$n$ & gas number density\\
$\Omega$ & Keplerian rotation frequency \\
$Q$ & Toomre Q parameter\\
$M_{\rm BH}$ & Black hole mass\\
$\sigma$ & velocity dispersion\\
$v_{\rm s}$ &sound speed\\
$\kappa$ &epicyclic frequency\\
$\Sigma$ & surface density\\
$f_{\rm g}$ &gas fraction\\
$h$ &scale height\\
$L_{\rm AGN}$ & total AGN luminosity\\
$\dot{\rho_{*}}$ &star formation rate per volume\\
$\nu$ &star formation efficiency\\
$t_{\rm dyn}$ &dynamical time\\
$t_{\rm cross}$ &crossing time ($\sim t_{\rm dyn}$)\\
$U_{\rm CR}$ &Cosmic-ray energy density\\
$L_{\rm CR}$ & Cosmic-ray luminosity\\
$t_{\rm pp}$ &proton-proton decay time\\
$t_{\rm wind}$ &wind escape time scale\\
$\zeta_{CR}$ & cosmic-ray ionization rate\\
$\zeta_{\rm X}$ &X-ray ionization raet\\
$\sigma_{\rm T}$ &Thomson cross section\\

\enddata
\end{deluxetable}

\begin{deluxetable}{ccccc}
\tablewidth{0pc}
\tabletypesize{\scriptsize}
\tablehead{
\colhead{disk model number} &\colhead{$<$1pc} &\colhead{1-3pc} &\colhead{3-100pc} &\colhead{$\rho (z)$/$\rho (0)$}}
\tablecaption{Radial Dependence of the Scale Height $h/r$. \label{diskst}}
\startdata
1 &0.01(r/pc)$^{2}$ &0.5 &0.01(r/pc) &exp(-$\frac{z^{2}}{2 h^{2}}$), CQM07\tablenotemark{a}~for 1-3pc\\
2 &0.01(r/pc)$^{2}$ &0.01(r/pc) &0.01(r/pc) &exp(-$\frac{z^{2}}{2 h^{2}}$)\\
3 &0.01(r/pc)$^{2}$ &0.5 &0.5 &exp(-$\frac{z^{2}}{2 h^{2}}$)\\
4 &0.01(r/pc)$^{2}$ &0.5 &0.5 &CQM07 for 1-100pc\\
\enddata
\tablenotetext{a}{Partial power-law dependence of density on height by fitting \citet{2007ApJ...662...94C} (CMQ07)}
\end{deluxetable}

\begin{deluxetable}{ccccccc}
\tablewidth{0pc}
\tabletypesize{\scriptsize}
\tablehead{
\colhead{Radius(pc)}  &\colhead{1}  &\colhead{3}  &\colhead{6}  &\colhead{16}  &\colhead{40}  &\colhead{100} \\ 
\colhead{$T_{\rm mid}$ (K)} &\colhead{750}  &\colhead{470}  &\colhead{300}  &\colhead{190}  &\colhead{120}  &\colhead{75}
 \\ 
 \colhead{$t_{\rm cross}$ (yr) } &\colhead{3$\times$10$^{3}$}  &\colhead{8$\times$10$^{3}$}  &\colhead{2$\times$10$^{4}$}  &\colhead{1$\times$10$^{5}$}  &\colhead{2$\times$10$^{5}$}  &\colhead{5$\times$10$^{5}$}} 
\tablecaption{Integrated Vertical Column Densities (cm$^{-2}$) for Model 1, $Q$ =1 at $t$ = $t_{\rm cross}$.\label{model1}}
\startdata
CO  &1.0e+19  &6.1e+20  &2.6e+20  &3.1e+20  &2.4e+20  &1.3e+20\\
C  &4.3e+16  &2.2e+19  &6.1e+17  &2.3e+18  &2.7e+19  &6.2e+19\\
CO$_2$  &1.1e+17  &6.0e+18  &2.0e+18  &1.5e+17  &7.9e+17  &2.9e+17\\
OH  &3.9e+17  &1.2e+19  &3.2e+18  &6.4e+18  &4.2e+18  &4.1e+18\\
H$_2$O  &1.7e+18  &3.3e+20  &1.2e+19  &3.8e+18  &1.8e+18  &1.6e+18\\
NH$_3$  &1.6e+13  &9.2e+16  &4.8e+16  &7.1e+16  &4.0e+16  &3.3e+16\\
HCN  &4.7e+16  &2.2e+18  &8.4e+17  &1.1e+17  &2.4e+17  &5.9e+16\\
HNC  &4.5e+15  &1.3e+17  &3.1e+17  &1.2e+17  &2.3e+17  &5.5e+16\\
CN  &1.7e+14  &2.4e+16  &1.9e+16  &2.9e+16  &2.8e+16  &1.8e+16\\
CS  &3.3e+10  &4.0e+13  &2.5e+13  &6.0e+13  &1.9e+14  &1.5e+15\\
SO  &2.4e+15  &7.4e+16  &7.2e+16  &2.0e+16  &6.6e+15  &1.6e+15\\
SO$_2$  &2.9e+14  &5.7e+16  &1.4e+16  &3.1e+15  &1.4e+15  &2.0e+14\\
C$_2$H$_2$  &7.3e+16  &5.8e+17  &1.9e+18  &2.2e+17  &5.4e+14  &6.7e+14\\
HC$_3$N  &2.8e+17  &9.9e+18  &1.1e+19  &1.1e+18  &6.7e+15  &1.8e+15\\
HCO$^+$  &2.2e+15  &1.9e+17  &1.0e+17  &1.6e+17  &3.3e+16  &2.7e+16\\
H$_3$O$^+$  &1.1e+16  &3.6e+17  &7.2e+16  &1.3e+17  &7.5e+16  &6.4e+16\\
HCNH$^+$  &1.1e+13  &1.4e+15  &1.5e+15  &9.8e+14  &2.5e+14  &2.1e+14\\
NH$_{4}^{+}$  &8.0e+10  &2.6e+14  &1.1e+14  &1.9e+14  &8.4e+13  &8.8e+13\\
H$_2$O$^+$  &1.3e+14  &2.3e+16  &1.2e+15  &3.3e+15  &7.7e+15  &6.1e+15\\
OH$^+$  &8.1e+13  &1.2e+17  &3.7e+15  &1.2e+16  &1.0e+16  &5.9e+15\\
C$_2$H  &5.8e+11  &1.0e+14  &8.0e+14  &2.4e+14  &3.7e+15  &1.6e+16\\
C$_3$H$_2$  &1.0e+17  &1.2e+17  &6.4e+17  &4.3e+17  &2.0e+14  &6.2e+14\\
NO  &1.7e+16  &2.8e+18  &1.6e+18  &2.6e+18  &4.2e+17  &4.1e+17\\
H$_3^{+}$  &1.5e+16  &3.9e+17  &2.4e+17  &3.5e+17  &3.1e+17  &5.6e+17\\
\enddata

\end{deluxetable}

\begin{deluxetable}{ccccccc}
\tablewidth{0pc}
\tabletypesize{\scriptsize}
\tablehead{
\colhead{Radius(pc)}  &\colhead{1}  &\colhead{3}  &\colhead{6}  &\colhead{16}  &\colhead{40}  &\colhead{100}\\ 
\colhead{$T_{\rm mid}$ (K)} &\colhead{750}  &\colhead{470}  &\colhead{300}  &\colhead{190}  &\colhead{120}  &\colhead{75}
 \\ 
 \colhead{$t_{\rm cross}$ (yr) } &\colhead{3$\times$10$^{3}$}  &\colhead{8$\times$10$^{3}$}  &\colhead{2$\times$10$^{4}$}  &\colhead{1$\times$10$^{5}$}  &\colhead{2$\times$10$^{5}$}  &\colhead{5$\times$10$^{5}$}} 
\tablecaption{Integrated Vertical Column Densities (cm$^{-2}$) for Model 2, $Q$ =1  at  $t$ = $t_{\rm cross}$.\label{model2}}
\startdata
CO  &4.5e+20  &3.4e+20  &2.2e+20  &1.8e+20  &1.8e+20  &1.9e+20\\
C  &6.9e+17  &2.7e+18  &1.5e+19  &2.4e+19  &1.5e+19  &1.9e+19\\
CO$_2$  &5.6e+18  &4.4e+18  &5.3e+17  &1.2e+18  &4.7e+17  &2.5e+17\\
OH  &4.9e+18  &4.4e+18  &1.0e+18  &1.9e+18  &3.0e+19  &2.9e+19\\
H$_2$O  &7.0e+20  &5.1e+19  &1.6e+20  &1.1e+20  &6.4e+19  &1.3e+20\\
NH$_3$  &1.1e+16  &9.7e+15  &4.0e+16  &1.5e+17  &1.1e+17  &7.4e+16\\
HCN  &4.7e+18  &3.5e+18  &1.6e+18  &5.5e+17  &2.1e+17  &1.1e+17\\
HNC  &1.2e+16  &2.3e+17  &5.1e+17  &2.9e+17  &8.4e+16  &2.5e+16\\
CN  &1.3e+15  &2.3e+15  &3.2e+14  &3.1e+15  &4.4e+15  &4.6e+15\\
CS  &2.2e+14  &3.7e+11  &2.5e+13  &1.7e+14  &1.3e+15  &5.5e+14\\
SO  &6.7e+16  &6.0e+16  &2.6e+16  &1.2e+16  &7.8e+15  &4.4e+15\\
SO$_2$  &4.0e+16  &4.0e+16  &3.1e+15  &1.7e+15  &1.8e+15  &6.9e+14\\
C$_2$H$_2$  &2.3e+18  &2.0e+17  &1.3e+18  &9.1e+15  &4.6e+14  &6.7e+14\\
HC$_3$N  &9.3e+18  &1.6e+19  &9.7e+18  &6.2e+15  &2.0e+15  &4.7e+14\\
HCO$^+$  &4.0e+16  &6.0e+15  &9.6e+16  &3.9e+16  &1.1e+17  &4.8e+17\\
H$_3$O$^+$  &9.1e+17  &9.6e+16  &6.5e+17  &3.2e+17  &9.3e+17  &1.9e+18\\
HCNH$^+$  &3.2e+16  &8.5e+13  &4.7e+15  &3.0e+15  &2.1e+15  &1.7e+15\\
NH$_4^+$  &5.9e+14  &1.7e+12  &3.1e+14  &5.0e+14  &4.3e+14  &3.9e+14\\
H$_2$O$^+$  &4.6e+13  &8.9e+15  &1.5e+15  &1.5e+16  &1.7e+16  &1.4e+16\\
OH$^+$  &9.9e+15  &7.6e+15  &7.0e+16  &5.0e+17  &9.1e+17  &6.3e+17\\
C$_2$H  &1.7e+13  &4.7e+13  &1.2e+14  &1.8e+14  &2.1e+14  &4.2e+14\\
C$_3$H$_2$  &9.0e+15  &1.4e+17  &6.2e+17  &3.8e+14  &2.0e+14  &3.1e+14\\
NO  &5.1e+18  &6.6e+16  &1.4e+18  &1.2e+18  &3.3e+18  &4.0e+18\\
H$_3^+$  &2.4e+17  &3.0e+16  &1.7e+17  &6.3e+16  &3.7e+17  &1.8e+18\\

\enddata
\end{deluxetable}

\pagebreak

\begin{deluxetable}{ccccccc}
\tablewidth{0pc}
\tabletypesize{\scriptsize}
\tablehead{
\colhead{Radius(pc)}  &\colhead{1}  &\colhead{3}  &\colhead{6}  &\colhead{16}  &\colhead{40}  &\colhead{100}\\ 
\colhead{$T_{\rm mid}$ (K)} &\colhead{750}  &\colhead{470}  &\colhead{300}  &\colhead{190}  &\colhead{120}  &\colhead{75}
 \\
 \colhead{$t_{\rm cross}$ (yr) } &\colhead{3$\times$10$^{3}$}  &\colhead{8$\times$10$^{3}$}  &\colhead{2$\times$10$^{4}$}  &\colhead{1$\times$10$^{5}$}  &\colhead{2$\times$10$^{5}$}  &\colhead{5$\times$10$^{5}$}} 
\tablecaption{Integrated Vertical Column Densities (cm$^{-2}$) for Model 3, $Q$ =5  at $t$ = $t_{\rm cross}$.\label{model3}}
\startdata
CO  &7.5e+21  &2.0e+21  &5.8e+20  &1.5e+20  &2.5e+19  &7.9e+18\\
C  &2.8e+20  &2.8e+19  &1.1e+20  &1.2e+20  &8.2e+19  &3.5e+19\\
CO$_2$  &3.1e+19  &1.7e+19  &4.6e+17  &9.3e+17  &1.5e+17  &2.5e+16\\
OH  &3.5e+20  &3.5e+19  &5.0e+18  &7.6e+17  &4.9e+17  &5.6e+16\\
H$_2$O  &4.8e+21  &1.5e+20  &4.1e+18  &7.2e+17  &2.4e+17  &5.4e+16\\
NH$_3$  &4.0e+18  &4.5e+17  &2.5e+16  &2.3e+16  &1.2e+15  &7.7e+14\\
HCN  &4.3e+19  &1.3e+19  &4.7e+17  &3.5e+17  &4.2e+16  &5.5e+15\\
HNC  &6.5e+18  &3.2e+18  &3.8e+17  &3.5e+17  &4.1e+16  &5.1e+15\\
CN  &1.5e+17  &9.3e+16  &3.3e+16  &1.5e+16  &8.0e+15  &8.2e+15\\
CS  &1.1e+15  &3.0e+14  &1.9e+15  &8.4e+15  &1.1e+16  &3.6e+15\\
SO  &6.5e+17  &4.4e+17  &1.3e+15  &6.1e+14  &2.9e+12  &4.4e+12\\
SO$_2$  &4.5e+17  &3.8e+16  &1.5e+13  &1.9e+13  &6.0e+09  &1.3e+10\\
C$_2$H$_2$  &2.1e+19  &2.1e+19  &2.3e+18  &6.6e+16  &1.1e+15  &1.6e+15\\
HC$_3$N  &1.7e+19  &4.8e+19  &5.9e+18  &8.7e+15  &9.6e+14  &1.2e+14\\
HCO$^+$  &6.1e+18  &9.0e+17  &2.0e+16  &1.8e+16  &1.2e+15  &5.2e+14\\
H$_3$O$^+$  &2.0e+19  &1.1e+18  &1.1e+17  &1.5e+16  &9.3e+15  &1.3e+15\\
HCNH$^+$  &5.3e+16  &8.0e+15  &1.2e+15  &3.4e+14  &9.5e+13  &3.3e+13\\
NH$_4^+$  &1.4e+16  &1.1e+15  &4.0e+13  &6.0e+13  &2.5e+12  &2.0e+12\\
H$_2$O$^+$  &1.9e+17  &7.3e+15  &1.4e+16  &6.3e+15  &1.6e+15  &4.7e+13\\
OH$^+$  &7.6e+17  &9.0e+16  &7.1e+16  &2.8e+16  &1.6e+15  &4.8e+13\\
C$_2$H  &7.0e+14  &2.6e+15  &6.6e+16  &2.7e+16  &1.2e+16  &6.1e+15\\
C$_3$H$_2$  &5.9e+18  &1.7e+19  &1.5e+19  &9.6e+15  &5.9e+14  &8.6e+14\\
NO  &9.6e+19  &1.1e+19  &5.1e+16  &1.8e+17  &2.3e+15  &3.9e+15\\
H$_3^+$  &9.9e+18  &3.1e+18  &2.0e+17  &1.2e+17  &8.2e+16  &2.1e+16\\

\enddata
\end{deluxetable}

\begin{deluxetable}{ccccccc}
\tablewidth{0pc}
\tabletypesize{\scriptsize}
\tablehead{
\colhead{Radius(pc)}  &\colhead{1}  &\colhead{3}  &\colhead{6}  &\colhead{16}  &\colhead{40}  &\colhead{100}\\ 
\colhead{$T_{\rm mid}$ (K)} &\colhead{750}  &\colhead{470}  &\colhead{300}  &\colhead{190}  &\colhead{120}  &\colhead{75}
 \\ 
 \colhead{$t_{\rm cross}$ (yr) } &\colhead{3$\times$10$^{3}$}  &\colhead{8$\times$10$^{3}$}  &\colhead{2$\times$10$^{4}$}  &\colhead{1$\times$10$^{5}$}  &\colhead{2$\times$10$^{5}$}  &\colhead{5$\times$10$^{5}$}} 
\tablecaption{Integrated Vertical Column Densities (cm$^{-2}$) for Model 4, $Q$ =5  at $t$ = $t_{\rm cross}$.\label{model4}}
\startdata
CO  &1.2e+16  &3.2e+19  &1.4e+19  &4.4e+18  &1.2e+18  &3.5e+17\\
C  &1.5e+19  &1.3e+19  &9.2e+18  &8.5e+18  &4.8e+18  &2.2e+18\\
CO$_2$  &2.6e+07  &9.3e+15  &3.8e+16  &2.8e+16  &6.4e+15  &1.2e+15\\
OH  &3.4e+16  &2.7e+18  &6.5e+17  &3.1e+17  &1.3e+17  &1.2e+16\\
H$_2$O  &7.8e+14  &1.1e+18  &1.9e+17  &6.8e+16  &2.5e+16  &4.4e+15\\
NH$_3$  &6.8e+05  &6.2e+15  &4.3e+15  &1.2e+14  &3.2e+13  &1.5e+13\\
HCN  &4.0e+09  &9.4e+15  &1.5e+16  &1.0e+16  &1.8e+15  &2.2e+14\\
HNC  &6.5e+07  &7.6e+15  &1.5e+16  &1.0e+16  &1.8e+15  &2.1e+14\\
CN  &2.9e+12  &1.1e+16  &2.2e+15  &9.7e+14  &6.9e+14  &2.4e+14\\
CS  &2.3e+05  &6.3e+12  &2.7e+14  &8.0e+14  &6.0e+14  &2.0e+14\\
SO  &1.6e+07  &3.5e+14  &1.8e+14  &6.2e+11  &1.3e+11  &8.7e+10\\
SO$_2$  &1.0e+02  &3.2e+13  &1.5e+13  &1.1e+09  &2.6e+08  &2.2e+08\\
C$_2$H$_2$  &8.4e+05  &2.7e+13  &1.2e+15  &1.6e+14  &7.6e+13  &3.4e+13\\
HC$_3$N  &4.6e-05  &4.8e+10  &4.0e+14  &2.0e+14  &3.8e+13  &5.2e+12\\
HCO$^+$  &3.3e+11  &3.3e+16  &4.7e+15  &4.8e+14  &8.3e+13  &2.4e+13\\
H$_3$O$^+$  &6.6e+13  &5.4e+16  &8.9e+15  &3.7e+15  &1.5e+15  &1.8e+14\\
HCNH$^+$  &1.1e+07  &2.7e+14  &4.0e+13  &9.5e+12  &3.9e+12  &1.1e+12\\
NH$_4^+$  &4.3e+02  &2.4e+13  &1.3e+13  &2.6e+11  &6.8e+10  &3.7e+10\\
H$_2$O$^+$  &2.4e+15  &4.1e+15  &4.9e+15  &3.1e+15  &2.6e+15  &3.1e+13\\
OH$^+$  &8.9e+16  &7.2e+16  &8.9e+16  &5.1e+16  &2.2e+16  &3.3e+13\\
C$_2$H  &3.4e+08  &7.5e+13  &7.5e+14  &1.2e+15  &7.8e+14  &2.5e+14\\
C$_3$H$_2$  &4.6e+00  &2.9e+12  &8.6e+13  &1.1e+14  &3.0e+13  &1.7e+13\\
NO  &1.4e+13  &5.2e+17  &7.9e+16  &1.2e+15  &2.2e+14  &5.9e+13\\
H$_3^+$  &1.3e+13  &1.2e+17  &3.5e+16  &2.2e+16  &1.1e+16  &4.0e+15\\
\enddata
\end{deluxetable}

\begin{deluxetable}{ccccccc}
\tablewidth{0pc}
\tabletypesize{\scriptsize}
\tablehead{
\colhead{Radius(pc)}  &\colhead{1}  &\colhead{3}  &\colhead{6}  &\colhead{16}  &\colhead{40}  &\colhead{100} \\ 
\colhead{$T_{\rm mid}$ (K)} &\colhead{750}  &\colhead{470}  &\colhead{300}  &\colhead{190}  &\colhead{120}  &\colhead{75}}
\tablecaption{Integrated Vertical Column Densities (cm$^{-2}$) for Model 1, $Q$ =1 at steady-state.\label{model1s}}
\startdata
CO  &9.4e+18  &5.2e+20  &3.7e+20  &3.3e+20  &2.8e+20  &2.0e+20\\
C  &2.3e+16  &2.4e+19  &5.8e+17  &1.5e+18  &3.2e+18  &3.7e+19\\
CO$_2$  &5.6e+16  &2.3e+18  &2.6e+18  &3.5e+17  &1.6e+17  &5.7e+16\\
OH  &5.6e+17  &1.6e+19  &2.7e+18  &7.2e+18  &1.1e+19  &6.6e+18\\
H$_2$O  &2.1e+19  &8.8e+20  &2.8e+20  &7.7e+18  &5.3e+18  &2.4e+18\\
NH$_3$  &1.7e+17  &5.9e+18  &7.3e+16  &9.7e+16  &7.0e+16  &3.5e+16\\
HCN  &1.3e+17  &4.2e+18  &3.9e+16  &4.0e+16  &4.4e+16  &1.2e+16\\
HNC  &4.1e+16  &1.5e+18  &1.5e+16  &2.9e+16  &3.1e+16  &9.0e+15\\
CN  &3.0e+13  &1.3e+16  &1.1e+16  &2.4e+16  &4.4e+16  &3.4e+16\\
CS  &3.9e+14  &1.2e+16  &1.8e+13  &4.6e+13  &5.4e+13  &6.3e+13\\
SO  &6.9e+14  &4.1e+16  &3.0e+16  &2.6e+16  &1.7e+16  &8.4e+15\\
SO$_2$  &4.3e+14  &2.3e+16  &5.7e+16  &2.7e+16  &9.4e+15  &2.7e+15\\
C$_2$H$_2$  &7.7e+17  &2.6e+19  &1.3e+15  &8.6e+14  &4.0e+14  &9.5e+14\\
HC$_3$N  &2.8e+15  &1.1e+17  &1.3e+12  &8.0e+11  &2.4e+11  &1.3e+11\\
HCO$^+$  &4.0e+15  &2.2e+17  &1.3e+17  &2.0e+17  &1.7e+17  &5.6e+16\\
H$_3$O$^+$  &2.7e+16  &8.0e+17  &1.2e+17  &1.9e+17  &2.2e+17  &1.1e+17\\
HCNH$^+$  &2.1e+13  &2.1e+15  &7.8e+14  &1.0e+15  &1.2e+15  &2.8e+14\\
NH$_4^+$  &2.4e+12  &5.2e+14  &1.7e+14  &2.3e+14  &1.7e+14  &7.8e+13\\
H$_2$O$^+$  &1.4e+14  &2.2e+16  &2.8e+14  &3.2e+15  &1.9e+15  &7.1e+15\\
OH$^+$  &4.7e+13  &1.1e+17  &2.7e+15  &6.7e+15  &1.6e+15  &7.5e+15\\
C$_2$H  &5.6e+10  &4.0e+13  &5.1e+13  &1.6e+14  &3.4e+15  &1.5e+16\\
C$_3$H$_2$  &1.3e+16  &2.7e+17  &4.6e+13  &4.2e+13  &1.5e+13  &9.8e+13\\
NO  &5.7e+16  &3.5e+18  &1.8e+18  &3.1e+18  &2.6e+18  &6.3e+17\\
H$_3^+$  &6.9e+15  &3.5e+17  &2.3e+17  &3.5e+17  &1.1e+18  &9.6e+17\\

\enddata
\end{deluxetable}

\begin{deluxetable}{ccccccc}
\tablewidth{0pc}
\tabletypesize{\scriptsize}
\tablehead{
\colhead{Radius(pc)}  &\colhead{1}  &\colhead{3}  &\colhead{6}  &\colhead{16}  &\colhead{40}  &\colhead{100} \\ 
\colhead{$T_{\rm mid}$ (K)} &\colhead{750}  &\colhead{470}  &\colhead{300}  &\colhead{190}  &\colhead{120}  &\colhead{75}}
\tablecaption{Integrated Vertical Column Densities (cm$^{-2}$) for Model 2, $Q$ =1 at steady-state.\label{model2s}}
\startdata
CO  &5.5e+20  &3.4e+20  &3.6e+20  &3.0e+20  &2.7e+20  &2.0e+20\\
C  &3.0e+18  &3.8e+17  &1.5e+18  &6.1e+18  &9.2e+18  &3.0e+19\\
CO$_2$  &1.5e+17  &5.6e+18  &2.0e+18  &2.4e+17  &1.1e+17  &3.1e+16\\
OH  &1.4e+18  &1.2e+19  &6.8e+18  &1.2e+19  &1.2e+19  &1.1e+19\\
H$_2$O  &7.9e+20  &7.1e+20  &2.6e+20  &1.2e+19  &5.7e+18  &3.6e+18\\
NH$_3$  &1.5e+19  &3.5e+18  &1.2e+17  &7.4e+16  &4.5e+16  &1.8e+16\\
HCN  &4.1e+18  &5.2e+18  &1.1e+17  &3.9e+16  &3.4e+16  &9.7e+15\\
HNC  &1.2e+17  &1.5e+18  &1.7e+16  &1.7e+16  &2.2e+16  &6.6e+15\\
CN  &6.8e+14  &8.2e+14  &9.4e+15  &2.2e+16  &3.8e+16  &3.0e+16\\
CS  &1.3e+15  &5.6e+15  &3.1e+13  &4.1e+13  &3.4e+13  &4.8e+13\\
SO  &6.4e+15  &2.5e+16  &1.5e+16  &1.7e+16  &1.0e+16  &4.7e+15\\
SO$_2$  &9.9e+14  &2.6e+16  &4.4e+16  &1.6e+16  &5.5e+15  &1.5e+15\\
C$_2$H$_2$  &1.3e+18  &1.6e+19  &8.2e+14  &4.9e+14  &3.0e+14  &9.4e+14\\
HC$_3$N  &1.1e+17  &4.4e+16  &9.3e+11  &3.9e+11  &1.1e+11  &1.0e+11\\
HCO$^+$  &1.0e+17  &1.5e+17  &2.2e+17  &2.0e+17  &1.4e+17  &5.1e+16\\
H$_3$O$^+$  &7.8e+17  &9.9e+17  &3.2e+17  &3.3e+17  &2.6e+17  &1.9e+17\\
HCNH$^+$  &6.9e+15  &1.5e+15  &1.6e+15  &9.6e+14  &9.2e+14  &2.4e+14\\
NH$_4^+$  &7.7e+14  &2.7e+14  &4.3e+14  &1.9e+14  &1.2e+14  &3.8e+13\\
H$_2$O$^+$  &3.8e+14  &1.7e+15  &1.2e+16  &1.3e+16  &1.5e+16  &1.8e+16\\
OH$^+$  &2.0e+16  &2.8e+14  &1.1e+16  &5.4e+16  &1.1e+17  &1.1e+17\\
C$_2$H  &1.8e+12  &9.0e+11  &2.5e+13  &9.7e+13  &2.2e+14  &1.2e+16\\
C$_3$H$_2$  &8.3e+15  &2.1e+17  &3.8e+13  &1.9e+13  &1.2e+13  &1.2e+14\\
NO  &1.9e+18  &2.4e+18  &3.2e+18  &2.8e+18  &1.9e+18  &3.8e+17\\
H$_3^+$  &1.7e+17  &2.9e+17  &2.8e+17  &5.4e+17  &5.9e+17  &1.1e+18\\

\enddata
\end{deluxetable}

\begin{deluxetable}{ccccccc}
\tablewidth{0pc}
\tabletypesize{\scriptsize}
\tablehead{
\colhead{Radius(pc)}  &\colhead{1}  &\colhead{3}  &\colhead{6}  &\colhead{16}  &\colhead{40}  &\colhead{100} \\ 
\colhead{$T_{\rm mid}$ (K)} &\colhead{750}  &\colhead{470}  &\colhead{300}  &\colhead{190}  &\colhead{120}  &\colhead{75}}
\tablecaption{Integrated Vertical Column Densities (cm$^{-2}$) for Model 3, $Q$ =5 at steady-state.\label{model3s}}
\startdata
CO  &7.8e+21  &2.7e+21  &9.3e+20  &3.4e+20  &1.2e+20  &4.8e+19\\
C  &5.1e+19  &2.2e+19  &2.9e+18  &1.8e+18  &3.4e+18  &4.9e+17\\
CO$_2$  &4.0e+18  &9.0e+18  &1.6e+18  &2.9e+17  &7.2e+16  &3.2e+16\\
OH  &1.1e+21  &4.4e+19  &8.3e+18  &2.3e+18  &1.5e+17  &2.1e+17\\
H$_2$O  &1.0e+22  &2.9e+21  &1.6e+19  &1.7e+18  &3.2e+17  &2.3e+17\\
NH$_3$  &6.9e+19  &1.3e+18  &2.2e+17  &9.5e+16  &3.7e+16  &2.5e+16\\
HCN  &3.9e+20  &1.0e+19  &5.2e+16  &6.3e+15  &1.9e+15  &2.3e+15\\
HNC  &6.5e+18  &5.4e+18  &3.9e+16  &5.4e+15  &1.6e+15  &2.1e+15\\
CN  &2.8e+17  &2.8e+16  &2.9e+16  &4.6e+15  &3.5e+15  &6.0e+14\\
CS  &7.7e+16  &1.9e+16  &6.9e+13  &3.8e+13  &2.0e+13  &3.4e+13\\
SO  &2.0e+17  &2.4e+17  &8.4e+16  &4.4e+16  &2.1e+16  &8.3e+15\\
SO$_2$  &3.1e+16  &2.1e+17  &1.2e+17  &3.1e+16  &7.9e+15  &1.9e+15\\
C$_2$H$_2$  &1.2e+20  &1.5e+19  &1.6e+15  &6.2e+14  &5.0e+14  &4.8e+14\\
HC$_3$N  &6.5e+17  &1.1e+17  &1.1e+12  &9.0e+10  &6.8e+10  &1.1e+11\\
HCO$^+$  &8.9e+18  &9.3e+17  &2.8e+17  &4.9e+16  &7.6e+15  &7.2e+15\\
H$_3$O$^+$  &4.4e+19  &2.7e+18  &2.1e+17  &5.0e+16  &3.7e+15  &4.6e+15\\
HCNH$^+$  &1.1e+18  &4.9e+15  &1.3e+15  &1.2e+14  &2.4e+13  &3.3e+13\\
NH$_4^+$  &6.4e+16  &2.0e+15  &3.8e+14  &1.2e+14  &3.4e+13  &4.4e+13\\
H$_2$O$^+$  &2.0e+17  &9.7e+15  &1.2e+16  &1.7e+15  &7.0e+14  &2.0e+14\\
OH$^+$  &2.8e+17  &9.9e+16  &1.7e+16  &2.1e+16  &3.3e+15  &1.9e+14\\
C$_2$H  &2.6e+18  &2.1e+13  &1.9e+14  &2.8e+13  &2.5e+14  &1.4e+13\\
C$_3$H$_2$  &1.2e+19  &3.8e+17  &5.8e+13  &1.8e+12  &1.3e+12  &9.8e+11\\
NO  &8.0e+19  &1.3e+19  &4.4e+18  &7.9e+17  &8.6e+16  &1.4e+17\\
H$_3^+$  &2.7e+20  &2.0e+18  &4.1e+17  &8.2e+16  &2.5e+16  &1.2e+16\\

\enddata
\end{deluxetable}

\newpage

\begin{deluxetable}{ccccccc}
\tablewidth{0pc}
\tabletypesize{\scriptsize}
\tablehead{
\colhead{Radius(pc)}  &\colhead{1}  &\colhead{3}  &\colhead{6}  &\colhead{16}  &\colhead{40}  &\colhead{100} \\ 
\colhead{$T_{\rm mid}$ (K)} &\colhead{750}  &\colhead{470}  &\colhead{300}  &\colhead{190}  &\colhead{120}  &\colhead{75}}
\tablecaption{Integrated Vertical Column Densities (cm$^{-2}$) for Model 4, $Q$ =5 at steady-state.\label{model4s}}
\startdata
CO  &1.2e+16  &3.3e+19  &2.6e+19  &1.5e+19  &6.4e+18  &2.7e+18\\
C  &1.5e+19  &1.4e+19  &3.0e+18  &1.2e+18  &1.0e+18  &1.4e+17\\
CO$_2$  &2.6e+07  &1.2e+16  &2.4e+16  &1.1e+16  &3.2e+15  &1.9e+15\\
OH  &3.4e+16  &1.8e+18  &6.3e+17  &1.3e+17  &2.9e+16  &6.9e+15\\
H$_2$O  &7.8e+14  &3.1e+18  &3.2e+17  &5.6e+16  &1.8e+16  &1.1e+16\\
NH$_3$  &6.8e+05  &7.8e+15  &6.5e+15  &3.8e+15  &1.9e+15  &1.3e+15\\
HCN  &4.0e+09  &1.5e+16  &1.2e+15  &3.5e+14  &6.4e+13  &1.0e+14\\
HNC  &6.5e+07  &2.9e+15  &8.3e+14  &2.6e+14  &6.0e+13  &9.4e+13\\
CN  &2.9e+12  &4.9e+15  &1.5e+15  &6.6e+14  &8.3e+13  &1.3e+13\\
CS  &2.3e+05  &3.6e+12  &3.3e+12  &1.3e+12  &1.4e+12  &2.3e+12\\
SO  &1.6e+07  &3.4e+14  &2.4e+15  &2.0e+15  &1.2e+15  &4.8e+14\\
SO$_2$  &1.0e+02  &4.0e+13  &1.7e+15  &1.3e+15  &4.2e+14  &1.0e+14\\
C$_2$H$_2$  &8.4e+05  &1.9e+14  &3.0e+13  &3.2e+13  &3.6e+13  &3.6e+13\\
HC$_3$N  &4.6e-05  &1.2e+11  &1.1e+10  &4.1e+09  &4.6e+09  &8.3e+09\\
HCO$^+$  &3.3e+11  &4.1e+16  &8.2e+15  &1.5e+15  &2.6e+14  &2.8e+14\\
H$_3$O$^+$  &6.6e+13  &5.9e+16  &1.2e+16  &1.5e+15  &2.9e+14  &1.8e+14\\
HCNH$^+$  &1.1e+07  &2.8e+14  &2.8e+13  &7.0e+12  &4.3e+11  &1.1e+12\\
NH$_4^+$  &4.3e+02  &3.5e+13  &1.3e+13  &3.8e+12  &1.4e+12  &2.0e+12\\
H$_2$O$^+$  &2.4e+15  &3.6e+15  &2.4e+15  &1.1e+15  &3.0e+14  &6.0e+13\\
OH$^+$  &8.9e+16  &6.8e+16  &5.0e+16  &3.9e+16  &1.1e+16  &1.2e+14\\
C$_2$H  &3.4e+08  &1.5e+13  &6.1e+12  &4.9e+12  &8.7e+13  &6.5e+12\\
C$_3$H$_2$  &4.6e+00  &1.2e+13  &5.0e+11  &6.6e+10  &6.6e+11  &8.3e+10\\
NO  &1.4e+13  &4.8e+17  &1.4e+17  &1.9e+16  &2.7e+15  &4.3e+15\\
H$_3^+$  &1.3e+13  &8.4e+16  &1.5e+16  &6.5e+15  &4.0e+14  &4.2e+14\\
\enddata
\end{deluxetable}

\begin{deluxetable}{ccccccc}
\tablewidth{0pc}
\tabletypesize{\scriptsize}
\tablehead{
\colhead{Species} &\colhead{Lines} &\colhead{Vel. Int. Peak Intensity} &\colhead{Vel. Int. Peak Intensity}&\colhead{Spatially Int. I} &\colhead{Beam size} &\colhead{Reference}\\
\colhead{} &\colhead{} &\colhead{(Jy/beam km/s)} &\colhead{(K km/s)}&\colhead{(Jy km/s)} &\colhead{} &\colhead{}}
\tablecaption{Observed Intensities in NGC 1068. Eastern knot is denoted as (E), and the western knot is denoted as (W). \label{table:obs}}
\startdata
$^{12}$CO & 1-0 &40(E), 19(W)  &760(E) &90(E), 40 (W) &1".0 $\times$ 0".8 & Krips et al. (2011) \\
&2-1 &70(E), 30(W) & 1980(E) &290(E),180(W) &1".0 $\times$ 0".8 &Krips et al. (2011) \\
&3-2 &470(E), 270(W) &5900(E) &1330(E), 720(W) &1".0 $\times$ 0".8 &Krips et al. (2011) \\
HCN & 1-0 & &121(E), 91(W) &19(E),15(W) &5" &Tacconi et al.(1994)\\
&2-1 & && & \\
&3-2 &51(E), 32(W) &1100(E) &110(E), 70(W) &0".53$\times$0".46 &Krips et al. (2011) \\
&4-3 & &13.9 &220 &14" &P.-B. et al. (2009)\\
HCO$^{+}$ & 1-0 & & & & &\\
&2-1 & & & & & \\
&3-2 &28 (E),14(W) &590(E) &52(E),40(W) &1".0 $\times$ 0".8 &Krips et al. (2011) \\
&4-3 &27(E) & &98(E) &1".0 $\times$ 0".8 &Krips et al. (2011) \\
CN & 1-0 & & & & &\\
&2-1 &10(E), 3.6(W)&200(E), 80(W)&&&\\
&3-2 &&&&&\\
\enddata
\end{deluxetable}

\pagebreak

\clearpage

\clearpage

\begin{figure*}
\centerline{\includegraphics[angle=0,width=8.5cm]{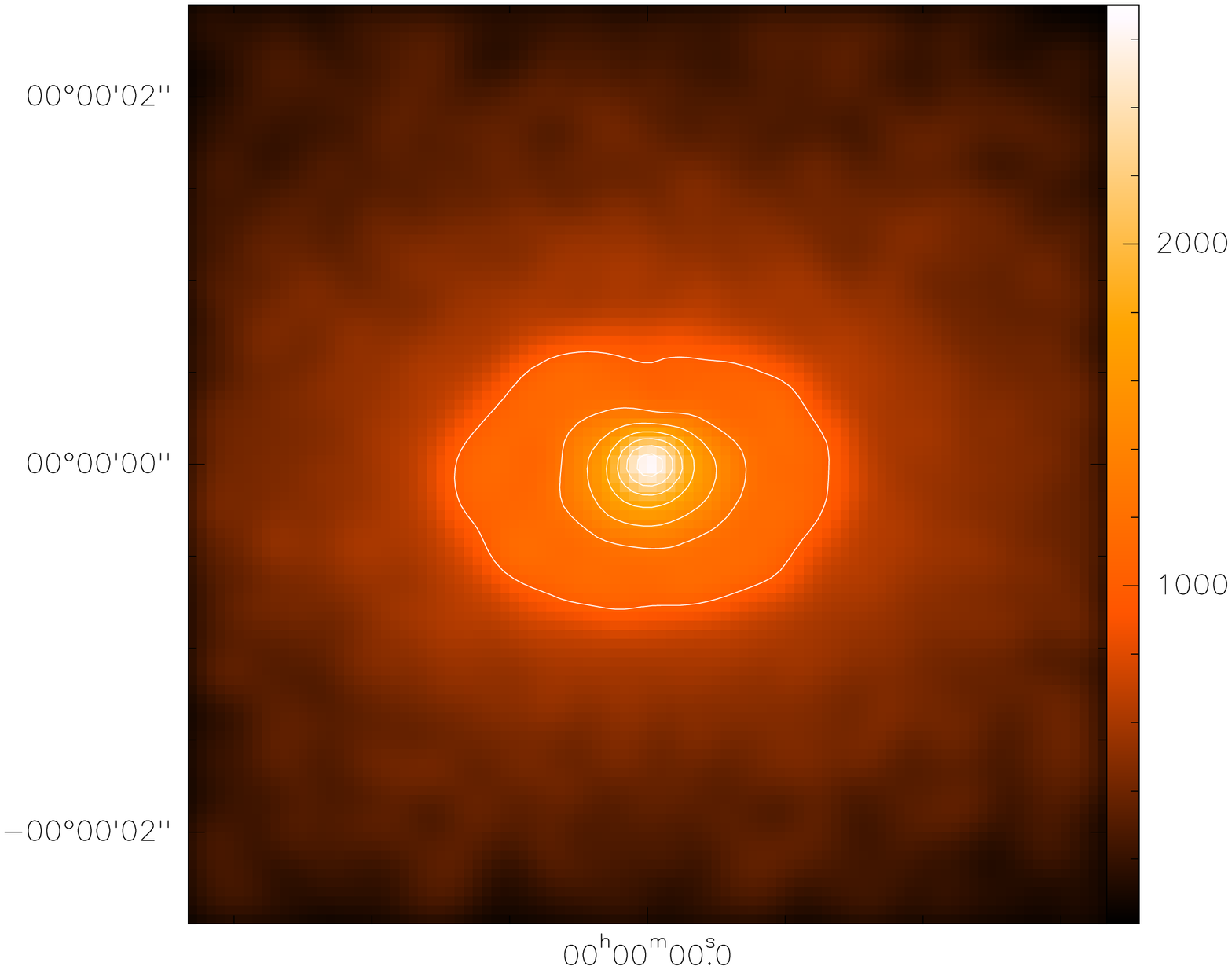}
\includegraphics[angle=0,width=8.5cm]{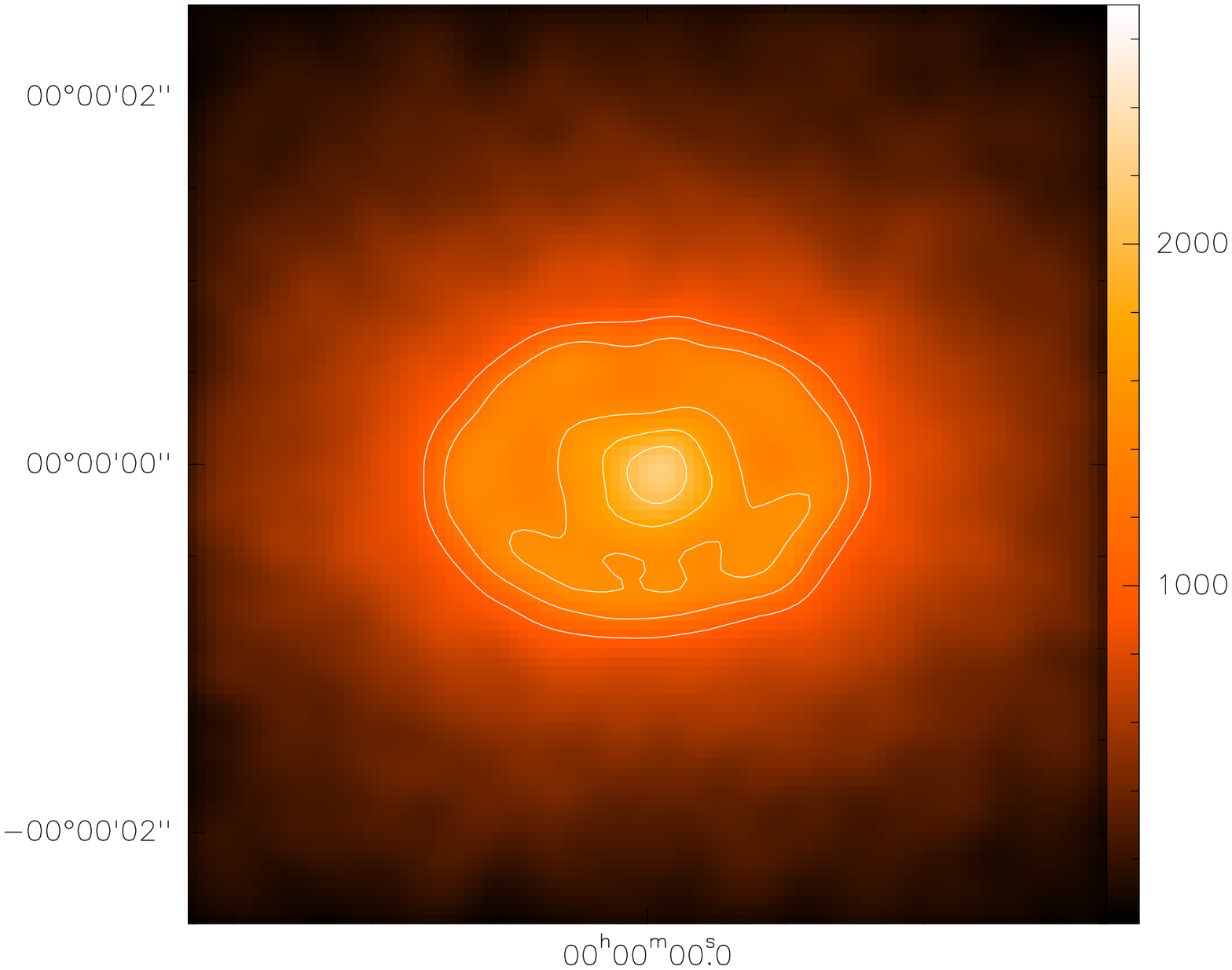}}
\centerline{\includegraphics[angle=0,width=8.5cm]{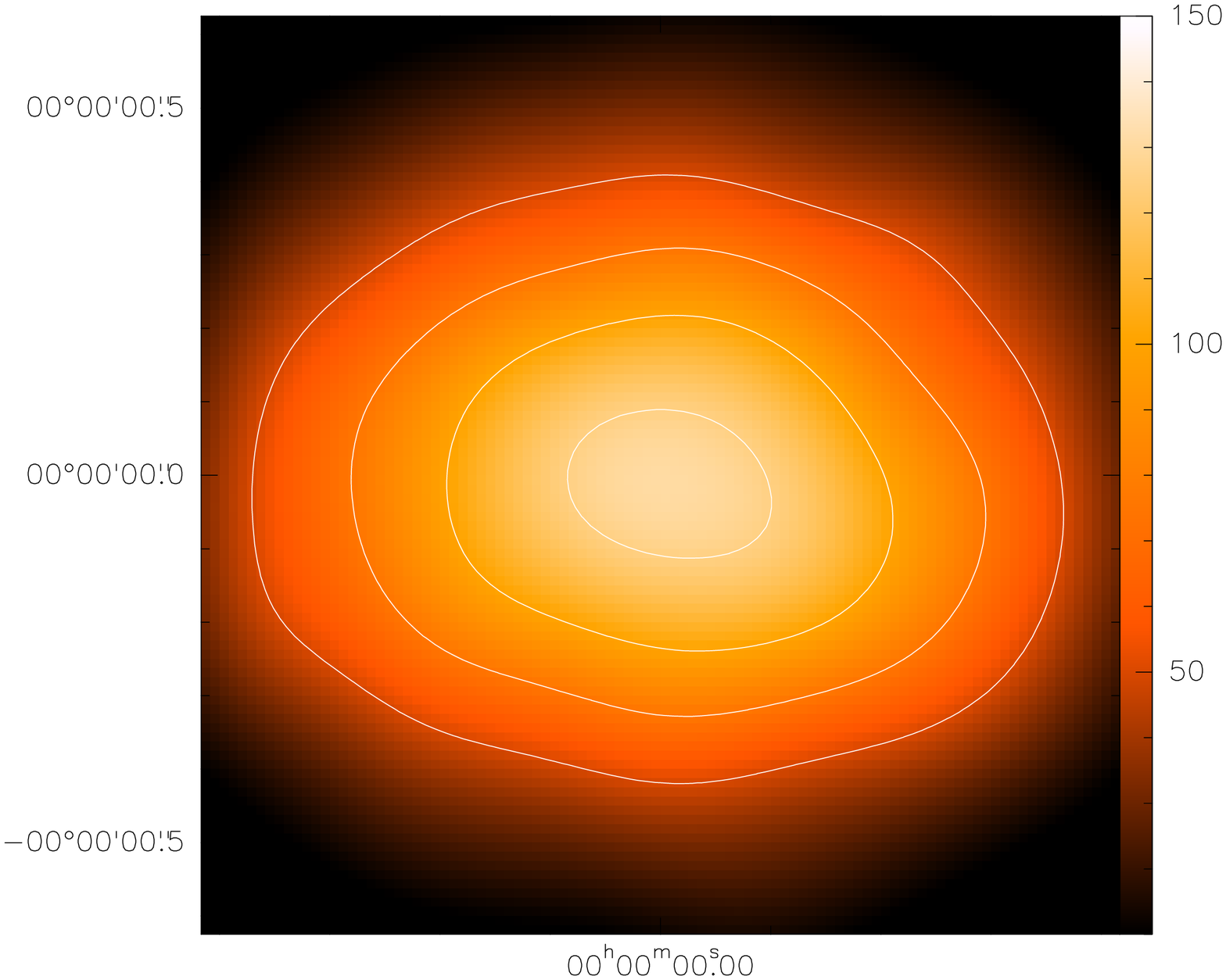}
\includegraphics[angle=0,width=8.5cm]{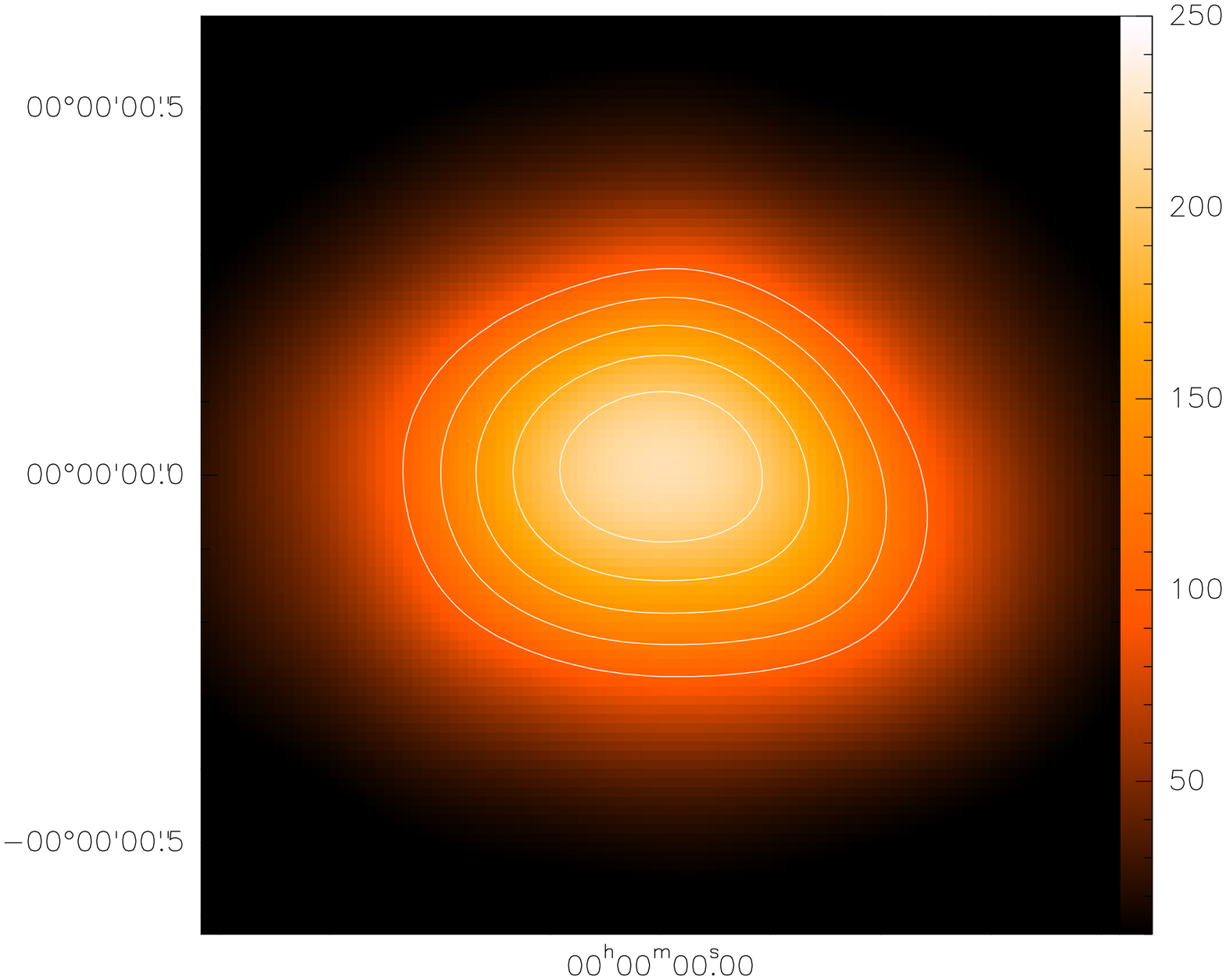}}
\caption{Velocity-integrated line intensities of HCN emission are shown.  {\it Upper left panel:} Model2 HCN(1-0). Contour levels are from 1000 to 2500 K km s$^{-1}$ with each step of 250 K km s$^{-1}$;
 {\it Upper right panel:} Model 2 HCN (3-2). Contour levels are from 1000 to 2000 K km s$^{-1}$ with each step of 250 K km s$^{-1}$;  {\it Lower left panel} Model 4 HCN (1-0). Contour levels are from 50 to 125 K km s$^{-1}$ with each step of 25 K km s$^{-1}$; {\it Lower right panel:} Model 4 HCN (3-2). Contour levels are from 100 to 225 K km s$^{-1}$ with each step of 25 K km s$^{-1}$}
\label{fig:hcn_int}
\end{figure*}

\begin{figure*}
\centerline{\includegraphics[angle=270,width=8.5cm]{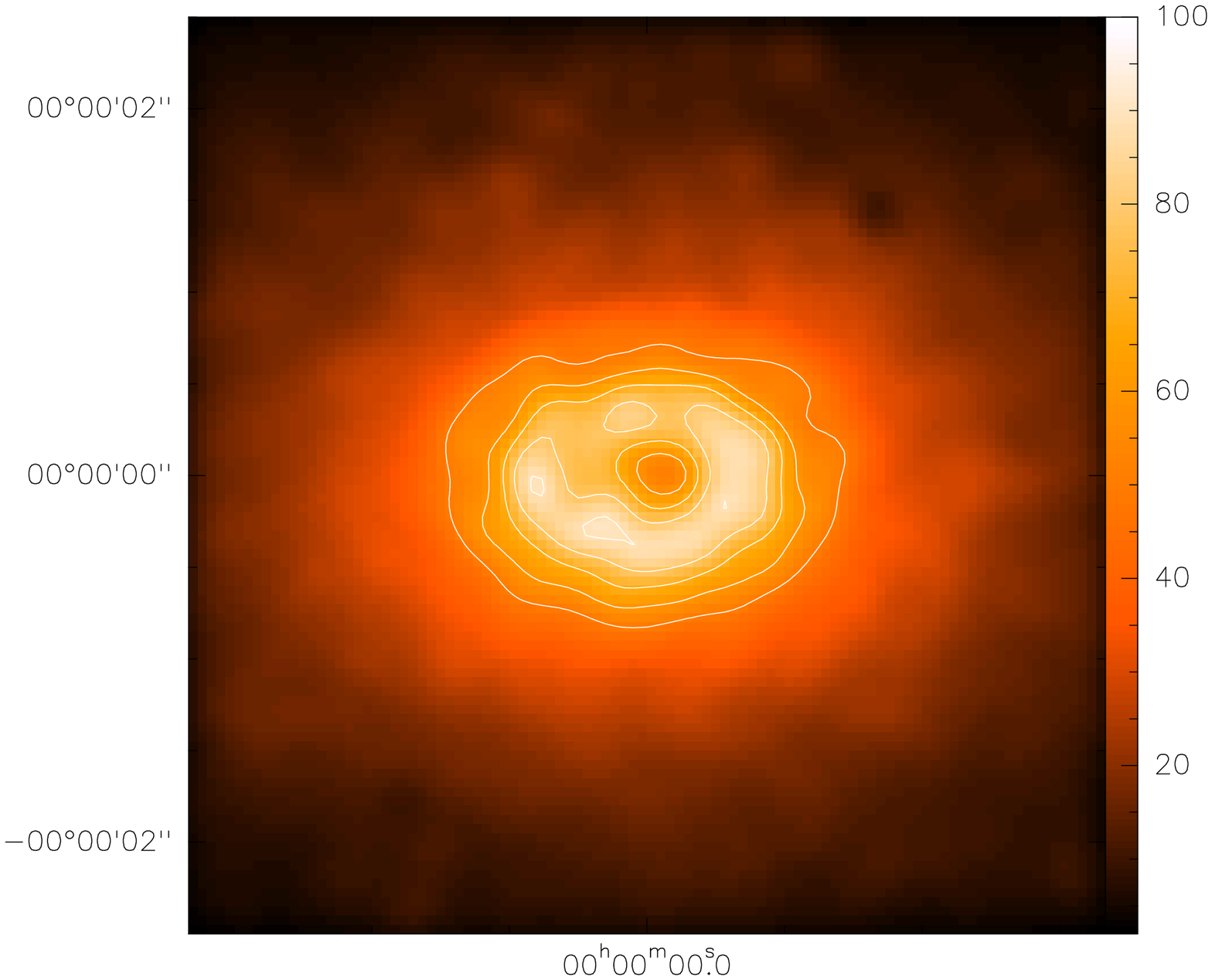}
\includegraphics[angle=270,width=8.5cm]{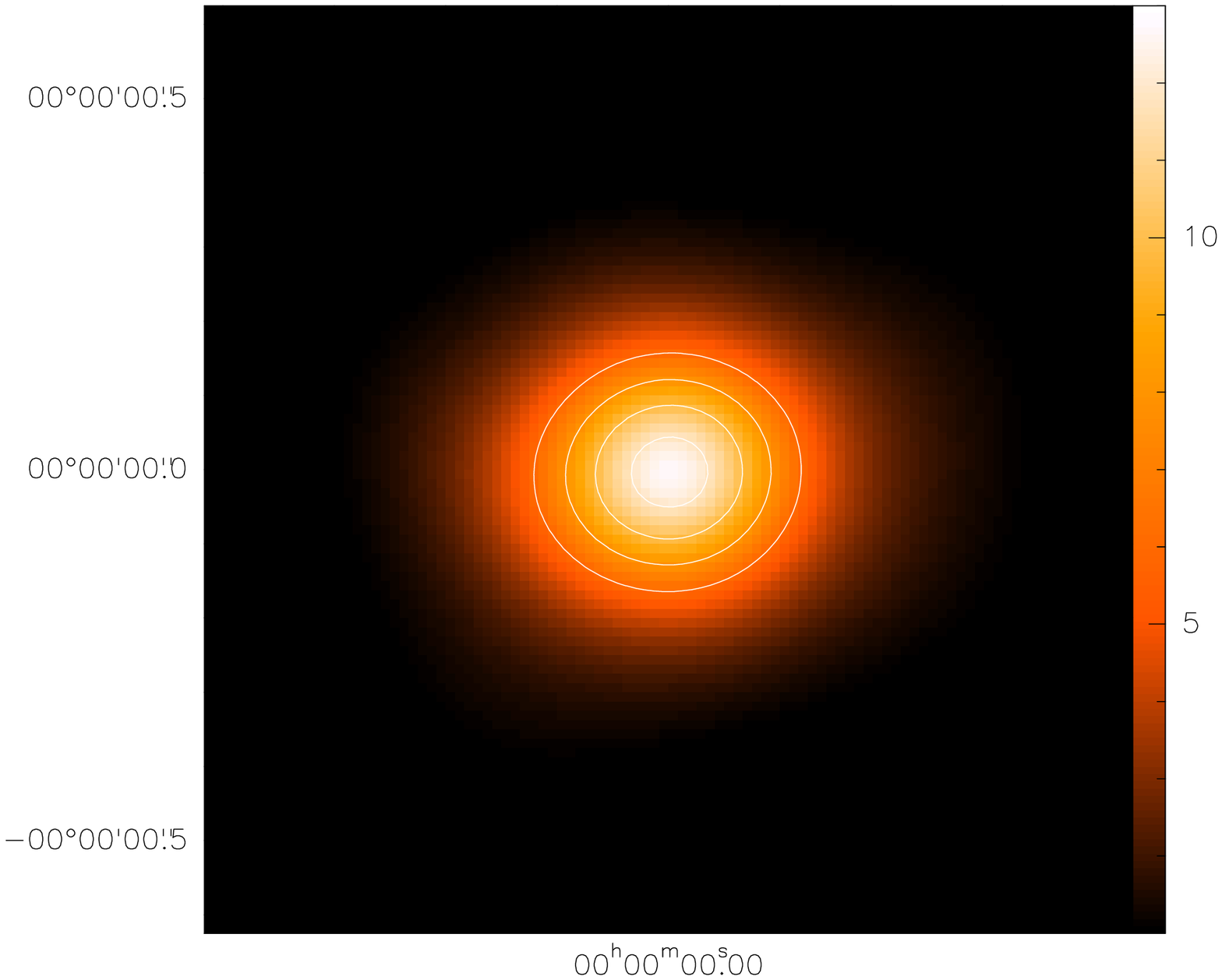}}
\caption{Velocity-integrated line intensities of CN emission are shown.  {\it Upper left panel:} Model2 CN(2$_{5/2}$-1$_{3/2}$). Contour levels are 50, 65, and 80 K km s$^{-1}$.;
 {\it Upper right panel:} Model 4 CN (2$_{5/2}$-1$_{3/2}$). Contour levels are 6, 8, 10, and 12 K km s$^{-1}$.}
\label{fig:hcn_int}
\end{figure*}

\begin{figure*}
\centerline{\includegraphics[angle=0,width=8.5cm]{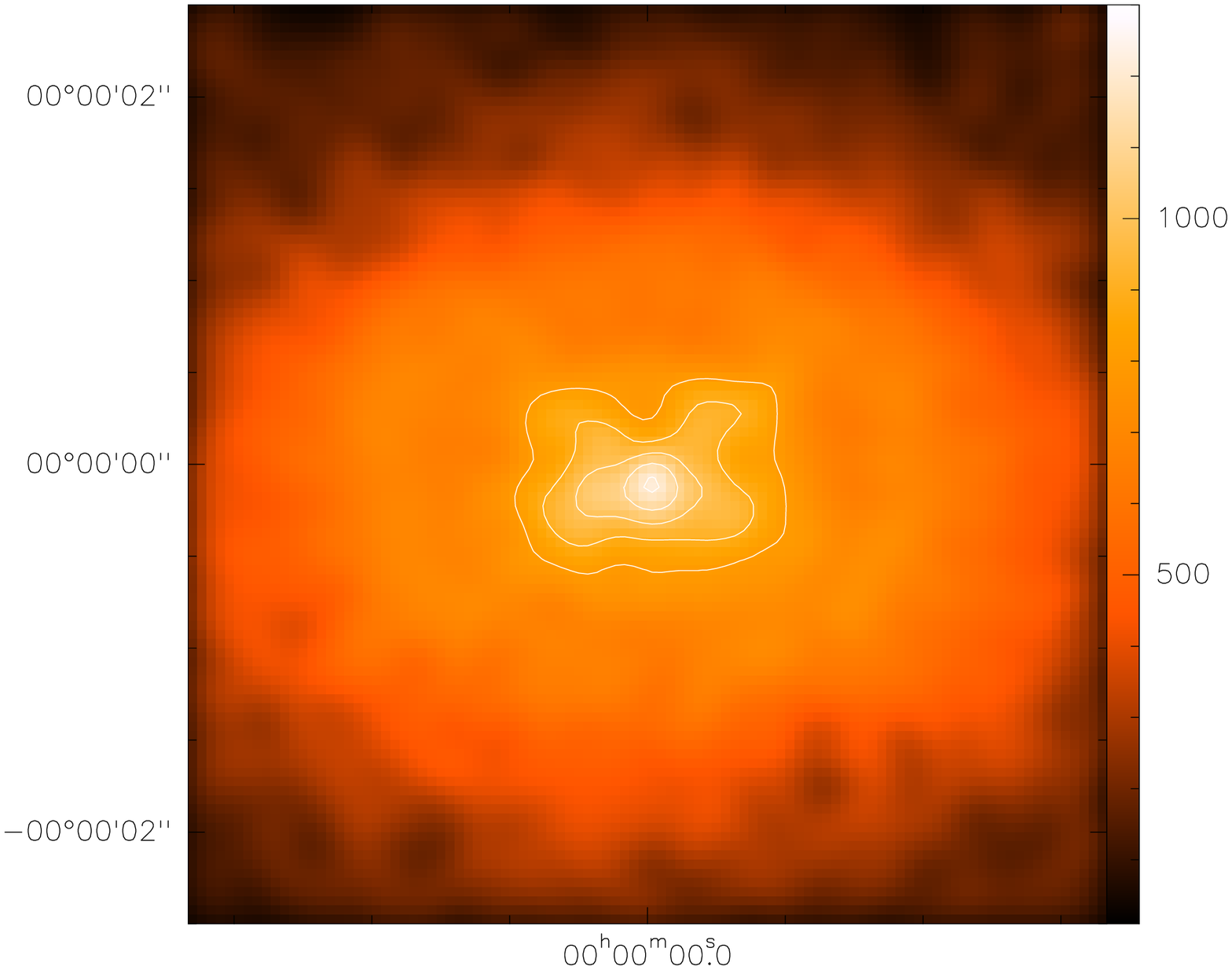}
\includegraphics[angle=0,width=8.5cm]{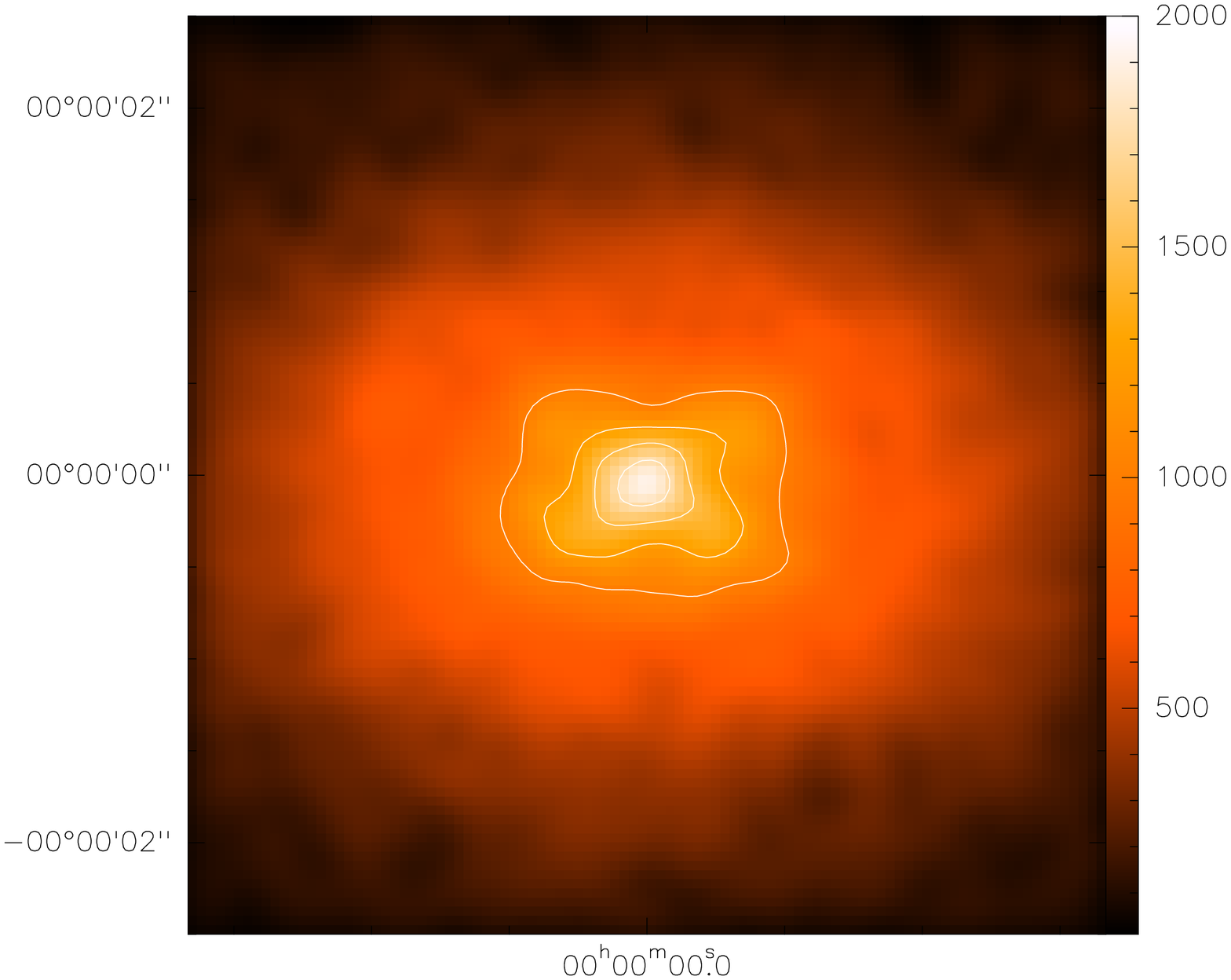}}
\centerline{\includegraphics[angle=0,width=8.5cm]{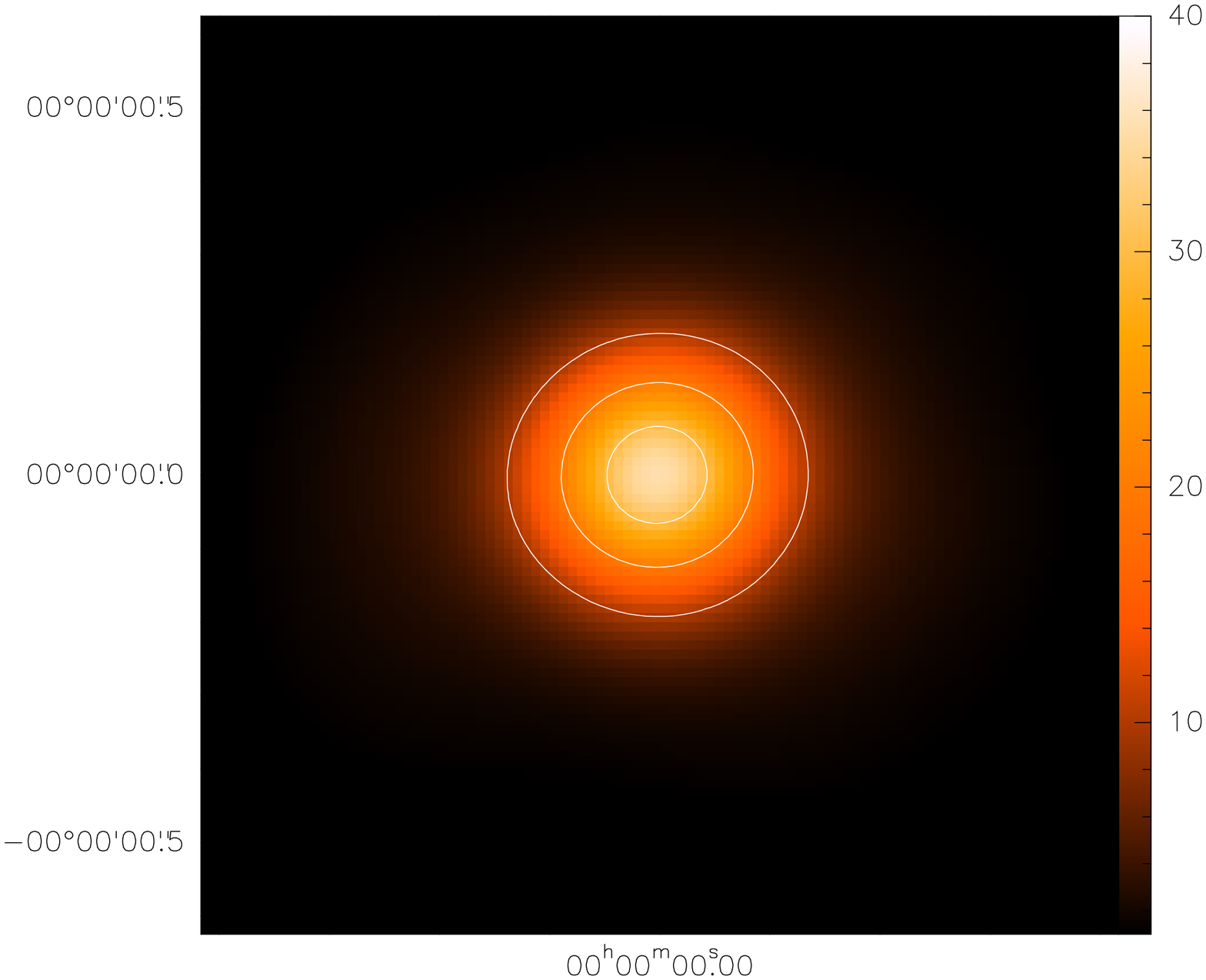}
\includegraphics[angle=0,width=8.5cm]{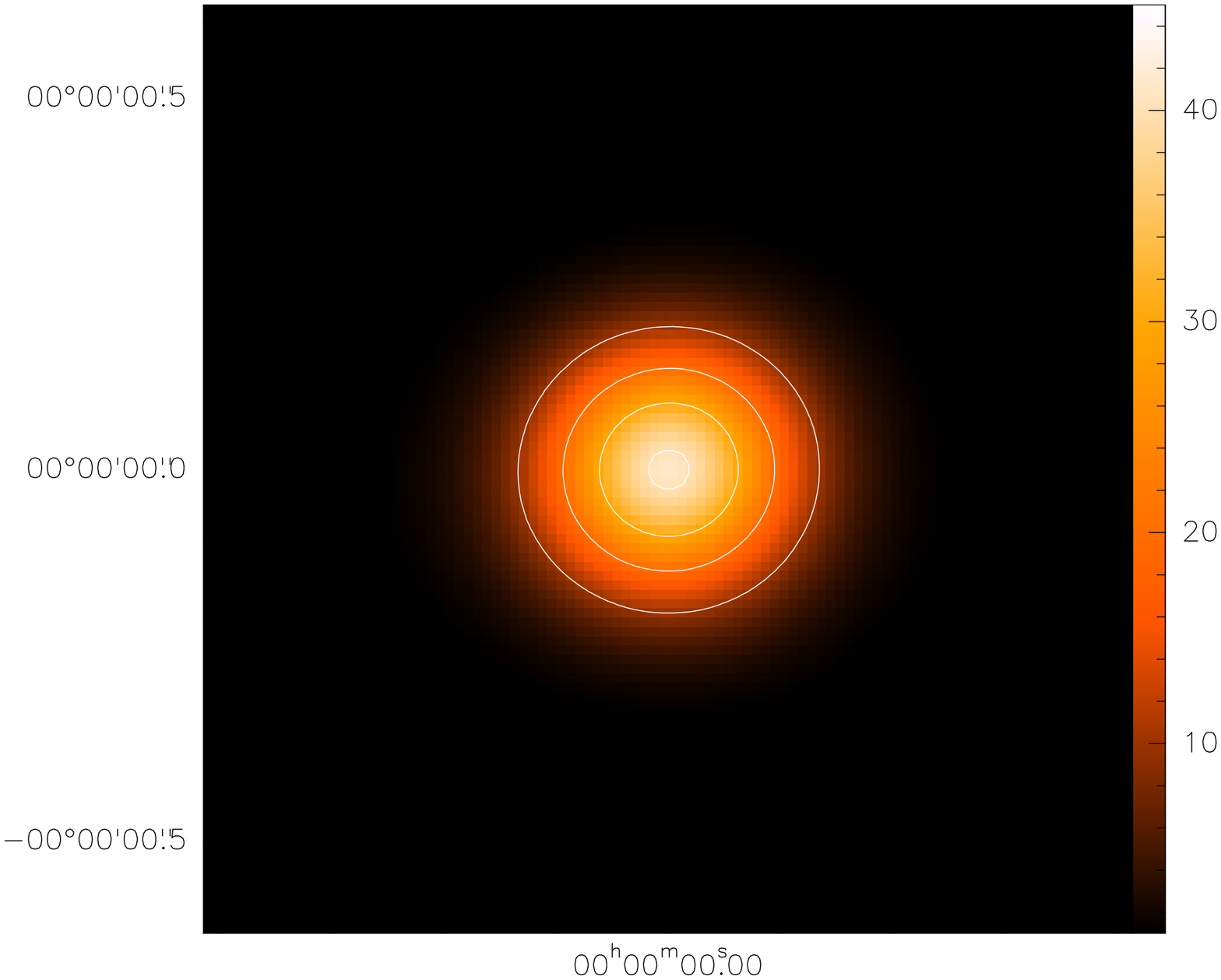}}
\caption{Velocity-integrated line intensities of \cop~emission are shown.  {\it Upper left panel:} Model2 \cop(1-0). Contour levels are from 800 to 1200 K km s$^{-1}$ with each step of 100 K km s$^{-1}$;
 {\it Upper right panel:} Model 2 \cop (3-2). Contour levels are from 1000 to 1750 K km s$^{-1}$ with each step of 250 K km s$^{-1}$;  {\it Lower left panel} Model 4 \cop(1-0). Contour steps are 10, 20, and 30 K km s$^{-1}$.; {\it Lower right panel:} Model 4 \cop(3-2) Contour steps are 10, 20, 30, and 40 K km s$^{-1}$.}
\label{fig:hcop_int}
\end{figure*}

\end{document}